\documentclass[prd,superscriptaddress,twocolumn,nopreprintnumbers,floatfix]{revtex4-1}
\pdfoutput=1

\usepackage{amssymb}
\usepackage{amsmath}
\usepackage{graphicx}
\usepackage{dcolumn}
\usepackage{hyperref}
\usepackage{color,units}
\usepackage{lineno}
\usepackage{xspace}
\usepackage{mathtools}
\usepackage{physics}
\usepackage{tensor}
\usepackage[utf8]{inputenc}
\usepackage{acronym}
\usepackage{multirow}
\usepackage{gensymb}

\newcommand{\bilby}{{\sc Bilby}\xspace}

\newcommand{\lalinference}{{\sc LALInference}\xspace}
\newcommand{\bayeswave}{{\sc BayesWave}\xspace}

\newcommand{\AEI}{\affiliation{Max Planck Institute for Gravitational Physics (Albert Einstein Institute), Am M\"uhlenberg 1, Potsdam 14476, Germany}}
\newcommand{\LIGOlabMIT}{\affiliation{LIGO Laboratory, Massachusetts Institute of Technology, 185 Albany St, Cambridge, MA 02139, USA}}
\newcommand{\MKI}{\affiliation{Department of Physics and Kavli Institute for Astrophysics and Space Research, Massachusetts Institute of Technology, 77 Massachusetts Ave, Cambridge, MA 02139, USA}}

\begin{document}

\title{Ready for what lies ahead? -- 
Gravitational waveform accuracy requirements for future ground based detectors}

\author{Michael P\"urrer}
\email{michael.puerrer@aei.mpg.de}
\AEI

\author{Carl-Johan Haster}
\email{haster@mit.edu}
\LIGOlabMIT \MKI

\date{\today}

\begin{abstract}
Future third generation (3G) ground-based gravitational wave (GW) detectors, such as the Einstein Telescope and Cosmic Explorer, will have unprecedented sensitivities enabling studies of the entire population of stellar mass binary black hole coalescences in the Universe, while the A+ and Voyager upgrades to current detectors will significantly improve over advanced LIGO \& Virgo design sensitivities. 
To infer binary parameters from a GW signal we require accurate models of the gravitational waveform as a function of black hole masses, spins, etc. 
Such waveform models are built from numerical relativity (NR) simulations and/or semi-analytical expressions in the inspiral. 
We investigate the limits of the current waveform models and study at what detector sensitivity these models will yield unbiased parameter inference for loud ``golden'' binary black hole systems, what biases we can expect beyond these limits, and what  implications such biases will have for GW astrophysics. 
For 3G detectors we find that the mismatch error for semi-analytical models needs to be reduced by at least \emph{three orders of magnitude} and for NR waveforms by \emph{one order of magnitude}.
In addition, we show that for a population of one hundred high mass precessing binary black holes, measurement errors sum up to a sizable population bias, about 10 -- 30 times larger than the sum of 90\% credible intervals for key astrophysical parameters.
Furthermore we demonstrate that the residual signal between the GW data recorded by a detector and the best fit template waveform obtained by parameter inference analyses can have significant signal-to-noise ratio and can lead to Bayes factors as high as $10^{11}$ between a coherent and an incoherent wavelet model for the population events. 
This coherent power left in the residual could lead to the observation of erroneous deviations from general relativity.
To address these issues and be ready to reap the scientific benefits of 3G GW detectors in the 2030s, waveform models that are significantly more physically complete and accurate need to be
developed in the next decade along with major advances in efficiency and accuracy of NR codes.
\end{abstract}


\maketitle

\acrodef{LSC}[LSC]{LIGO Scientific Collaboration}
\acrodef{aLIGO}{Advanced Laser Interferometer Gravitational wave Observatory}
\acrodef{aVirgo}{Advanced Virgo}
\acrodef{LIGO}[LIGO]{Laser Interferometer Gravitational-Wave Observatory}
\acrodef{IFO}[IFO]{interferometer}
\acrodef{LHO}[LHO]{LIGO-Hanford}
\acrodef{LLO}[LLO]{LIGO-Livingston}
\acrodef{O2}[O2]{second observing run}
\acrodef{O1}[O1]{first observing run}
\acrodef{BH}[BH]{black hole}
\acrodef{BBH}[BBH]{binary black hole}
\acrodef{BNS}[BNS]{binary neutron star}
\acrodef{NS}[NS]{neutron star}
\acrodef{BHNS}[BHNS]{black hole--neutron star binaries}
\acrodef{NSBH}[NSBH]{neutron star--black hole binary}
\acrodef{PBH}[PBH]{primordial black hole binaries}
\acrodef{CBC}[CBC]{compact binary coalescence}
\acrodef{GW}[GW]{gravitational wave}
\acrodef{CWB}[cWB]{coherent WaveBurst}
\acrodef{SNR}[SNR]{signal-to-noise ratio}
\acrodef{FAR}[FAR]{false alarm rate}
\acrodef{IFAR}[IFAR]{inverse false alarm rate}
\acrodef{FAP}[FAP]{false alarm probability}
\acrodef{PSD}[PSD]{power spectral density}
\acrodef{ASD}[ASD]{amplitude spectral density}
\acrodef{GR}[GR]{general relativity}
\acrodef{NR}[NR]{numerical relativity}
\acrodef{PN}[PN]{post-Newtonian}
\acrodef{EOB}[EOB]{effective-one-body}
\acrodef{ROM}[ROM]{reduced-order-model}
\acrodef{IMR}[IMR]{inspiral-merger-ringdown}
\acrodef{PDF}[PDF]{probability density function}
\acrodef{PE}[PE]{parameter estimation}
\acrodef{CL}[CL]{credible level}
\acrodef{EOS}[EOS]{equation of state}
\acrodef{LAL}[LAL]{LSC Algorithm Library}
\acrodef{ET}[ET]{Einstein Telescope}
\acrodef{CE}[CE]{Cosmic Explorer}
\acrodef{MAP}[MAP]{maximum a posteriori}
\acrodef{KDE}[KDE]{kernel density estimate}
\acrodef{CDF}[CDF]{cumulative distribution function}
\acrodef{CCE}[CCE]{Cauchy characteristic extraction}
\acrodef{GPR}[GPR]{Gaussian process regression}
\acrodef{PN-NR}[PN-NR]{post-Newtonian - numerical relativity}

\newcommand{\PN}[0]{\ac{PN}\xspace}
\newcommand{\BBH}[0]{\ac{BBH}\xspace}
\newcommand{\BNS}[0]{\ac{BNS}\xspace}
\newcommand{\BH}[0]{\ac{BH}\xspace}
\newcommand{\NR}[0]{\ac{NR}\xspace}
\newcommand{\GW}[0]{\ac{GW}\xspace}
\newcommand{\SNR}[0]{\ac{SNR}\xspace}
\newcommand{\aLIGO}[0]{\ac{aLIGO}\xspace}
\newcommand{\PE}[0]{\ac{PE}\xspace}
\newcommand{\IMR}[0]{\ac{IMR}\xspace}
\newcommand{\PDF}[0]{\ac{PDF}\xspace}
\newcommand{\GR}[0]{\ac{GR}\xspace}
\newcommand{\PSD}[0]{\ac{PSD}\xspace}
\newcommand{\EOS}[0]{\ac{EOS}\xspace}

\newcolumntype{d}[1]{D{.}{.}{#1}}

\section{Introduction}
\label{sec:intro}

\begin{figure}[h]
  \centering
    \includegraphics[width=.5\textwidth]{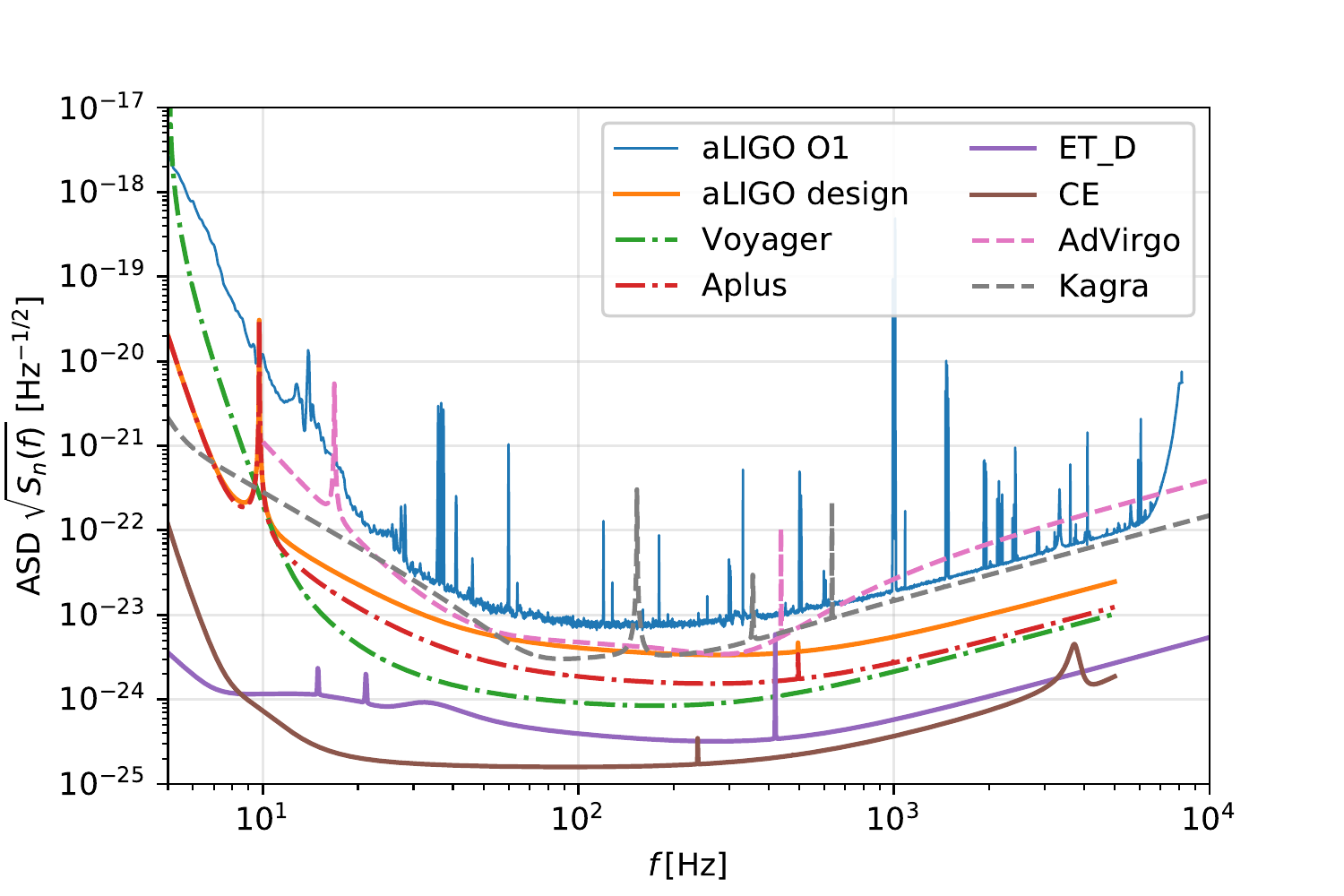}
  \caption{Evolution of sensitivity of GW interferometric detectors.
  Text data files for the \acp{PSD} or \acp{ASD} can be found in~\cite{ASD_curves18, aLIGO_design_updated} under the names given in this table.
  }
  \label{fig:plots_ASDs_evolution}
\end{figure}

Observations of \acp{GW} from coalescing compact object binaries have revolutionized our knowledge about the Universe and provided access to astrophysics previously outside our grasp~\cite{LIGOScientific:2018mvr}.
These observations were made possible by the construction and operation of a network of \ac{GW} detectors, Advanced LIGO~\cite{TheLIGOScientific:2014jea}, Advanced Virgo~\cite{TheVirgo:2014hva} and KAGRA~\cite{Aso:2013eba}.
As expected from a relatively young field of observational astrophysics, there are still significant technological improvements within reach~\cite{InstrumentSciencePaper,ObservingScenarios,AplusGrant} increasing the sensitivity of both current generation \ac{GW} detectors over the next $\sim 5$ years~\footnote{Funding has already been awarded to upgrade the LIGO detectors as part of the A+ project~\cite{AplusGrant}}
 in addition to paving the way for next-generation ground based facilities~\cite{Punturo:2010zz,Evans:2016mbw,Reitze:2019iox,Reitze:2019dyk} as well as space-based observatories~\cite{Audley:2017drz, Luo:2015ght}, all planned to be operational in the early 2030s.
This will allow for direct observation of all stellar-mass \ac{BBH} mergers throughout the cosmological history of the Universe~\cite{Kalogera:2019sui} and will enable unprecedented and unique science in extreme gravity and fundamental physics~\cite{Sathyaprakash:2019yqt}.
Whereas the vast majority of observed \ac{GW} signals will have originated at large cosmological distances~\cite{Vitale:2018yhm, Farr:2019twy}, binaries from the currently observable volume of the Universe will, as the detector sensitivities improve (see Fig.~\ref{fig:plots_ASDs_evolution}), be observable with increasing fidelity.
\ac{GW} models are crucial for elucidating the astrophysical properties of compact binaries. For this increase in the information available for a typical \ac{BBH} observation, these models need to satisfy stricter accuracy requirements.
The currently available model waveforms, which approximates the solutions to the 2-body problem in \ac{GR}, have been shown to be sufficiently accurate to not cause any systematic biases in the recovered parameters (masses, spins, location in the Universe etc.) for \acp{BBH} observed so far~\cite{Abbott:2016wiq}.
These ``golden binaries'', stellar-mass \acp{BBH} like GW150914 (the first direct \ac{GW} detection~\cite{Abbott:2016blz}) observed at high \ac{SNR} and with high fidelity, have also allowed to perform previously inaccessible tests of \ac{GR}~\cite{LIGOScientific:2019fpa} and the robustness of the waveform models used.
As the \ac{GW} detectors improve, the expected \ac{SNR} for \acp{BBH} from the local Universe will increase correspondingly, thus reducing the statistical uncertainties in the recovered source parameters.
As the statistical uncertainties approach the inherent systematic uncertainties of the \ac{GW} approximant models, parameter biases will eventually appear and reduce the reliability of future \ac{GW} observations.

In this study we want to investigate the appearance and significance of these parameter biases, their connection to the accuracy of the \ac{GW} models used and what relevance the biases will have on future astrophysical statements based on high \ac{SNR} observations of \ac{BBH} systems~\cite{Cutler:2007mi}.
In addition to possible biases in the source parameters of the \ac{BBH}, the use of \ac{GW} observations as means to test \ac{GR} puts even more stringent requirements on \ac{GW} model accuracy~\cite{Yunes:2009ke,LIGOScientific:2019fpa,Berti:2018cxi,Berti:2018vdi}.
If there are effects of beyond-\ac{GR} theories embossed on the ``raw'' \ac{GR} waveform, these effects will be scrambled by any residual signal left by an inaccurate \ac{GW} model and thus will limit the strength of the \ac{GR} test.
Even worse, inaccurate \ac{GW} models may lead to erroneous results claiming deviations from \ac{GR}.
Many of these analyses, both \ac{PE} studies and tests of \ac{GR}, are strongly dependent on robust observation of the two polarisation states of a \ac{GW} signal as described by \ac{GR}~\cite{Eardley:1974nw}.
This is primarily done by requiring a coherent observation of a given \ac{GW} signal in more than one detector~\cite{Vitale:2016icu,Vitale:2016avz,Vitale:2018nif}, which is also crucial for accurate and reliable localisation of the \ac{GW} source in the Universe.
Whereas the \acp{BBH} investigated here are not expected to produce any observable counterpart~\cite{Abbott:2016gcq, Doctor:2018ray,LIGOScientific:2019gag}, precise localisation is crucial for cosmology studies~\cite{Schutz:1986gp,Abbott:2019yzh, Soares-Santos:2019irc, Gray:2019ksv, Chen:2017rfc} as well as for inferring the parameters of the \acp{BBH} in their source frame~\footnote{\ac{GW} observations measure a signal that has been redshifted during its propagation, thus changing the recovered \ac{BH} masses. The source-frame parameters are inferred by taking the measured luminosity distance, assuming a cosmology and applying the appropriate redshift factor to the measured detector-frame parameters}.

In addition to biases caused by inaccurate \ac{GW} models, the data generated by the detectors themselves carry inherent uncertainties originating from the calibration process applied to the raw detector output~\cite{Cahillane:2017vkb} as well as imprecise modelling assumptions for the noise processes of a given detector system~\cite{Littenberg:2014oda}.
These types of uncertainties can however already be quantified, and thus incorporated into the \ac{PE} infrastructure allowing their effects to be marginalised out from the final inferred parameter distributions~\cite{Veitch:2014wba,TheLIGOScientific:2016pea,LIGOScientific:2018mvr}.
The marginalisation over calibration uncertainties, and similarly the marginalisation over eventual uncertainties in the noise estimation,
is primarily expected to broaden the recovered posterior distributions thus effectively absorbing any misestimate in the \ac{GW} amplitude or phase irrespective of whether it originated from uncertainties in the data itself or from the assumed waveform model.

The rest of the paper presents the details of our study.
In Sec.~\ref{sec:methods} we describe the \ac{GW} models used, together with details on the analysis methods.
In Sec.~\ref{sec:results-golden-binaries} we report our findings on the analysis of individual ``golden binary'' \ac{BBH} signals, including requirements on the accuracy of the \ac{GW} models and the consequences any inaccuracies will entail.
In Sec.~\ref{sec:population} we explore a population of \ac{BBH} observations, and what effects \ac{GW} model accuracy will have on the properties of the inferred population.
Finally in Sec.~\ref{sec:discussion} we discuss our findings and present an outlook for how to tackle the issues we have presented.

\section{Methodology}
\label{sec:methods}

In this section we introduce the methods we use to study the impact of waveform inaccuracies on measurements of compact binary parameters from \acp{GW} in Secs.~\ref{sec:results-golden-binaries} and~\ref{sec:population}.
We first discuss common data analysis tools for \ac{GW} waveforms in Sec.~\ref{sub:tools} and
\ac{NR} waveforms, the most accurate waveforms we have available for the solutions of the 2-body problem in \ac{GR}, in Sec.~\ref{sub:numerical_relativity_waveforms}. 
In addition to \ac{NR} waveforms, we also use \ac{PN-NR} hybrid waveforms as described in Sec.~\ref{sub:hybrid_waveforms} to more fully fill the band of more sensitive detectors to lower frequencies as mock signals in this study.
In Sec.~\ref{sub:waveform_models}, we discuss fast, but approximate semi-analytic models of the \acp{GW} emitted from compact binaries. 
These models are crucial to infer binary properties with Bayesian parameter estimation methods for single events and populations of binaries, as discussed in Sec.~\ref{sub:bayesian_parameter_estimation}.

\subsection{Gravitational waveforms and overlaps} 
\label{sub:tools}

Gravitational waveforms are often decomposed in a basis of spherical harmonics 
$_{-2}Y_{lm}$. The two \ac{GW} polarizations can be expanded into modes as
\begin{equation}
  \label{eq:GWmodes}
  h_+ - i h_\times = \sum_{\ell,m} h_{\ell m} \; {}_{-2}Y_{lm}.
\end{equation}

We define the overlap, or match, between two waveforms $h_1$ and $h_2$ as
\begin{equation}
  \label{eq:overlap}
  \mathcal{O}(h_1, h_2) := \frac{\langle h_1 | h_2 \rangle}
                                {\sqrt{\langle h_1 | h_1 \rangle \langle h_2 | h_2 \rangle}},
\end{equation}
where 
\begin{equation}
  \label{eq:inner-product}
  \langle h_1 | h_2 \rangle = 4 \mathrm{Re} \int_{f_\mathrm{low}}^{f_\mathrm{high}} 
                              \frac{\tilde h_1(f) \tilde h_2^*(f)}{S_n(f)} df
\end{equation}
represents a noise-weighted inner product with \ac{PSD} $S_n$, and ${}^*$ denotes complex conjugation.
We also maximize the overlap over a time and phase-shift between the two waveforms.
We often quote the mismatch, $1 - \mathcal{O}(h_1, h_2)$ instead of the overlap.

\subsection{Numerical relativity waveforms} 
\label{sub:numerical_relativity_waveforms}

We use two \ac{NR} simulations of binary black hole coalescences by the SXS Collaboration~\cite{Boyle:2019kee} using the Spectral Einstein Code (SpEC)~\cite{SpEC} code.
The first simulation, SXS\_BBH\_0308~\cite{SXS_BBH_0308,Boyle:2019kee}, was performed at parameters inferred from the LIGO \ac{PE} analysis of GW150914 with semi-analytic waveform models~\cite{Abbott:2016blz,TheLIGOScientific:2016wfe} and was subsequently used to study possible effects of waveform systematics on the inferred parameters~\cite{Abbott:2016wiq,Abbott:2016apu}. 
The waveform describes a nearly equal mass binary with small effective aligned spin and moderate precession (see Table~\ref{tab:signal-configurations}). 
The waveform accumulates $12.6$ orbits and a length of $2822 M$ in time before the formation of a common horizon. 
The mismatch between simulations at different resolutions at the total mass of GW150914 with aLIGO design sensitivity is $\sim 2 \times 10^{-4}$. 
We use the highest resolution available, Lev5.

The second simulation, SXS\_BBH\_0104~\cite{SXS_BBH_0104,Mroue:2013xna}, is at mass-ratio 1:3 and has some effective aligned and precession spin. 
Systems at this mass-ratio still lie within the population posterior for the mass-ratio that has been found in the LIGO \& Virgo O1 \& O2 analysis~\cite{LIGOScientific:2018mvr}.
The waveform accumulates $21.9$ orbits and a length of $5192 M$ before the formation of a common horizon. 
There is only a single resolution, Lev5, available for this simulation. 
An estimate for the mismatch for a simulation using similar technology (SXS\_BBH\_0053~\cite{SXS_BBH_0053,Mroue:2013xna}) gives $\sim 10^{-3}$ at the total mass of GW150914 with aLIGO design sensitivity.

These waveforms are for quasi-circular inspirals and mergers of \acp{BBH}. 
Since initial conditions are not exactly known, there is a low amount of residual eccentricity in these simulations. 
For SXS\_BBH\_0308 eccentricity at the relaxed time is estimated to be $\sim 0.0005$ while for SXS\_BBH\_0104 it is $\sim 0.001$. 
We do not consider the effect of eccentricity in the waveform in this study.


\subsection{Hybrid waveforms} 
\label{sub:hybrid_waveforms}

We use an extension of the \texttt{GWFrames}~\cite{GWFrames,Boyle:2013nka} package to hybridize \ac{NR} with \ac{PN} waveforms. 
First we read in an \ac{NR} waveform and its horizon data (i.e. the spins and orbital track data computed from the apparent horizon finder). 
We generate a \ac{PN} waveform at the physical parameters of the \ac{NR} configuration and align it by shifting in time and attitude to match the \ac{NR} waveform. 
The waveform modes and the quaternions describing the motion of the inertial frame are then blended over a hybridization region in time. 
More details about the procedure and the hybrid waveforms are given in App.~\ref{app:hybrids}.

Estimates of the accuracy of hybrid PN-NR waveforms are difficult to obtain.
Hybrid errors are expected to be significantly higher than for pure \ac{NR} waveforms due to errors in the \ac{PN} part of the waveform and additional errors from smoothly combining the \ac{PN} and \ac{NR} waveform modes over a blending window in time~\cite{Hannam:2010ky,MacDonald:2011ne,Boyle:2011dy,Ajith:2012az}. 
We show in App.~\ref{app:hybrids} that hybridization errors are lower than \ac{NR} error 
estimates for the simulations considered in this study.
Semi-analytical waveform models usually have good accuracy in the inspiral and are less 
accurate near merger. Therefore, the \ac{PN}-\ac{NR} hybrids used as mock signals in this 
study should be much more accurate than the semi-analytic waveform models described in
Sec.~\ref{sub:waveform_models} which we use as template waveforms.


\subsection{Waveform models} 
\label{sub:waveform_models}

In this study we use two fast frequency domain waveform models as template waveforms. 
These are the \texttt{IMRPhenomPv2}~\cite{Hannam:2013oca,PhenomPv2_tech_rep} and \texttt{SEOBNRv4\_ROM}~\cite{Bohe:2016gbl} \ac{IMR} models. 

\texttt{IMRPhenomPv2} uses the aligned-spin \texttt{IMRPhenomD}~\cite{Khan:2015jqa, Husa:2015iqa} model as a base waveform in the co-precessing frame and twists up its $(2,\pm 2)$ modes with a \ac{PN} prescription of the Euler angles that describe the motion of the inertial frame for precessing black hole binaries, thus generating all $\ell = 2$ modes~\cite{Schmidt:2012rh,Schmidt:2014iyl}. 
The model also assumes that the opening angle of the precession cone is small~\cite{PhenomPv2_tech_rep} which make it most suitable for binaries with small to moderate precession and moderate mass-ratios. 
The model has been shown to be smooth ~\cite{Smith:2016qas} up to mass-ratio $q \sim 1/4$.

\texttt{SEOBNRv4\_ROM} is a frequency domain reduced order model of the time domain \texttt{SEOBNRv4} effective-one-body model~\cite{Bohe:2016gbl} using the methodology developed in~\cite{Purrer:2014fza,Purrer:2015tud}. 
The model describes the $(2,\pm 2)$ modes for non-precessing binaries and can be used for a wide range in mass-ratio and BH spin magnitudes up to maximal spin.

Both \texttt{IMRPhenomD} which underlies \texttt{IMRPhenomPv2} and \texttt{SEOBNRv4} have been tuned to \ac{NR} waveforms in the non-precessing sector.
While more complete models in terms of precession are available~\cite{Pan:2013rra, Babak:2016tgq, Knowles:2018hqq} we were not able to use them for this computationally demanding study because we could not obtain converged posterior distributions in time. 
Models that also include higher harmonics~\cite{SEOBNRv4P} or are computationally more efficient~\cite{Khan:2018fmp} are now becoming available. 

In the population study described in Sec.~\ref{sec:population} we use \texttt{NRSur7dq2} to represent the population of astrophysical signals~\cite{Blackman:2017pcm}. 
These signals were stochastically drawn and thus we could not use \ac{NR} simulations which are only available at specific points in parameter space.
The \texttt{NRSur7dq2} \ac{NR}-surrogate model is however a very good approximation to \ac{NR} waveforms. 
It describes generic precessing systems with mass-ratios up to $q = 1/2$ and spin magnitudes of $0.8$. 
\texttt{NRSur7dq2} is built from multiple surrogates that model waveform mode combinations in the co-orbital frame, the averaged frequency of the $(2,\pm 2)$ modes in the co-precessing frame, and the frame motion through the right hand sides of the precession equations.
We intended to also use \texttt{NRSur7dq2} as a template waveform for the study discussed in Sec.~\ref{sec:results-golden-binaries}, but, while being very accurate, this model has a limited length and this severely limits the mass space that can be explored to high mass systems and high starting frequencies.


\subsection{Bayesian parameter estimation} 
\label{sub:bayesian_parameter_estimation}

The inference of the source parameters $\vec{\theta}$ of a \ac{GW} signal is expressed as a posterior \ac{PDF} $p(\vec{\theta}|d(t))$ as part of a \ac{PE} analysis given the data $d(t)$ recorded from the detectors.
Through application of Bayes' theorem, $p(\vec{\theta}|d(t))$ is directly proportional to the likelihood $\mathcal{L}(d(t)|\vec{\theta})$ of observing the data given an assumed waveform model $h(t; \vec{\theta})$, in turn characterized by the source parameters $\vec{\theta}$, together with the prior probability $\pi(\vec{\theta})$.

For the analysis of the ``golden binaries'' in Sec.~\ref{sec:results-golden-binaries} this prior is defined to be uniform over the two-dimensional space defining the masses of the binary objects, $m_1$ and $m_2$ (with $m_1 \geq m_2$), as observed in the rest frames of the \ac{GW} detectors.
The dimensionless spins of the \acp{BH} are assumed to follow a prior uniform in spin magnitude (between 0 and 1) allowing for isotropic and uncorrelated directions of the two black hole spins.
We also assume an isotropic prior for the location of the \ac{GW} on the sky, and a distance prior corresponding to a homogenous rate density in the nearby Universe. 
For these analyses, we disregard any cosmological corrections to the rate density which for the redshifts explored ($z\sim 0.1$) are expected to be negligible. 
The orientation of the binary follows a prior probability uniform in the polarisation angle $\psi$ and in the cosine of $\theta_{JN}$, the angle between the total angular momentum $\mathbf{J}$ and the line of sight $\mathbf{N}$.
The parameter space defined by $\pi(\vec{\theta})$ is, for the golden binaries analysed in Sec.~\ref{sec:results-golden-binaries}, explored stochastically using a Markov Chain Monte Carlo code implemented as part of the \lalinference package~\cite{Veitch:2014wba,LALInference-code} available as part of the LSC Algorithm Library (LAL)~\cite{lalsuite}.

For the analysis of the \ac{BBH} population in Sec.~\ref{sec:population} the \bilby inference package was used~\cite{Ashton:2018jfp,Bilby-code} exploring the parameter space using the Nested Sampling algorithm \textsc{dynesty}~\cite{Dynesty}.
Here, similar parameterizations and prior assumptions as for the analysis in Sec.~\ref{sec:results-golden-binaries} were made.
The analyses however differ in their assumptions over \ac{BH} masses, here using a prior uniform in the binary chirp mass $\mathcal{M} = (m_1 m_2)^{3/5}/(m_1 + m_2)^{1/5}$ and the asymmetric mass ratio $q = m_2/m_1$ as well as assuming a prior on distance that is uniform in co-moving volume.
The different prior choices between Sec.~\ref{sec:results-golden-binaries} and Sec.~\ref{sec:results-golden-binaries} are not expected to have significant impact on the recovered parameters, or on the conclusions about waveform accuracy requirements based on this inference.

In a multi-detector \ac{PE} analysis we project the signal and template waveforms on the interferometric \ac{GW} detectors and compute the strain from the waveform polarizations ($+$ and $\times$) and their corresponding detector antenna pattern functions~\cite{Cutler:1994ys}
\begin{equation}
  \label{eq:detector-response}
  h(t; \vec{\theta}) = h_+(t; \vec{\theta}) F_+(\mathrm{ra}, \mathrm{dec}, \psi) + h_\times(t; \vec{\theta}) F_\times(\mathrm{ra}, \mathrm{dec}, \psi).
\end{equation}
As the focus of this study is on effects of accuracy of the waveform themselves, the signal waveforms representing the true \ac{GW} signals are added to a time-series containing no noise, 
as the standard assumption of Gaussian noise could introduce random biases in the recovered parameters.
The true \ac{GW} strain $h(t; \vec{\theta})$ as emitted by the \ac{GW} source may however differ from $h^M(t; \vec{\theta})$, the strain measured by the detectors, due to uncertainties in the calibration of the detectors and their recording of the \ac{GW} strain~\cite{Cahillane:2017vkb, Viets:2017yvy}.
We can model the relation between the measured and true strain as
\begin{align}
  \label{eq:miscalibrated-strain}
\tilde{h}^M(f; \vec{\theta}) = \tilde{h}(f; \vec{\theta})\left[1+ \delta A(f; \vec{\theta}^{\rm cal})\right] \exp{i\, \delta\phi(f; \vec{\theta}^{\rm cal})}\,,
\end{align}
for $\tilde{h}(f; \vec{\theta})$ and $\tilde{h}^M(f; \vec{\theta})$ where the tilde denotes the Fourier transforms of the time-domain strain $h(t; \vec{\theta})$ and $h^M(t; \vec{\theta})$ respectively.
The uncertainty in the strain amplitude and phase, caused by uncertainties in the detector calibration, are characterized by the terms $\delta A(f; \vec{\theta}^{\rm cal})$ and $\delta\phi(f; \vec{\theta}^{\rm cal})$ that are nominally expected to vary both across the bandwidth used for the observation as well as over time from observation to observation.
The frequency-dependent correction factors are modelled as cubic splines with nodes spaced uniformly in $\log f$, each with an independent $\delta A$ and $\delta\phi$ parameter~\cite{SplineCalMarg-T1400682} which are then numerically marginalised over.
For this study, we assume zero-mean Gaussian priors on $\delta A$ with a standard deviation of $1\%$ ($5\%$ for the O1 analysis) and for $\delta\phi$ a standard deviation of $1^{\circ}$ ($5^{\circ}$ for the O1 analysis).
The magnitude of these amplitude and phase uncertainties are consistent with the performance of the LIGO detectors during O1~\cite{TheLIGOScientific:2016pea, Abbott:2016jsd}\footnote{For LIGO's second observing run, the detectors had calibration uncertainties of $\delta A \sim 3\%$ and $\delta\phi \sim 2^{\circ}$~\cite{LIGOScientific:2018mvr}}, and the predicted calibration uncertainties for future detector configurations~\cite{Kissel_calibration_GWDAW_17,Kissel_calibration_GWDAW_18}.

\subsubsection{Hierarchical inference} 
\label{ssub:hierarchical_inference}

For the population study detailed in Sec.~\ref{sec:population}, the \bilby inference package~\cite{Ashton:2018jfp,Bilby-code} was used for both the analysis of individual \acp{BBH} as well as the subsequent inference on their population parameters.
Following the analysis of each individual \ac{GW} signal assumed to be part of the observed population, their joint population properties, here a single parameter $\alpha$, can be inferred as a hyper-posterior~\cite{Thrane:2018qnx}
\begin{equation}
  p_\mathrm{tot}(\alpha | \vec d) = 
  \frac{\mathcal{L}_\mathrm{tot}(\vec d | \alpha) \, \pi(\alpha)}
  {\int d\alpha \, \mathcal{L}_\mathrm{tot}(\vec d | \alpha) \, \pi(\alpha)},
\end{equation}
where $\mathcal{L}_\mathrm{tot}(\vec d | \alpha)$ is the hyper-likelihood, $\pi(\alpha)$ is the hyper-prior, $\vec d$ is a collection of data for $N$ independent events drawn from the injection distribution. 
We write the injection prior as $\pi(\theta|\alpha)$ and our goal is to estimate the hyper-posterior which in turn relies on a hyper-likelihood that can be written as
\begin{equation}
  \mathcal{L}_\mathrm{tot}(\vec{d} | \alpha) = \prod_i^N \frac{\mathcal{Z}_{\o}(d_i)}{n_i}
  \sum_k^{n_i} \frac{\pi(\theta^k_i| \alpha)}{\pi(\theta^k_i|{\o})},
\end{equation}
where $\pi(\theta^k_i|{\o})$ denotes the default prior that is used to perform single event parameter estimation and $\mathcal{Z}_{\o}(d_i)$ is the evidence obtained for event $i$.
The integral is then approximated in a Monte-Carlo sense, using the single event posterior samples that have been obtained previously.



\section{Results for Golden Binaries}
\label{sec:results-golden-binaries}

In this section we give predictions about parameter biases that would arise if we used current \ac{BBH} semi-analytic waveform models to infer the properties of high mass \acp{BBH} in a sequence of past, current and future ground based detector networks.

We select two exceptionally loud ``golden binaries'': one binary with parameters mimicking GW150914 and one binary at mass-ratio 1:3. 
Both systems contain \acp{BH} with spins misaligned with the orbital angular momentum vector causing the systems to be moderately precessing. 
We hold the luminosity distance of the systems constant, so that more sensitive detector networks will observe them with higher \acp{SNR} and obtain more precise measurements. 
Parameters for these systems are given in Table~\ref{tab:signal-configurations}. 
As signal waveforms we use \ac{NR} simulations from the SXS~\cite{Boyle:2019kee} catalog computed with the SpEC code~\cite{SpEC}, as described in Sec.~\ref{sub:numerical_relativity_waveforms}.
Since these waveforms are too short to fill the frequency band of future interferometers which extends well below $20$ Hz, we hybridize the \ac{NR} waveforms with \ac{PN} approximants in the inspiral, including higher order modes up to $\ell=8$. 
We use the effective precession spin \ac{IMR} waveform model \texttt{IMRPhenomPv2} for our main results and also quote complementary results for the non-precessing \texttt{SEOBNRv4\_ROM} model.
\texttt{IMRPhenomPv2} includes $\ell=2, m=\pm 2$ modes in the co-precessing frame, and a \ac{PN} description of the motion of the co-precessing frame with an approximation for small precession angles~\cite{Hannam:2013oca,Khan:2015jqa,Husa:2015iqa,PhenomPv2_tech_rep}.

\subsection{Indistinguishability} 
\label{sub:indistinguishability}

We want to find an estimate that predicts beyond which \ac{SNR} a particular waveform model that is used as a template in \ac{PE} yields biased posterior distributions for the above \ac{BBH} signals. 
We can find the answer by calculating the posterior distribution using Bayesian inference. 
However, this method is fairly costly for the very sensitive future detectors (see Fig.~\ref{fig:plots_ASDs_evolution}) where the signals have \acp{SNR} up to several thousands. 
Therefore, we compare against and extend a simpler metric for predicting the presence of biases.

If two waveforms $h_1$ and $h_2$ fulfill the criterion~\cite{Flanagan:1997kp, Lindblom:2008cm, McWilliams:2010eq, Chatziioannou:2017tdw}
\begin{equation}
  \label{eq:indistinguishability-criterion}
  1 - \mathcal{O}(h_1, h_2) < D / (2\rho^2)
\end{equation}
for a given \ac{PSD} and \ac{SNR} $\rho$ then they are deemed \emph{indistinguishable}, i.e, $\langle \delta h | \delta h \rangle < 1$ and the posterior \ac{PDF} should be unbiased in the sense that systematic errors from waveform inaccuracies are smaller than $1-\sigma$ statistical errors.

While this criterion is simple to evaluate, there are several problems that affect its usefulness in practise:
The criterion is only sufficient, but not necessary and as a result it tends to be too \emph{conservative}. 
Namely, if it is violated, biases \emph{can}, but \emph{need not} arise.
In addition, the pre-factor $D$ is not known precisely. 
It can be derived as the number of (intrinsic) parameters whose measurability is affected by model inaccuracy~\cite{Chatziioannou:2017tdw}.
The criterion also applies only in the high \ac{SNR} limit as is the case for the Fisher information matrix~\cite{Vallisneri:2007ev}.

To enhance the usefulness of the indistinguishability criterion we use the following procedure to tune the pre-factor $D$.
\begin{enumerate}
  \item We compute posterior distributions for a sequence of detector networks on the above synthetic signals.
  \item From the posterior distributions we compute statistical and systematic errors for key parameters (chirp-mass, mass-ratio, effective aligned spin, and effective precession spin).
  \item We estimate the network (balance) \ac{SNR} $\rho_b$ at which the computed systematic and statistical errors become comparable.
  \item We compute the mismatch 
  $1 - \mathcal{O}(h_\mathrm{model},h_\mathrm{true})(\theta_\mathrm{true})$ 
  between the template waveform and the signal at signal parameters for a representative detector sensitivity.
  \item Finally, we calculate 
  \begin{equation}
    \label{eq:D_equation}
    D = 2 \rho_b^2 \left[ 1 - \mathcal{O}(h_\mathrm{model}, 
                                          h_\mathrm{true})(\theta_\mathrm{true})\right]
  \end{equation}
\end{enumerate}

We present results of applying this procedure to the selected golden binaries in Sec.~\ref{sub:predicted_waveform_accuracy_requirements}. 
First we discuss some assumptions we make in applying it.

When computing the balance \ac{SNR} and the mismatch we have to assume a \acf{PSD}.
We find empirically that systematic and statistical errors become comparable at network \acp{SNR} of $\sim 60$ for the above sources. 
This \ac{SNR} is found at aLIGO design sensitivity for SXS\_BBH\_0308 and at about A+ sensitivity for SXS\_BBH\_0104.
The mismatch is only sensitive to the shape of the \ac{PSD} and the frequency range of the overlap integral. 
We pick aLIGO design sensitivity~\cite{aLIGO_design_updated} as a reference \ac{PSD} since this is close to the sensitivity where the balance SNR is found, and it is in its vicinity that the tuned indistinguishability criterion should be most accurate. 
In general we expect that mismatches will degrade as we approach future detectors since they will be sensitive to lower frequencies and will have significantly more waveform cycles in band.
The network \ac{SNR} determines the discerning power of a network of detectors since we analyze the signal coherently. 
We neglect that the interferometers that make up detector networks usually have different sensitivities and pick a representative \ac{PSD}. 
We use this \ac{PSD} to compute the single interferometer mismatch in the indistinguishability criterion.

We use mock signals as a proxy for the true waveform obtained from exactly solving the two body problem in General Relativity. 
Hence we also assume that \ac{GR} is the correct theory of gravity. 
Ideally our mock signals would be pure \ac{NR} waveforms. 
This is in general not feasible since the cost of computing \ac{BBH} coalescences with \ac{NR} simulations scales very steeply with the initial frequency, so that in practise only part of the detector band can be filled by the \ac{NR} signal for high mass \acp{BBH}. 
Therefore, we hybridize \ac{NR} signals with \ac{PN} inspiral waveforms. 

\ac{NR} simulations are only approximations of true \ac{GR} waveforms. 
\ac{NR} accuracy depends on the choice of configuration (e.g. more unequal mass-ratios and higher spin systems are harder to simulate accurately as the size of the apparent horizon of the \acp{BH} decreases) and on the size of the grid used to discretize Einstein's equations. 
In reality, \ac{NR} simulations use multiple domains and a particular discretization method (finite differences~\cite{Campanelli:2005dd,Bruegmann:2006at,Husa:2007hp}, multi-domain spectral collocation methods~\cite{SpEC} or more advanced methods, such as discontinuous Galerkin~\cite{Hesthaven:190448, Vincent:2019qpd}).
While we can obtain a good estimate of the \ac{NR} waveform error by computing mismatches for the same physical configuration but different grid sizes to decrease the truncation error and wave extraction errors, it is difficult to estimate the error in a hybrid waveform. We discuss this
further in App.~\ref{app:hybrids}.

In the above procedure for estimating the pre-factor $D$ we need to find the \ac{SNR} at which the
systematic and statistical errors are comparable. We know that parameters are in general
correlated and thus we should take these correlations into account when estimating these errors.
The indistinguishability criterion also makes this assumption.
When quoting parameter estimation results we rely on errors computed from one and two-dimensional marginal posterior distributions, which are straightforward to compute and present. 
Therefore, we also compute the statistical and systematic errors from 1D marginal posteriors.
A more conservative measure of the error is to compute where the injection lies in the posterior distribution, or a marginal \ac{PDF} thereof. 
We obtain the percentile of the credible level of the injected parameters in the full posterior by performing parameter estimation with all sampling parameters fixed, except for the time and phase of coalescence. Detailed measurements of the latter are of no astrophysical interest, and as they can very strongly affect the likelihood, we prefer to marginalize over them.

We also consider a third method where we take into account the correlations in a set of key parameters only. 
To do this, we compute a \ac{KDE} of the marginal posterior distribution in the parameters of interest, compute the posterior probability value at the injection parameters and find its credible level in the marginal posterior.
We compute a Gaussian \ac{KDE} $\mathcal{K}(\tilde\theta) = \mathrm{KDE}[p(\tilde\theta | d)]$ of the marginal posterior distribution $p(\tilde\theta | d)$ and then solve numerically the equation $Q(\mathcal{K}(\theta^{(i)}); p) = \mathcal{K}(\theta_s)$ to find at which percentile $100 p$ the true parameters $\theta_s$ of the signal lie in the marginal posterior. 
Here $Q$ is the quantile function $Q(\mathrm{PDF}; p) = \mathrm{CDF}^{-1}(p)$ for a given \ac{PDF} and its \ac{CDF}. 
In practise we work with the logarithm of the \ac{PDF} to reduce the dynamic range.
We discuss results from these procedures in the next section.


\subsection{Predicted Waveform Accuracy Requirements} 
\label{sub:predicted_waveform_accuracy_requirements}

\begin{figure*}[t]
  \centering
\includegraphics[width=.45\textwidth]{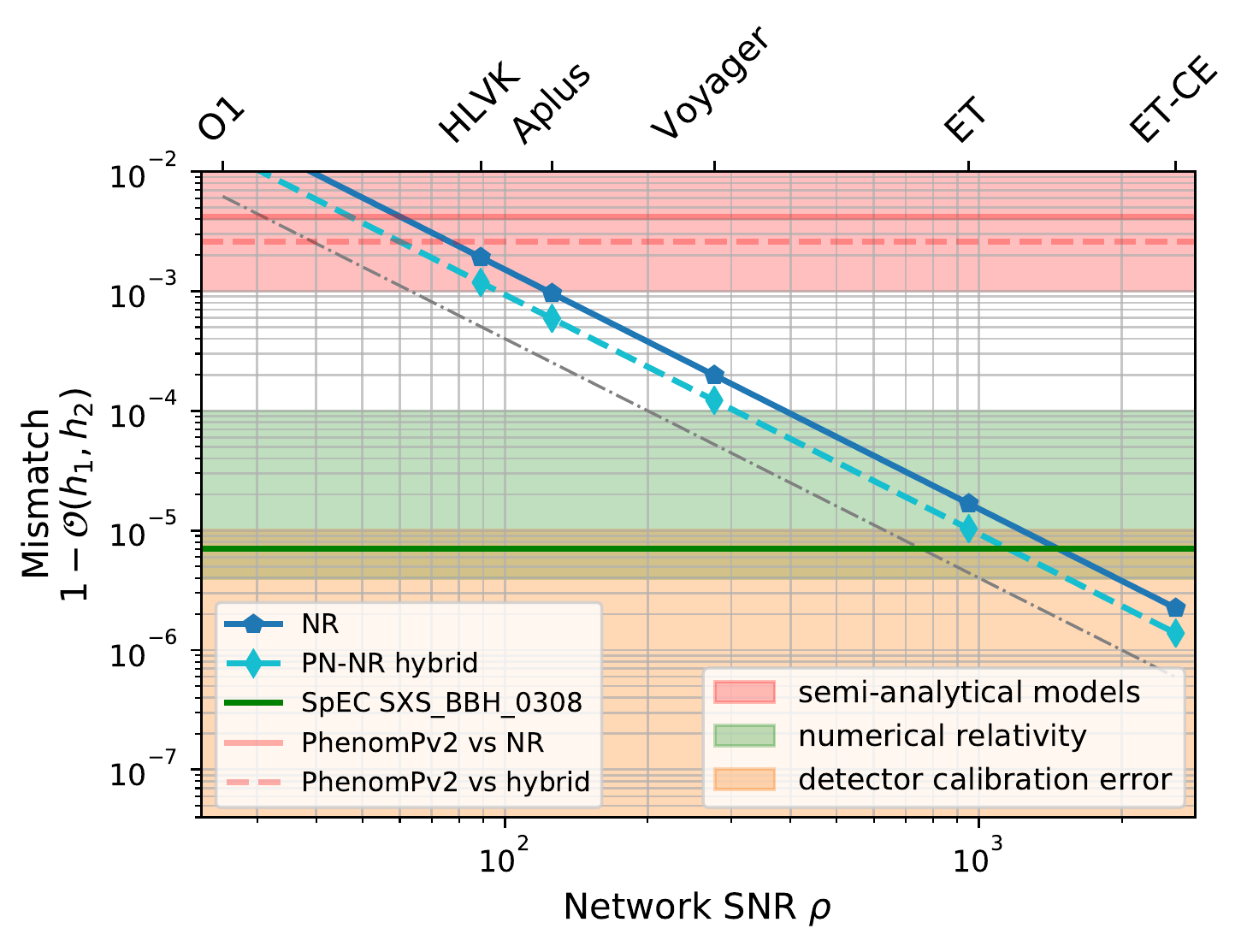} \includegraphics[width=.45\textwidth]{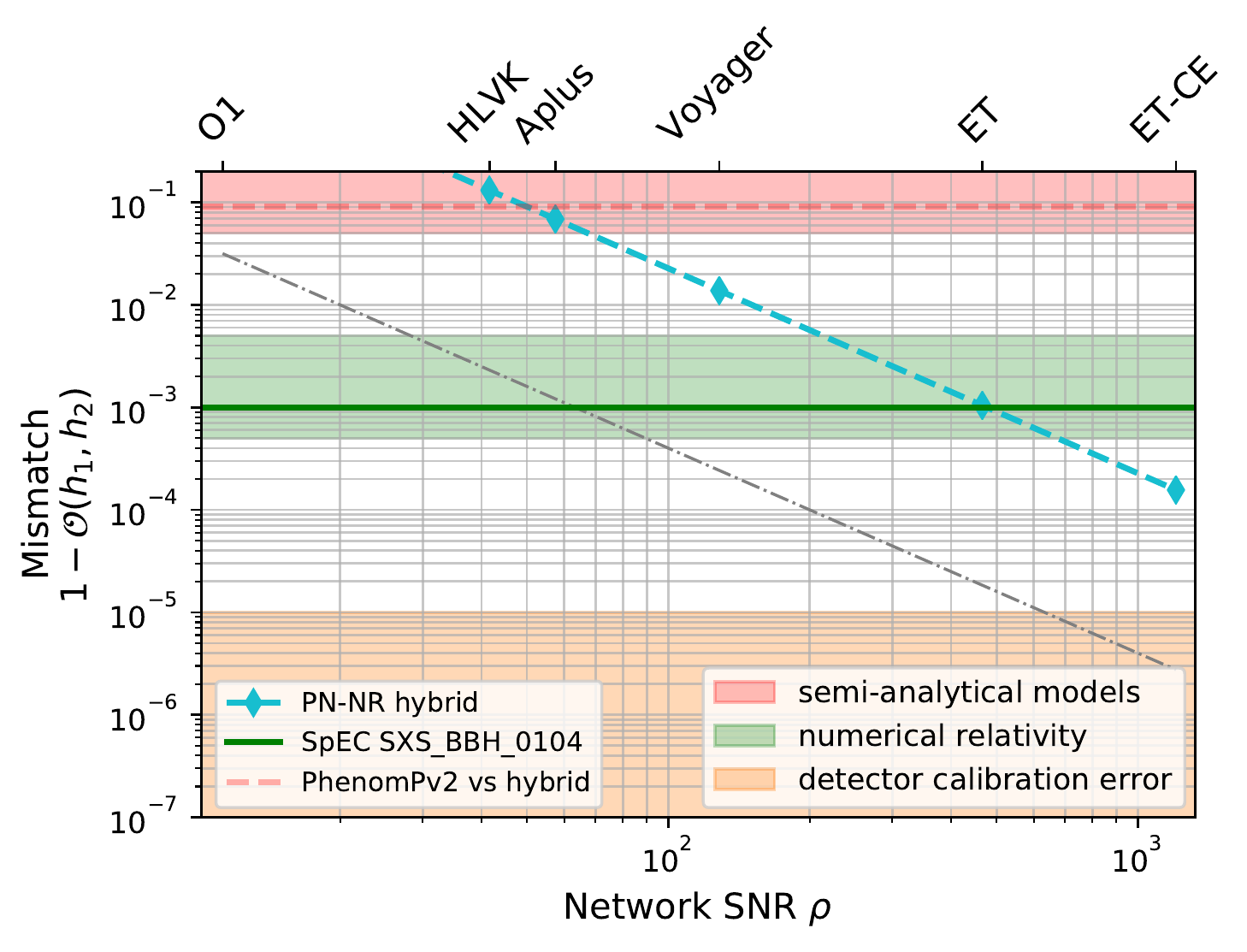}
  \caption{Predicted waveform accuracy requirements for second and third generation ground based detector networks.
  We show results for two binaries \textit{left:} SXS\_BBH\_0308, and \textit{right:} SXS\_BBH\_0104 (see Table~\ref{tab:signal-configurations}).
  Each panel shows mismatch against network \ac{SNR} and on the top x-axis the detector network (see Table~\ref{tab:detector-networks}) in which the signal had the \ac{SNR} shown in the bottom x-axis. 
  Solid lines indicate results for pure \ac{NR} signals, while dashed lines come from \ac{NR} signals hybridized against \ac{PN} waveforms in the inspiral.
  The blue lines and data points show how the mismatch falls with rising \ac{SNR} according to the indistinguishability criterion Eq.~\eqref{eq:indistinguishability-criterion} with the prefactor $D$ tuned according to Eq.~\eqref{eq:D_equation}, as $D / (2\rho^2)$. 
The dash-dotted gray line shows the prediction of Eq.~\eqref{eq:indistinguishability-criterion} with $D=8$.
  Horizontal red lines show the mismatch of the signal against the \texttt{IMRPhenomPv2} template waveform at the signal parameters (also called ``unfaithfulness'') for aLIGO design sensitivity. 
The horizontal green line shows the mismatch between \ac{NR} waveforms obtained for different grid resolutions for the same signal configuration.
  Shaded regions provide rough estimates of the accuracy of current semi-analytic waveform models and current \ac{NR} waveforms for the particular binary systems and the level of expected detector calibration error in terms of mismatch.
  Waveform error estimates are higher for the more challenging unequal mass spinning SXS\_BBH\_0104 configuration compared to SXS\_BBH\_0308.
  }
  \label{fig:plots_indistinguishability}
\end{figure*}

\begin{table*}
\begin{tabular}{l|l|l|l|l|l|l|l|l}
  \hline
  \hline
 Configuration & $M_\mathrm{tot}^\mathrm{src} / M_\odot$ & $\mathcal{M}^\mathrm{src} / M_\odot$ & $q$ & $\vec\chi_1$ & $\vec\chi_2$ & $\chi_\mathrm{eff}$ & $\chi_\mathrm{p}$ & $\theta_\mathrm{JN}$\\
  \hline
  \hline
  SXS\_BBH\_0308 & 66.4555 & 28.7443 & 0.8143 & $(-0.1407, 0.0225, 0.3053)$ & $(-0.2209, 0.3075, -0.5580)$ & -0.0822 & 0.2994 & 2.7454\\
  \hline
  SXS\_BBH\_0104 & 66.4555 & 24.3406 & 0.3333 & $(-0.0550, -0.0144, 0.4966)$ & $( -0.2737, -0.4173, 0.0112)$ & 0.3753 & 0.1442 & 1.0839\\
  \hline
  \hline
\end{tabular}
\caption{
    \label{tab:signal-configurations}
    Binary configurations studied in Sec.~\ref{sec:results-golden-binaries}.
    We indicate the SXS ID~\cite{Boyle:2019kee} of the SpEC \ac{NR} simulations, the total mass and chirp mass in the source frame, the mass-ratio $q = m_2 / m_1 \leq 1$, the dimensionless spin vectors $\vec\chi_i = \vec S_i / m_i^2$ of the \acp{BH}, the effective aligned spin and effective precession spin and the inclination angle between the total angular momentum $\vec J$ and the line of sight $\vec N$.
    Signals are hybridized with \texttt{SpinTaylorT1}.
    Spin vectors are defined at a reference frequency of 30 Hz.
    We select the remaining common parameters to be $\mathrm{ra} = 1.949725$, $\mathrm{dec} = -1.261573$ (radians), a luminosity distance of $d_L = 562.59 \mathrm{Mpc}$ (which corresponds to a redshift of about $z = 0.115$), a polarization angle $\psi = 1.4289$. 
The GPS time at the geocenter was $1126259642.413$ s and coalescence phase $\phi_\mathrm{coa} = 0$.
    }
\end{table*}

\begin{table*}
\begin{tabular}{l|l|l| d{4.1} | d{4.1}}
  \hline
  \hline
 Network & List of Interferometers and \acp{PSD} & $f_\mathrm{low} [\mathrm{Hz}]$ & 
 \multicolumn{2}{c}{Network SNR}\\
 &  &  & \multicolumn{1}{c|}{$0308$} & \multicolumn{1}{c}{$0104$}\\
  \hline
  \hline
  O1      & H1, L1 (\texttt{O1}) & 30 & 25.4 & 11.2\\
  HLVK    & H1, L1 (\texttt{aLIGODesign\_2018\_T1800044}), V1 (\texttt{AdVirgo}), K1 (\texttt{KAGRA}) & 10 & 88.9 & 41.6\\
  A+   & H1, I1 (\texttt{A}$+$) & 10 & 125.7 & 57.5\\
  Voyager & H1, L1 (\texttt{Voyager}) & 10 & 276.3 & 128.4\\
  ET      & [E1, E2, E3] (\texttt{ET\_D}) & 5 & 950.9 & 466.3\\
  ET-CE   & [E1, E2, E3] (\texttt{ET\_D}), H1 (\texttt{CE}) & 5 & 2598.8 & 1205.2\\
  \hline
  \hline
\end{tabular}
\caption{
    \label{tab:detector-networks}
List of ground-based detector networks used in this study. 
The networks are defined by the positions of the detectors on the Earth and their \acp{PSD} in parentheses. 
We also indicate the frequency $f_\mathrm{low}$ at which we start integrating the likelihood integral and the network \ac{SNR} of the PN-NR hybrid signals in these networks (see Table~\ref{tab:signal-configurations} for the parameters).
Detector locations are indicated by H1: LIGO Hanford, L1: LIGO Livingston, V1: Virgo, K1: KAGRA, I1: LIGO India, E1, E2, E3: the interferometers of the triangular \ac{ET} detector~\cite{LALDetectors}.
  Text data files for the \acp{PSD} or \acp{ASD} can be found in~\cite{ASD_curves18, aLIGO_design_updated} under the names given in this table.
    }
\end{table*}


We now apply the procedure presented in Sec.~\ref{sub:indistinguishability} to posterior probability distributions and mismatches obtained for the two mock \ac{BBH} signals shown in Table~\ref{tab:signal-configurations} for a series of detector networks. 
The networks are defined by the positions of the detectors on the Earth and their \acp{PSD} as listed in Table~\ref{tab:detector-networks}.

Fig.~\ref{fig:plots_indistinguishability} shows the main results. 
According to Eq.~\ref{eq:indistinguishability-criterion} the general takeaway is that as long as the mismatch for a given semi-analytical waveform model against the mock signal (red lines) lies below the \emph{tuned} indistinguishability curve (light or dark blue lines) we do not expect parameter recovery to be biased. 
One can think of the indistinguishability curve showing the ``acceptable error'' for a waveform model for a particular \ac{SNR}.
Without tuning, the predicted \ac{SNR} above which we would see biases (assuming that 6 intrinsic model parameters are affected) is about 25 for the SXS\_BBH\_0308 \ac{NR} signal (and \ac{SNR} 11 for the hybrid). 
For SXS\_BBH\_0104 it is predicted to be an \ac{SNR} of $\sim 6$. 
As we will see in Sec.~\ref{sub:parameter_estimates} these predictions are certainly way too conservative for the hybrid signals when compared with the parameter estimation results and the assumption that 6 parameters are biased is not correct either.

A first observation is that semi-analytic models (here the representative \texttt{IMRPhenomPv2}
and \texttt{SEOBNRv4\_ROM} models) were sufficiently accurate to analyze GW150914 during aLIGO's first observing run. 
This is hardly a surprise and has been studied in depth by comparing against \ac{NR} simulations and waveform models by the LVC~\cite{Abbott:2016wiq,Abbott:2016apu}.
Fig.~\ref{fig:plots_indistinguishability} also predicts that semi-analytical models will lead to biased parameter recovery at and beyond HLVK sensitivity for SXS\_BBH\_0308 and at and beyond the A+ network for SXS\_BBH\_0104. 
Moreover, current \ac{NR} waveforms will not be guaranteed to be sufficiently accurate for unbiased parameter recovery beyond the Voyager network (where the dark blue line intersects the dark green line). 
Clearly then current waveform models will not be accurate enough for 3G ground based detectors such as \ac{ET} and \ac{CE} which are currently being planned. We will require waveform models
to be at least \emph{three orders of magnitude} more accurate, and improvements of \emph{one order of magnitude} for \ac{NR} waveforms.

Fig.~\ref{fig:plots_indistinguishability} presents a simplified picture to convey the main message that current waveform models are not accurate enough for planned 3G detectors. 
We now come back to some of the assumptions we have mentioned in Sec.~\ref{sub:indistinguishability} and shed some light on details.
The shape of the \acp{PSD} and the range in frequency over which particular interferometers are sensitive varies with the networks and influences the value of the mismatch that enters the indistinguishability criterion. 
The horizontal lines shown in Fig.~\ref{fig:plots_indistinguishability} provide a simplified representative measure of the error.
For SXS\_BBH\_0308 mismatches against \texttt{IMRPhenomPv2} range from 0.002 (aLIGO O1) to 0.02 (CE) for the pure \ac{NR} signal which is in band from 20Hz and above, and from 0.002 (aLIGO O1) to 0.008 (CE) for the hybrid signal.
Starting frequencies are given in Table~\ref{tab:detector-networks}.
Mismatches for aLIGO, AdVirgo, KAGRA and A+ are very similar to those for the aLIGO O1 results. 
For the non-precessing \texttt{SEOBNRv4\_ROM} model mismatches range from 0.003 (aLIGO O1) to 0.02 (CE) for the pure \ac{NR} signal and from 0.005 (aLIGO O1) to 0.03 (CE) for the hybrid signal.
For the SXS\_BBH\_0104 hybrid signal the mismatches against \texttt{IMRPhenomPv2} range from 0.06 (CE) to 0.09 (aLIGO O1, aLIGO design). Here, mismatches against \texttt{SEOBNRv4\_ROM} are surprisingly slightly better 0.04 (CE) to 0.07 (AdvVirgo).

We want to stress that the mismatches depend very sensitively on the inclination angle under which the signal is seen. If we were to change the inclination for SXS\_BBH\_0308 from near face-off, $2.7454$, which is compatible with GW150914, to $\pi/3$ which emphasizes more harmonics content beyond the dominant $(2, \pm 2)$ mode, then the mismatch is about an order of magnitude worse. If biases were to appear at the same SNR for this changed inclination this would make $D$ an order of magnitude larger and the left panel of Fig.~\ref{fig:plots_indistinguishability} would look markedly different and have stronger implications for how much waveform models need to be improved. 

We have indicated in Fig.~\ref{fig:plots_indistinguishability} the estimated accuracy of waveform models and \ac{NR} simulations for the particular binary configurations by colored regions that are independent of the detector networks. 
These regions are supposed to give a rough sense of how accurate currently available models or codes are in the neighborhood of the \ac{BBH} configurations considered here.
Similar considerations as for the mismatches quoted above apply for the bounds of these regions.
For simplicity we have bounded these very rough estimates by constant mismatch.
Finally, detector calibration error depends on the detector network and is expected to improve over time up to a level that is believed to be attainable from current understanding. 
In Fig.~\ref{fig:plots_indistinguishability} the estimate of the mismatch error due to detector calibration errors uses a realistic estimate for future detectors and assumes 1\% relative error in amplitude, $1^{\circ}$ error in phase~\cite{Kissel_calibration_GWDAW_17, Kissel_calibration_GWDAW_18} and the additional assumption that the functional form of the
dephasing from detector calibration errors can be modeled by a quadratic function which 
decreases towards high frequencies.
Ultimately, the noise floor that comes from detector calibration error will only become problematic for 3G detector networks if we are not otherwise dominated by waveform errors.

\subsubsection{Balancing accuracy using the full posterior} 
\label{ssub:balancing_using_full_posterior}

The above results used 1D marginal posterior distributions to calculate statistical and systematic errors and find the \ac{SNR} at which they are comparable. 
We now discuss results where we take into account correlations between binary parameters and how they compare to the above.
Irrespective of how many parameters we choose to include in the marginal posterior distribution we can always ask the question at which credible level the injection lies in the posterior distribution. 
We want to estimate when this is close to the 68th percentile. 
Since we only have data for fixed networks we need to interpolate the percentile values to estimate the SNR at which errors are balanced.

For the SXS\_BBH\_0308 PN-NR hybrid signal we find the injection in the full posterior at the 2nd percentile for the O1 network and at the 100th percentile for HLVK, marginalizing over relative time and phase. 
For the marginal posterior in $(\mathcal{M}, q, \chi_\mathrm{eff}, \chi_\mathrm{p})$ we find the injection at the $12$th and $100$th percentile in O1 and HLVK, respectively.
For the 1D marginal distributions in these parameters we find that the injection lies between the $4$th to $50$th percentile for O1 and between the $78$th and $99$th percentile in HLVK.
Therefore, for this configuration we find a similar balance \ac{SNR} of 60 for these different ways of computing the error balance.
This estimate is somewhat uncertain, since we do not have any datapoints in between the O1 and HLVK networks. 
In terms of the prefactor $D$, we would expect to have $D \sim 8$ if the key parameters are biased, but we find $D \sim 20$ if the errors are balanced at \ac{SNR} 60. 
We note that the chirp mass and the effective precession spin are quite biased for this signal. 
For the NR only signal we find $D \sim 30$ because the mismatch is worse in the late inspiral and merger part.

For the SXS\_BBH\_0104 source we find the injection in the full posterior at the $7$th and $100$th percentiles for the O1 and HLVK networks, respectively. 
For the marginal posterior in $(\mathcal{M}, q, \chi_\mathrm{eff}, \chi_\mathrm{p})$ we find the injection below the 40th percentile for O1, HLVK, A+, and Voyager, and it lies at the 100th percentile for the ET and ET-CE networks. 
For the 1D marginal distributions in these parameters we find that the injection lies between the $3$rd to $43$rd percentile for O1 and between the $12$th and $40$th percentile in HLVK.
The balance SNR is then estimated to be $\sim 250$. 
In combination with the large mismatch between the signal and template waveforms, it results in an enormous prefactor of $D \sim 10^4$. 
The reason is that there is almost no bias in $(\mathcal{M}, q, \chi_\mathrm{eff}, \chi_\mathrm{p})$, as can be seen in Fig. \ref{fig:0104_posterior_PDFs} discussed in Sec.~\ref{sub:parameter_estimates}. 
If we add inclination and distance parameters then there is a noticeable bias and we find the injection at the $50$th percentile for O1 and at the $100$th percentile for the HLVK network and beyond. 
This results in a more reasonable balance \ac{SNR} of $\sim 22$ and a prefactor of $D \sim 90$. 
Using the 1D marginal errors we find a balance \ac{SNR} of roughly $50$ and a prefactor of $D \sim 450$. 
The naive indistinguishability criterion with $D = 8$ predicts biased recovery at \ac{SNR} $10$, which is close to the \ac{SNR} of the signal in the O1 network.


\subsection{Parameter estimation results} 
\label{sub:parameter_estimates}

\begin{figure*}[t]
  \centering
  \includegraphics[width=.45\textwidth, trim=-0.2mm +1.3mm +0.2mm -1.3mm]{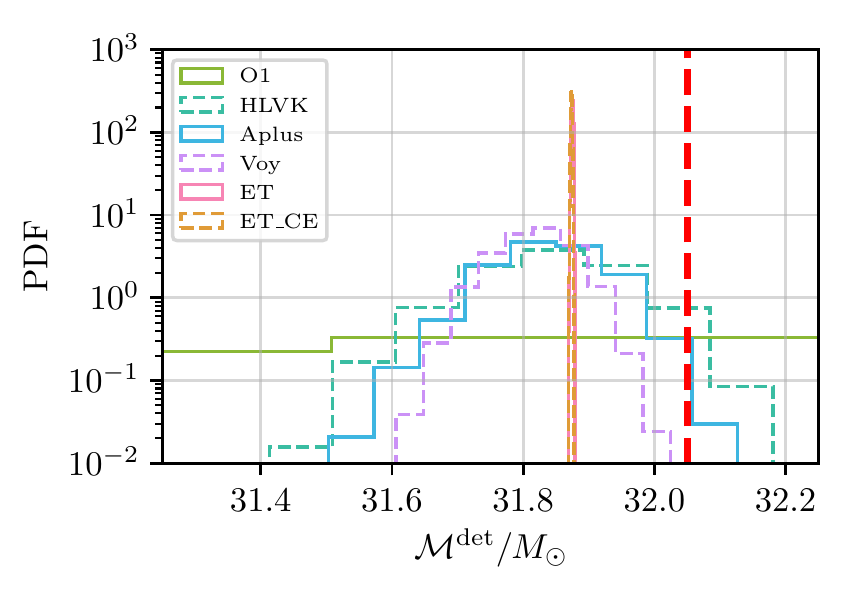}
  \includegraphics[width=.45\textwidth]{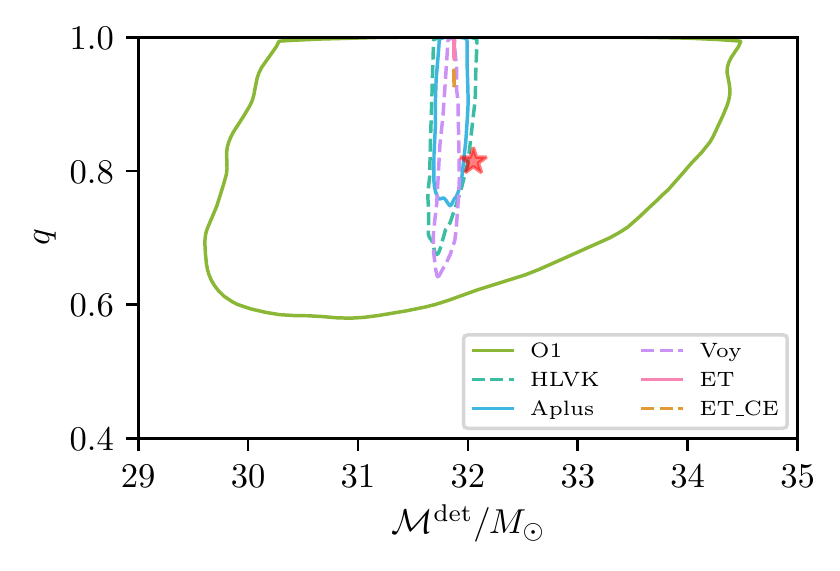}
  \includegraphics[width=.45\textwidth, trim=-0.0mm -2.5mm +0.0mm +2.5mm]{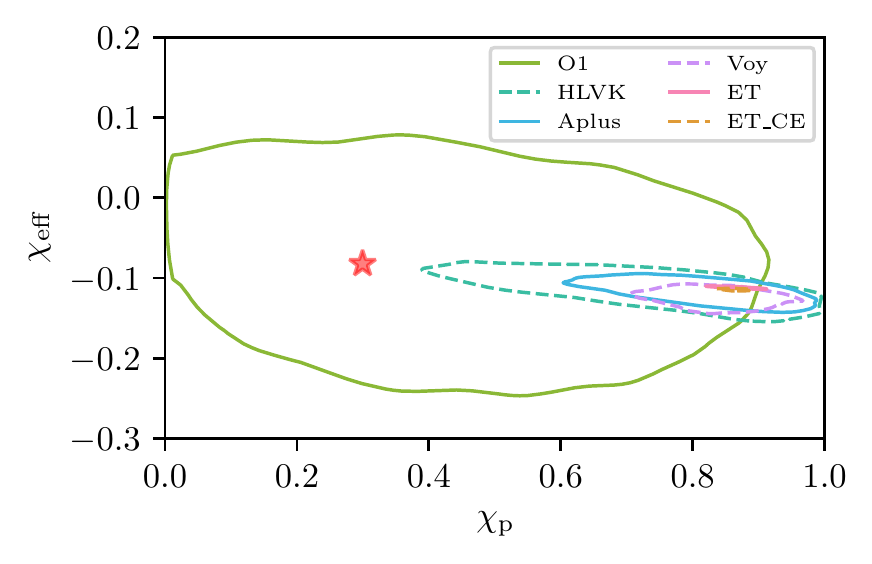}
  \includegraphics[width=.45\textwidth, trim=-0.0mm -0.5mm +0.0mm +0.5mm]{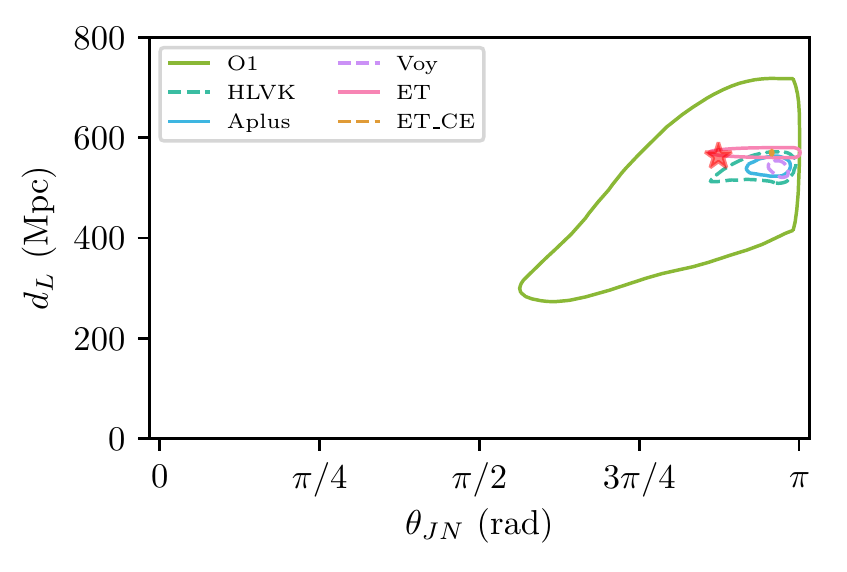}
  \caption{Posterior PDFs for the SXS\_BBH\_0308 PN-NR hybrid signal (see Table~\ref{tab:signal-configurations}) using \texttt{IMRPhenomPv2} as the template waveform for a sequence of detector networks (see Table~\ref{tab:detector-networks}).
  We show either histograms of 1-dimensional PDFs or contours indicating 90\% 
  credible regions for 2-dimensional PDFs.
  \textit{Top left panel:} histograms for detector-frame chirp mass
  \textit{Top right panel:} contours for chirp-mass and mass-ratio
  \textit{Bottom left panel:} contours for effective precession spin $\chi_\mathrm{p}$ 
  and effective aligned spin $\chi_\mathrm{eff}$
  \textit{Bottom right panel:} contours for inclination angle $\theta_\mathrm{JN}$ and luminosity distance $d_L$.
  True parameter values of the source binary are indicated as red dashed lines or a red asterisk.
  }
  \label{fig:0308_posterior_PDFs}
\end{figure*}

\begin{figure*}[t]
  \centering
  \includegraphics[width=.472\textwidth, trim=-0.0mm +1.5mm +0.0mm -1.5mm]{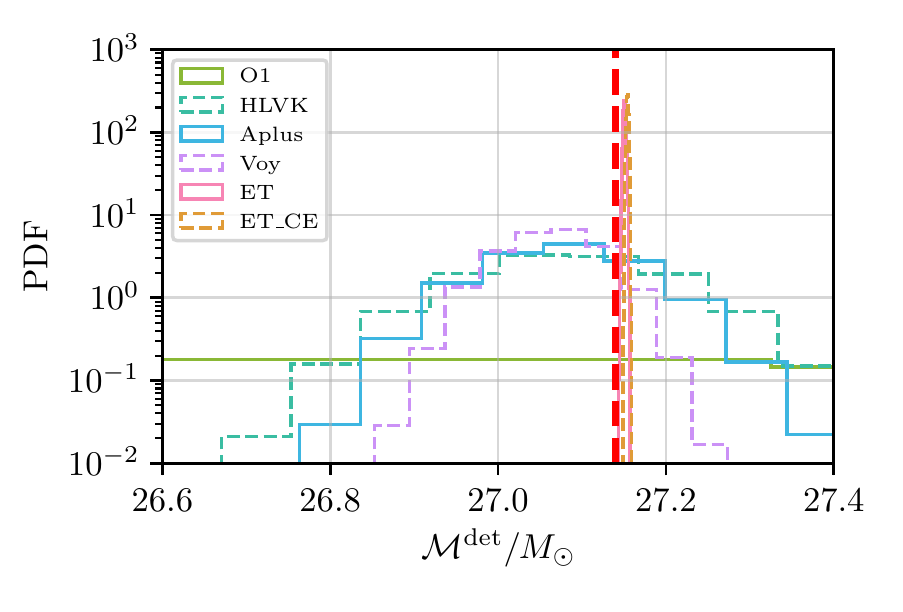}
  \includegraphics[width=.45\textwidth]{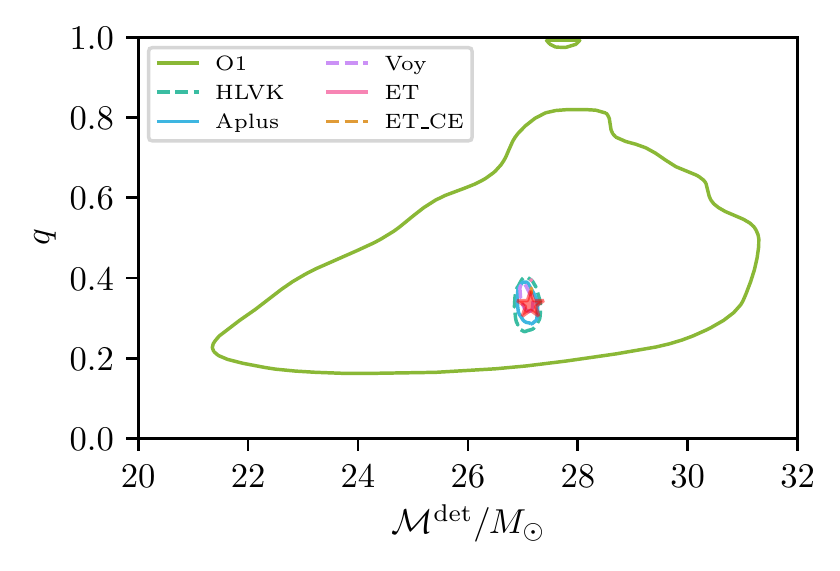}
  \includegraphics[width=.45\textwidth]{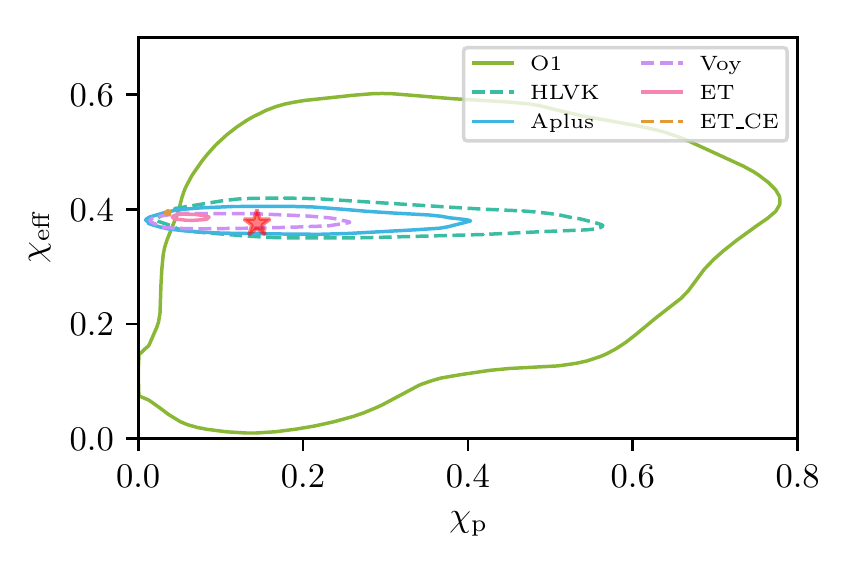}
  \includegraphics[width=.46\textwidth, trim=-0.5mm +0.8mm +0.5mm -0.8mm]{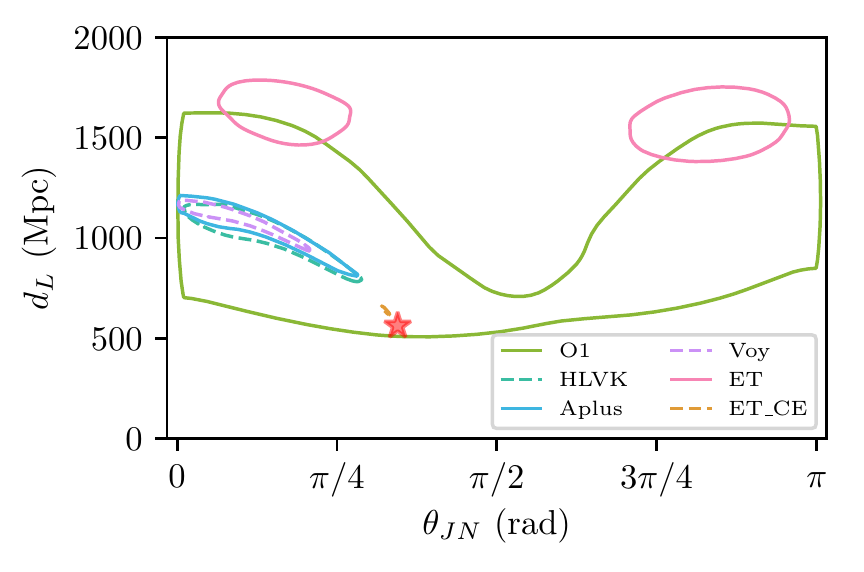}
  \caption{Posterior PDFs for the SXS\_BBH\_0104 PN-NR hybrid signal (see Table~\ref{tab:signal-configurations}) using \texttt{IMRPhenomPv2} as the template waveform for a sequence of detector networks (see Table~\ref{tab:detector-networks}).
  We show either histograms of 1-dimensional PDFs or contours indicating 90\% 
  credible regions for 2-dimensional PDFs.
  \textit{Top left panel:} histograms for detector-frame chirp mass
  \textit{Top right panel:} contours for chirp-mass and mass-ratio
  \textit{Bottom left panel:} contours for effective precession spin $\chi_\mathrm{p}$ 
  and effective aligned spin $\chi_\mathrm{eff}$
  \textit{Bottom right panel:} contours for inclination angle $\theta_\mathrm{JN}$
  and luminosity distance $d_L$.
  True parameter values of the source binary are indicated as red dashed lines or a red asterisk.
  }
  \label{fig:0104_posterior_PDFs}
\end{figure*}

We now turn to looking directly at posterior distributions for the analysis of the two mock \ac{BBH} signals from Table~\ref{tab:signal-configurations} for a series of detector networks.
Histograms and 90\% credible regions for key parameters are shown in Figs.~\ref{fig:0308_posterior_PDFs} and~\ref{fig:0104_posterior_PDFs} for the SXS\_BBH\_0308 and SXS\_BBH\_0104 sources, respectively. Here we show \texttt{IMRPhenomPv2} posteriors since this
model includes approximate precession effects, in contrast to the non-precessing 
\texttt{SEOBNRv4\_ROM} model.

\subsubsection{Results} 
\label{ssub:results}

The posterior distributions of the detector-frame chirp mass shown in the top left panel of
Figs.~\ref{fig:0308_posterior_PDFs} and~\ref{fig:0104_posterior_PDFs} become progressively tighter as we go to more sensitive networks, their widths scaling roughly inversely with the \ac{SNR}. 
This is the expected behavior for a multi-modal Gaussian which the posterior distribution is expected to follow in the high \ac{SNR} limit, although the 90\% credible regions for some marginal 2D posteriors shown in the other panels are clearly not Gaussian. 
Only part of the chirp mass posterior is shown for O1 sensitivity so that we can more clearly see the posteriors for networks operating at higher sensitivities. 
The measurement precision in chirp mass in terms of the width of the 90\% credible interval increases from $\sim 5 M_\odot$ (O1), to $0.4 M_\odot$ (HLVK), and $0.004 M_\odot$ (ET+CE). 
The massive increase in precision for 3G detectors is expected due to the improved sensitivity
and the significantly larger number of waveform cycles in the detector frequency band. 
For instance, for SXS\_BBH\_0308 there are 64 cycles in band from 10 Hz, compared to 217 cycles from 5 Hz and 1025 cycles from 2Hz.
For SXS\_BBH\_0308 the O1 posterior is unbiased, with the true chirp mass value (red dashed line) near its peak. 
For the HLVK network and beyond the posteriors peak away from the true chirp mass. 
The true chirp mass is found at the 98th percentile for HLVK and for the Voyager network and beyond at the 100th percentile. 
For ET and ET-CE the chirp mass is underestimated by $0.18 M_\odot$.
For SXS\_BBH\_0104, there is again no visible bias at O1 sensitivity. 
For HLVK, the true value lies at one sigma away from the peak, and at the 92nd percentile for the Voyager network. 
Recovery is very accurate for the ET and ET-CE networks with a bias of $-0.01 M_\odot$.

In the remaining panels of Figs.~\ref{fig:0308_posterior_PDFs} and~\ref{fig:0104_posterior_PDFs} we show 90\% credible regions for marginal 2D posteriors for several key parameters, to give a sense of the correlations between binary parameters, starting with chirp mass and mass-ratio. 
Compared to chirp mass, the mass-ratio is much more difficult to measure, resulting in very wide posteriors. 
This is especially true for the near equal mass SXS\_BBH\_0308 source. 
There, the one-sided 10\% percentile of the mass-ratio PDFs is roughly at $0.7$ for 2G detectors. 
For 3G detectors the measurement is much more precise, again due to more inspiral cycles being observable, but in this case the mass-ratio is estimated to be too close to equal-mass with a bias $q_\mathrm{true} - q_\mathrm{MAP} \approx -0.15$.
For the unequal mass SXS\_BBH\_0104 source, the mass-ratio is much better measured. 
The measurement precision in terms of the 90\% interval increases from $0.5$ (O1) to $0.1$ (HLVK) and $0.01$ (ET-CE). 
Biases only appear for 3G detectors, where they are about $-0.05$.

While there are 6 spin degrees of freedom in a generic precessing \ac{BBH}, most of them are very difficult to measure. 
The aligned-spin degrees of freedom, in particular a mass-weighted linear combination called $\chi_\mathrm{eff}$ is the best measured spin parameter which is also degenerate with the mass-ratio~\cite{Poisson:1995ef,Cutler:1994ys,Baird:2012cu,Purrer:2015nkh,Purrer:2013xma,Vitale:2014mka,Vitale:2016avz,Abbott:2016wiq,Ng:2018neg}. 
For SXS\_BBH\_0308 the 90\% interval for $\chi_\mathrm{eff}$ shrinks from $0.26$ (O1) to $0.05$ (HLVK), and $0.003$ (ET-CE), while for SXS\_BBH\_0104 it shrinks from $0.45$ (O1) to $0.05$ (HLVK), and $0.004$ (ET-CE). 
Beyond O1 sensitivity the biases are below 0.05 for SXS\_BBH\_0308 and 0.02 for SXS\_BBH\_0104, only becoming significant for 3G detectors.

During LIGO and Virgo's O1 \& O2 observing runs precession effects have so far eluded measurement from compact binaries~\cite{LIGOScientific:2018mvr}. 
In terms of the effective precession spin parameter $\chi_\mathrm{p}$ ~\cite{Hannam:2013oca,Schmidt:2014iyl} the posterior distributions shown in GWTC-1 have not provided new information compared to the prior distribution.
We expect this situation to change with the improved sensitivity of future detectors~\cite{Vitale:2016icu,Vitale:2016avz} and the analysis of these sources is a case in point that we will be able to measure precession effects with future detectors.
For SXS\_BBH\_0308 the 90\% interval for $\chi_\mathrm{p}$ shrinks from $0.7$ (O1) to $0.5$ (HLVK), $0.2$ (Voyager) and $0.04$ (ET-CE), while for SXS\_BBH\_0104 it shrinks from $0.6$ (O1) to $0.4$ (HLVK), $0.2$ (Voyager) and $0.004$ (ET-CE).
Beyond O1 sensitivity where the measurement is uninformative, SXS\_BBH\_0308 posteriors are severely biased, overestimating $\chi_\mathrm{p}$ by about $0.6$. 
The system is thus seen as close to maximally precessing while the averaged in-plane spin is only  $\sim 0.3$. 
For SXS\_BBH\_0104 the $\chi_\mathrm{p}$ measurements are much more reliable and only offset by $\sim 0.1$.

Finally we show results for the marginal posteriors in luminosity distance $d_L$ and the inclination angle $\theta_\mathrm{JN}$ between the total angular momentum $J$ of the binary and the line of sight vector $N$ under which the binary is seen from the detector network.
These two parameters are especially degenerate in how they affect the amplitude of the source and the 2D posteriors are in general not Gaussian which limits the usefulness of 1-dimensional interval estimates and biases. 
The inclination posterior can have a single mode as for SXS\_BBH\_0308 which is seen close to the face-off inclination of the source, with some overestimation in $\theta_\mathrm{JN}$ and underestimation in the distance, or it can be bi-modal as for SXS\_BBH\_0104 for networks with (close to) co-located detectors (O1, ET) which have a harder time constraining it to the correct mode. 
Networks with better coverage of the Earth (HLVK, A+, Voyager, ET-CE) obtain the correct mode, but
the inclination angle is substantially underestimated along with overestimating the distance by a
factor of about 2. The only network to recover the inclination and distance with good accuracy is
the ET-CE network.


\subsubsection{Discussion} 
\label{ssub:Discussion}

\begin{table*}[hptb]
\begin{ruledtabular}
\begin{tabular}{l l c c c c c c}
Event & Waveform model & $\mathcal{M}^\mathrm{det}/M_\odot$ & $q$ & $\chi_\mathrm{eff}$ & $\chi_\mathrm{p}$ & $d_L/\mathrm{Mpc}$ & $\theta_\mathrm{JN}$ \\[0.05cm]
\hline \\[-1.0em]
\multirow{2}{*}{SXS\_BBH\_0308} & IMRPhenomPv2  & $31.959_{-2.043}^{+1.838}$ & $0.842_{-0.216}^{+0.141}$ & $-0.072_{-0.145}^{+0.116}$ & $0.37_{-0.29}^{+0.45}$ & $514_{-191}^{+132}$ & $2.698_{-1.365}^{+0.337}$\\[+0.1em]
                                & SEOBNRv4\_ROM & $31.524_{-2.290}^{+2.077}$ & $0.831_{-0.237}^{+0.151}$ & $-0.076_{-0.163}^{+0.122}$ & N/A & $469_{-200}^{+160}$ & $2.501_{-1.242}^{+0.469}$ \\[+0.1em]
\hline
\\[-1.0em]
\multirow{2}{*}{SXS\_BBH\_0104} & IMRPhenomPv2  & $26.703_{-3.427}^{+3.599}$ & $0.352_{-0.120}^{+0.425}$ & $0.330_{-0.236}^{+0.209}$ & $0.30_{-0.21}^{+0.38}$ & $1040_{-452}^{+485}$ & $0.815_{-0.640}^{+2.083}$\\[+0.1em]
                                & SEOBNRv4\_ROM & $27.897_{-3.622}^{+4.086}$ & $0.406_{-0.154}^{+0.423}$ & $0.390_{-0.209}^{+0.249}$ & N/A & $1088_{-516}^{+591}$ & $0.939_{-0.732}^{+1.899}$ \\
\end{tabular}
\end{ruledtabular}
\vspace{0.5cm}
%
\begin{ruledtabular}
\begin{tabular}{l l c c c c c c}
Event & Waveform model & $\mathcal{M}^\mathrm{det}/M_\odot$ & $q$ & $\chi_\mathrm{eff}$ & $\chi_\mathrm{p}$ & $d_L/\mathrm{Mpc}$ & $\theta_\mathrm{JN}$ \\[0.05cm]
\hline \\[-1.0em]
\multirow{2}{*}{SXS\_BBH\_0308} & IMRPhenomPv2  & $31.845_{-0.182}^{+0.169}$ & $0.894_{-0.174}^{+0.095}$ & $-0.113_{-0.027}^{+0.027}$ & $0.75_{-0.31}^{+0.17}$ & $540.6_{-29.8}^{+21.0}$ & $2.970_{-0.224}^{+0.109}$ \\[+0.1em]
                                & SEOBNRv4\_ROM & $32.060_{-0.132}^{+0.136}$ & $0.793_{-0.102}^{+0.141}$ & $-0.071_{-0.017}^{+0.021}$ & N/A & $519.8_{-109.3}^{+66.7}$ & $2.636_{-0.339}^{+0.359}$ \\[+0.1em]
\hline
\\[-1em]
\multirow{2}{*}{SXS\_BBH\_0104} & IMRPhenomPv2  & $27.083_{-0.184}^{+0.188}$ & $0.331_{-0.047}^{+0.056}$ & $0.383_{-0.027}^{+0.026}$ & $0.23_{-0.14}^{+0.27}$ & $1038_{-215}^{+93}$ & $0.407_{-0.271}^{+0.417}$ \\[+0.1em]
                                & SEOBNRv4\_ROM & $27.045_{-0.175}^{+0.165}$ & $0.374_{-0.044}^{+0.046}$ & $0.389_{-0.065}^{+0.038}$ & N/A & $949_{-328}^{+188}$ & $0.607_{-0.442}^{+0.435}$ \\
\end{tabular}
\end{ruledtabular}
\vspace{0.5cm}
%
\begin{ruledtabular}
\begin{tabular}{l l c c c c c c}
Event & Waveform model & $\mathcal{M}^\mathrm{det}/M_\odot$ & $q$ & $\chi_\mathrm{eff}$ & $\chi_\mathrm{p}$ & $d_L/\mathrm{Mpc}$ & $\theta_\mathrm{JN}$ \\[0.05cm]
\hline 
\\[-1.0em]
\multirow{2}{*}{SXS\_BBH\_0308} & IMRPhenomPv2  & $31.874_{-0.002}^{+0.002}$ & $0.939_{-0.010}^{+0.010}$ & $-0.114_{-0.001}^{+0.001}$ & $0.86_{-0.02}^{+0.02}$ & $569.6_{-3.6}^{+3.8}$ & $3.008_{-0.005}^{+0.004}$ \\[+0.1em]
                                & SEOBNRv4\_ROM & $31.894_{-0.002}^{+0.002}$ & $0.770_{-0.005}^{+0.005}$ & $-0.100_{-0.001}^{+0.001}$ & N/A & $461.2_{-23.0}^{+44.6}$ & $2.405_{-0.074}^{+0.152}$ \\[+0.1em]
\hline
\\[-1em]
\multirow{2}{*}{SXS\_BBH\_0104} & IMRPhenomPv2  & $27.154_{-0.002}^{+0.002}$ & $0.369_{-0.004}^{+0.005}$ & $0.394_{-0.002}^{+0.002}$ & $0.036_{-0.002}^{+0.002}$ & $639_{-16}^{+17}$ & $1.025_{-0.016}^{+0.014}$ \\[+0.1em]
                                & SEOBNRv4\_ROM & $27.179_{-0.002}^{+0.002}$ & $0.415_{-0.003}^{+0.003}$ & $0.421_{-0.002}^{+0.002}$ & N/A & $179_{-3}^{+3}$ & $1.731_{-0.003}^{+0.003}$ \\
\end{tabular}
\end{ruledtabular}
\caption{
  \label{tab:parameter_estimates}
  Medians and 90\% credible intervals for selected source parameters for the O1 network (top), the HLVK network (middle) and the the ET-CE network (bottom).
  We show the detector-frame chirp mass $\mathcal{M}^\mathrm{det}$, the mass-ratio $q = m_2 / m_1 \leq 1$, the effective aligned spin $\chi_\mathrm{eff}$, effective precession spin $\chi_\mathrm{p}$, the luminosity distance $d_L$ and the inclination angle $\theta_\mathrm{JN}$ between the total angular momentum of the binary and the line of sight.
  Only \texttt{IMRPhenomPv2} provides a posterior for the effective precession spin $\chi_\mathrm{p}$, since \texttt{SEOBNRv4\_ROM} is a non-precessing model.
  }
\end{table*}

\begin{table*}[hptb]
\begin{ruledtabular}
\begin{tabular}{l l c c c c c c}
Event & Waveform model & $\mathcal{M}^\mathrm{det}/M_\odot$ & $q$ & $\chi_\mathrm{eff}$ & $\chi_\mathrm{p}$ & $d_L/\mathrm{Mpc}$ & $\theta_\mathrm{JN}$ \\[0.05cm]
\hline \\[-1.0em]

\multirow{2}{*}{SXS\_BBH\_0308} & IMRPhenomPv2  & -0.16 & -0.05 & -0.23  & -0.47 & -0.63  & -0.38 \\
                                & SEOBNRv4\_ROM & -0.34 & 0.27 & -0.55  & N/A   & 1.04 & 0.09 \\[+0.1em]
\hline
\\[-1em]
\multirow{2}{*}{SXS\_BBH\_0104} & IMRPhenomPv2  & 0.80 & 0.02  & 0.76 & -0.88 & -2.19 & 0.88 \\
                                & SEOBNRv4\_ROM & -0.25 & 0.09 & 0.05  & N/A  & -2.39 & 1.12 \\
\end{tabular}
\end{ruledtabular}
\vspace{0.5cm}
%
\begin{ruledtabular}
\begin{tabular}{l l c c c c c c}
Event & Waveform model & $\mathcal{M}^\mathrm{det}/M_\odot$ & $q$ & $\chi_\mathrm{eff}$ & $\chi_\mathrm{p}$ & $d_L/\mathrm{Mpc}$ & $\theta_\mathrm{JN}$ \\[0.05cm]
\hline \\[-1.0em]

\multirow{2}{*}{SXS\_BBH\_0308} & IMRPhenomPv2  & 2.21 & -1.28 & 2.73  & -4.32 & 1.10  & -2.30 \\
                                & SEOBNRv4\_ROM & 0.98 & -0.23 & 0.62  & N/A   & -0.02 & -0.37 \\[+0.1em]
\hline
\\[-1em]
\multirow{2}{*}{SXS\_BBH\_0104} & IMRPhenomPv2  & -0.08 & 0.93  & -1.09 & -1.29 & -5.91 & 3.91 \\
                                & SEOBNRv4\_ROM &  1.37 & -0.82 & 0.18  & N/A   & -1.52 & 1.10 \\
\end{tabular}
\end{ruledtabular}
\vspace{0.5cm}
%
\begin{ruledtabular}
\begin{tabular}{l l c c c c c c}
Event & Waveform model & $\mathcal{M}^\mathrm{det}/M_\odot$ & $q$ & $\chi_\mathrm{eff}$ & $\chi_\mathrm{p}$ & $d_L/\mathrm{Mpc}$ & $\theta_\mathrm{JN}$ \\[0.05cm]
\hline \\[-1.0em]

\multirow{2}{*}{SXS\_BBH\_0308} & IMRPhenomPv2  & 154.80 & -20.21 & 39.16 & -45.78 & -2.73 & -98.98 \\
                                & SEOBNRv4\_ROM &  153.90 & 15.59  & 55.09 & N/A    & 4.56  & 4.26 \\[+0.1em]
\hline
\\[-1em]
\multirow{2}{*}{SXS\_BBH\_0104} & IMRPhenomPv2  & -10.34 & -13.23  & -13.51  & 83.57 & -7.23 & 6.02 \\
                                & SEOBNRv4\_ROM & -30.67 & -42.55 & -48.59 & N/A   & 215.28 & -333.11 \\
\end{tabular}
\end{ruledtabular}
\caption{
  \label{tab:normalized_biases_PE}
  Normalized biases from 1D marginal posteriors for selected source parameters for the O1 network (top), the HLVK network (middle), and the ET-CE network (bottom). We show the bias
  $\theta_{s} - \theta_\mathrm{MAP}$ in the binary parameter $\theta$ divided by the standard deviation of $p(\theta | d)$
  $s$ refers to the value for the mock source and \ac{MAP} is the maximum a posteriori value of the posterior distribution, i.e. $\max_\theta p(\vec\theta | d)$.
  }
\end{table*}

In this section we provide a comparison between results obtained from two different waveform models, the agreement between these models and the source \ac{PN}-\ac{NR} waveforms and discuss the importance of limitations of the models in interpreting the parameter estimation results.

In this study we perform parameter estimation on signals in \emph{zero noise}. 
This is a particular noise realization that can be interpreted as the average over all possible Gaussian noise realizations.
It is an appropriate choice when one wants to focus on the effect of waveform systematics on posterior distributions. 
Therefore, any discrepancy we see between the posterior estimates and the true source parameters of the mock signals must be due to disagreements between the source and template waveforms or due to prior effects. 
Given that we use high accuracy \ac{NR} or \ac{PN}-\ac{NR} waveforms as the signal, which are good
approximations of \ac{GR} waveforms, and we analyse high \ac{SNR} events these disagreements are
assumed to come from approximations to \ac{GR} waveforms made in the waveform models we use as
templates.

We performed the parameter estimation analyses with the \texttt{IMRPhenomPv2} and \texttt{SEOBNRv4\_ROM} \ac{IMR} models. 
The assumptions made in these models are described in Sec.~\ref{sec:methods}. 
In Sec.~\ref{ssub:results} we presented results from the effective precessing \texttt{IMRPhenomPv2} model. 
Here we juxtapose these results against the posterior distributions obtained for the aligned-spin \texttt{SEOBNRv4\_ROM} model.
In Table~\ref{tab:parameter_estimates} we show medians and 90\% credible intervals for selected source parameters and the two \ac{BBH} sources for the O1, HLVK, and ET-CE networks. 
To gauge measurement accuracy we show absolute biases divided by the standard deviation in Table~\ref{tab:normalized_biases_PE}.
We find that the two models give overall similar results for the parameter estimates. 
Noticeable differences are as follows: 
The chirp mass for the HLVK network is recovered more accurately for \texttt{SEOBNRv4\_ROM} for SXS\_BBH\_0308 compared to \texttt{IMRPhenomPv2}. 
Similarly, \texttt{SEOBNRv4\_ROM} recovers the mass-ratio, effective aligned spin, luminosity distance and inclination angle with better accuracy than \texttt{IMRPhenomPv2} for SXS\_BBH\_0308 in the HLVK network. 
For SXS\_BBH\_0104 in the HLVK network, \texttt{SEOBNRv4\_ROM} does not recover the chirp mass very accurately, but finds the other selected source parameters with better accuracy than \texttt{IMRPhenomPv2}.
For the O1 network all parameters except distance and inclination are unbiased.
At HLVK sensitivity several parameters exceed unity in the modulus of the normalized bias, which indicates that the difference between true and \ac{MAP} parameter value is larger than one standard deviation. 
The largest biases are found in the luminosity distance and inclination for SXS\_BBH\_0104 recovered by the \texttt{IMRPhenomPv2} model and for the effective precession spin $\chi_\mathrm{p}$ for SXS\_BBH\_0308 found by \texttt{IMRPhenomPv2}.

Turning towards the ET-CE network we see in the size of the 90\% intervals that measurement precision has increased dramatically, for instance the chirp mass is measured to $\pm 0.002 M_\odot$, two orders of magnitude more accurately than for HLVK. 
The precision for the mass-ratio has increased by about one order of magnitude to roughly $\pm 0.005$ and similarly the effective aligned spin is measured to $\pm 0.002$ and the effective precession spin better than $\pm 0.02$.
In the ET-CE network all parameters shown here have normalized biases exceeding unity in their absolute value. 
All of these parameters are estimated to lie outside one standard deviation for the two waveform models employed here, making it clear that waveform models need to be improved for analyses with 3G detectors.

The \ac{PN}-\ac{NR} signal waveforms we used to represent the \acp{GW} emitted by the source binaries contain higher harmonics beyond $\ell = 2$, but the waveform models used as templates only include the dominant quadrupolar modes. 
In fact, the models do also not include all of the $m$ modes at $\ell = 2$, but merely the $\ell = 2, m=\pm 2$ contributions in the co-precessing frame. 
This begs the question how much the missing higher modes affect the analyses. In terms of \ac{SNR} $\rho$ for SXS\_BBH\_0308, 99.5\% of $\rho^2$ is found in the $(2, \pm 2)$ mode (ignoring precession), while for SXS\_BBH\_0104 95.6\% of the total $\rho^2$ is found in the $(2,\pm 2)$ mode, and 3.8\% in the $(3, \pm 3)$ mode. 
The above percentages are stated in terms of $\rho^2$ as \ac{SNR} adds in quadrature.
The overlap between a signal with and without higher harmonics at the signal parameters is $0.9997$ for SXS\_BBH\_0308 and $0.96$ for SXS\_BBH\_0104, which illustrates that higher modes only become important for higher mass-ratios. To compute these numbers we used the SEOBNRv4\_ROM~\cite{Bohe:2016gbl,Purrer:2014fza,Purrer:2015tud} and a SEOBNRv4HM\_ROM~\cite{Cotesta:2019inprep} waveform models and aLIGO design sensitivity with a starting frequency of 10 Hz. Computing the detector response~\eqref{eq:detector-response} and
optimizing over the polarization angle for the template waveform while keeping sky location 
fixed yields overlaps of $0.9993$ and $0.96$, instead.
For SXS\_BBH\_0308 (SXS\_BBH\_0104), the overlap between an \ac{NR} waveform that includes all $\ell=2$ modes vs a waveform that only includes the $(2,\pm 2)$ modes in the co-precessing frame is $0.99996$ ($0.99992$), or $0.9992$ when optimizing over the polarization angle. This shows that the for these configurations the $(2, \pm 1)$ modes in the co-precessing frame are very weak.

As we have seen in Sec.~\ref{sub:predicted_waveform_accuracy_requirements},
overlaps between the \ac{PN}-\ac{NR} signals and the two waveform models at the signal parameters for aLIGO design are significantly lower than the overlaps which include or leave out higher modes. They are $0.97$ ($0.91$) for the SXS\_BBH\_0308 hybrid signal and $0.91$ ($0.94$) for SXS\_BBH\_0104 using \texttt{IMRPhenomPv2} (\texttt{SEOBNRv4\_ROM}).
This indicates that the disagreement between the signal and template waveforms comes predominantly from modeling error in the co-precessing frame $(\ell, m) = (2, \pm 2)$ mode.
Some disagreement could also come from the approximate description of precession.

For all analyses presented in this study, we have assumed that the zero noise data still carries an inherent uncertainty in its calibration, as described in Sec.~\ref{sub:bayesian_parameter_estimation}.
This uncertainty is modelled as a cubic spline, enforcing a smooth variation across the bandwidth of the analysis.
In principle, by allowing this additional degree of freedom which could absorb some of the mismatch between the \ac{PN}-\ac{NR} hybrids and the approximate waveform models used in the \ac{PE} analysis the observed biases could be expected to be reduced.
Comparing the 1D posterior distributions shown in Fig.~\ref{fig:0308_posterior_PDFs} to distributions from analyses where the marginalisation over calibration uncertainties has been disabled, the observed biases remain. 
It should be noted that the analyses which includes marginalisation over calibration uncertainties systematically recovers a slightly higher \ac{SNR} accumulated over the detector network, but as this increase is of order $\lesssim 1/1000$ this is not expected to affect the conclusions with any significance.



\section{Population study}
\label{sec:population}

We have seen in Sec.~\ref{sec:results-golden-binaries} that in the HLVK design network we already expect biases with current waveform models for loud \acp{BBH} such as GW150914. 
Even small biases found for weaker single events could still manifest themselves when estimating properties of the population of \acp{BBH}~\cite{Kovetz:2016kpi,Wysocki:2018mpo, Talbot:2018cva,LIGOScientific:2018jsj}. 
In this section we perform a \ac{PE} analysis for a population consisting of one hundred high mass precessing \ac{BBH} events.
On the one hand we study the distribution and correlation of parameter biases and compute the overall bias over the population. 
On the other hand we analyze the residual between the signal and the best matching template waveform, in terms of its \ac{SNR}, power in the time frequency plane, and in terms of Bayes factors between analyses assuming coherent and incoherent signals across the detector network as implemented with \bayeswave~\cite{Cornish:2014kda,Littenberg:2015kpb}.
Finally, we compute the population posterior for the power law index of the primary mass of the source binaries.

\subsection{Setup} 
\label{sub:setup}

The events in this population study were drawn from the following distribution of source parameters: 
The primary mass has a \ac{PDF} $p(m_1) \propto m_1^{-\alpha}$ with $\alpha = 1.3$ and $m_1 \in [45, 50] M_\odot$ and the mass-ratio is distributed as $q \sim U(0.5, 1)$. 
The chirp mass $\mathcal{M} = M_\mathrm{tot} \eta^{3/5}$ is computed from $(m_1, q)$, where $M_\mathrm{tot} = m_1 + m_2$ and $\eta = q / (1+q)^2$.
The remaining parameters are distributed as follows: spin magnitudes $a_i \sim U(0, 0.8)$, spin tilts $\cos t_i \sim U(-1, 1)$, the azimuthal angle between the spin vectors $\phi_{12} \sim U(0, 2\pi)$, the angle between the total and orbital angular momentum $\phi_\mathrm{JL} \sim U(0, 2\pi)$, the inclination angle $\cos\theta_\mathrm{JN} \sim U(-1, 1)$, the polarization angle $\psi \sim U(0, \pi)$.
The luminosity distance, geocenter time, sky location, and phase were fixed at the parameters given in Table~\ref{tab:signal-configurations}.

Since \ac{NR} waveforms are only available at isolated points in parameter space and thus cannot well represent the above distribution we choose the \texttt{NRSur7dq2} \ac{NR} surrogate model for the signal waveforms~\cite{Blackman:2017pcm}. 
This choice implies restrictions to mass-ratio $q \geq 1/2$, spin magnitudes $a_i \leq 0.8$ and, due to the relatively short waveform length, the constraint on the primary mass $m_1 \geq 45 M_\odot$, so that waveforms representing the \ac{BBH} population start at or below 20 Hz. 
We perform \ac{PE} analyses with the \bilby code~\cite{Ashton:2018jfp,Bilby-code} with signals in zero noise and \texttt{IMRPhenomPv2} templates for the HLVK network.


\subsection{Bias} 
\label{sub:bias_results}

We can learn about how population parameters will be affected by studying correlations between biases in key source parameters for events drawn from a population and to what degree single event biases average out over the population. 
In Fig.~\ref{fig:absolute_biases_pop} we show absolute biases, defined as the difference between the true source parameters $\theta_{s}^i$ and a point estimate of the posterior distribution
$\theta_\mathrm{p}^i$ for event $i$
\begin{equation}
  \label{eq:abs_bias}
  \mathcal{B}_i := \theta_{s}^i - \theta_\mathrm{p}^i.
\end{equation}
As a default we use the \ac{MAP} value of the posterior distributions as the point estimate.

We see that biases are large when the signal is represented by \texttt{NRSur7dq2} waveforms and the template by the \texttt{IMRPhenomPv2} model. 
In contrast, when the signal and template are represented by the same \texttt{IMRPhenomPv2} waveforms there is only a small discrepancy between the \ac{MAP} and the true signal parameters which is expected to arise from stochastic sampling and prior effects. 
Here the posterior distribution is dominated by the likelihood since the signal SNRs are high. 
We find that log-likelihood values come close, but are a bit lower than, the peak value of the
log-likelihood at the signal parameters. While the \ac{MAP} (or equivalently maxL) parameters
are a bit different than the signal parameters, the deviations in the waveform are tiny and the 
SNR in the residual is on the order of one.
The spread is a factor 7 smaller in chirp mass, a factor 4 smaller in effective aligned spin and a factor 2 smaller in mass-ratio.
For both types of signals we observe pronounced correlations between these parameters which we expect on physical grounds due to how these parameters enter the inspiral waveform~\cite{Poisson:1995ef,Cutler:1994ys,Baird:2012cu,Purrer:2015nkh,Purrer:2013xma,Vitale:2014mka,Vitale:2016avz,Abbott:2016wiq,Ng:2018neg}. 
We find Pearson correlation coefficients of $R_{\mathcal{M}, \chi_\mathrm{eff}} \sim 0.8 (0.8)$ and $R_{\mathcal{M}, q} \sim 0.5 (0.3)$ for \texttt{NRSur7dq2} (\texttt{IMRPhenomPv2}) signals. 
For NR-surrogate signals the chirp mass shows a clear tendency to be \emph{overestimated}. 
This is also true for effective spin and mass-ratio. In contrast, for \texttt{IMRPhenomPv2} signals the distribution of the single event biases is more symmetrical. 
We also see that for NR-surrogate signals heavy binaries are prone to \emph{overestimation} of $\chi_\mathrm{eff}$ as indicated by the luminosity of the red pentagons.

In Fig.~\ref{fig:absolute_biases_pop_distance_cos_inclination} we see that bias in the luminosity distance tends to be negative, and with the definition in Eq.~\ref{eq:abs_bias} this implies that the distance is overestimated in inference as a rule. The distance bias is reduced by about half for \texttt{IMRPhenomPv2} signals, compared to \texttt{NRSur7dq2} signals, but it is still sizable. In contrast we find relatively small biases of about $10^{\degree}$ in the inclination angle.

To get a better sense of how much these biases matter we discuss the distribution of the ratio $\mathcal{R}$ of absolute biases and 90\% credible intervals 
\begin{equation}
  \label{eq:bias-ratio}
  \mathcal{R}_i := 
  \frac{\theta_{s}^i - \theta_\mathrm{MAP}^i}{\mathrm{CI}_{90}[p(\theta^i|d^i)]}
\end{equation}
shown in Fig.~\ref{fig:absolute_biases_over_CI90}, where we divide the bias by the extent of the 90\% credible interval for each event and parameter. 
For NR-surrogate signals $|\mathcal{R}|$ reaches unity for the chirp-mass, takes values up to $2$ for the effective aligned spin, and about $1.5$ for the mass-ratio which indicates that the parameter recovery is strongly biased. 
The choice of comparing to the 90\% interval is more conservative than to $1-\sigma$ which is assumed in the indistinguishability criterion.
In contrast, for \texttt{IMRPhenomPv2} signals $|\mathcal{R}|$ is smaller than $0.4$ for all parameters and the majority of events are found with very good accuracy $|\mathcal{R}| \lesssim 0.1$.

We show the overall bias over the population in Table~\ref{tab:pop-biases}. 
For NR-surrogate signals the largest population bias is seen for the \ac{MAP}. Using the mean or
median as a point estimate the overall bias is significantly lower than when using the \ac{MAP}.
This is not the case for \texttt{IMRPhenomPv2} signals, where the largest bias is found for the mean.
We also show the sum of ratios of the biases over the 90\% intervals, $\sum_i \mathcal{R}_i$.
The size of this quantity shows more clearly how severe the biases are overall averaged over the
population. Again the sum of the biases is much larger for the NR-surrogate signals, 
about 10 -- 30 times larger than the sum of 90\% credible intervals for key astrophysical
parameters.
The magnitude of $\sum_i \mathcal{R}_i$ for \texttt{IMRPhenomPv2} signals is about $\sim 2$ indicating that there is no significant bias when combining all events in the population.
We will revisit the question of how population estimates are affected in Sec.~\ref{sub:population_inference}.


\subsection{Residuals} 
\label{sub:residuals_results}

We previously discussed biases found for events in the \ac{BBH} population study. 
The biases stem from a disagreement between the signal NR-surrogate waveforms $h_\mathrm{s}(t; \theta_s)$ and \texttt{IMRPhenomPv2} template waveforms $h_\mathrm{m}(t; \theta_s)$ used in the analysis at the source parameters $\theta_s$. 
This disagreement will also lead to some residual power being left over after subtracting the data containing the signal from the best fit template waveform, $h_\mathrm{m}(t; \theta_\mathrm{MAP})$. 
Here we discuss how this residual power can be characterized in terms of \ac{SNR} and power in the time frequency plane. 
We also perform an analysis with \bayeswave.

\begin{figure*}[h]
  \centering
  \includegraphics[width=.45\textwidth]{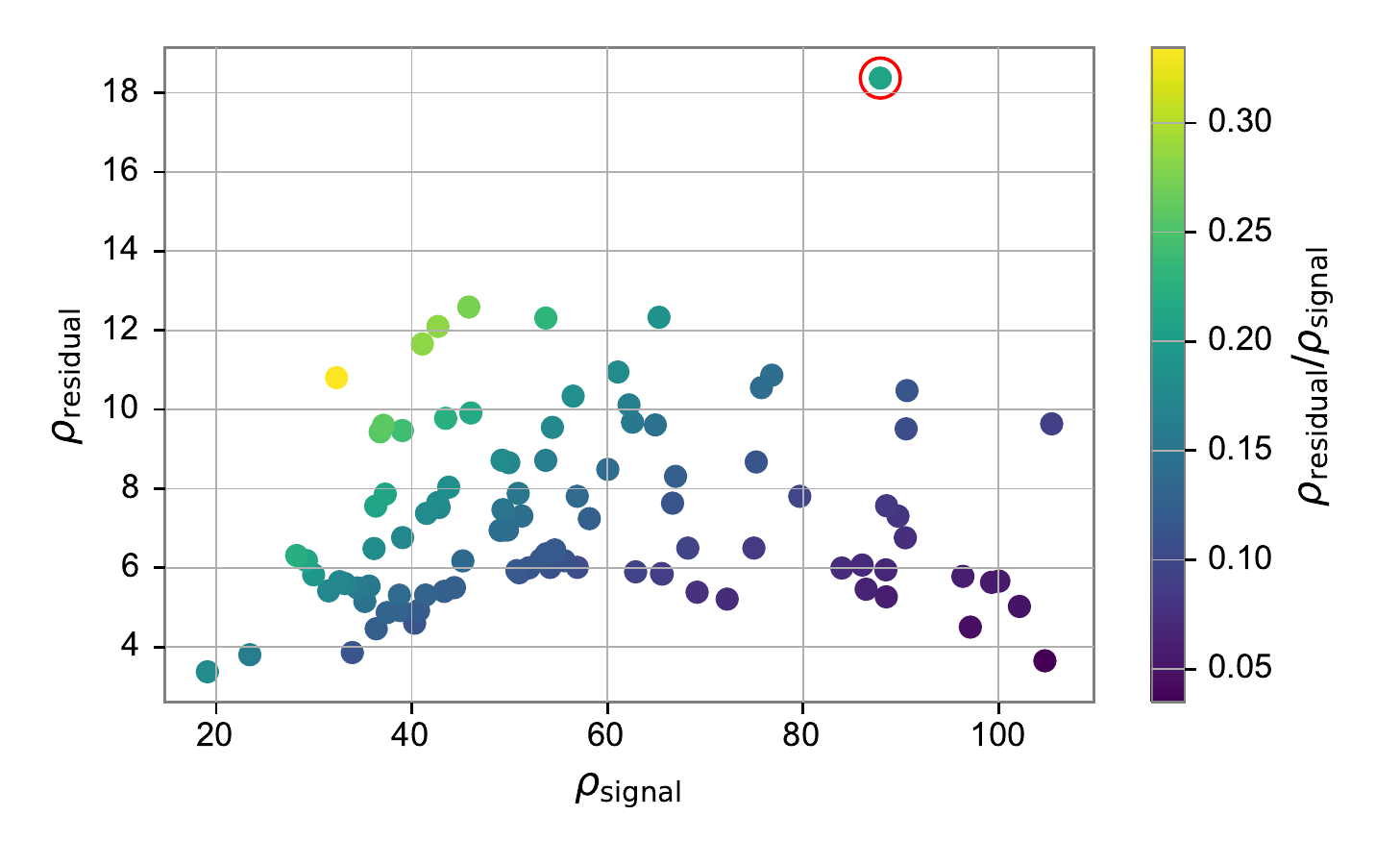}
  \includegraphics[width=.45\textwidth]{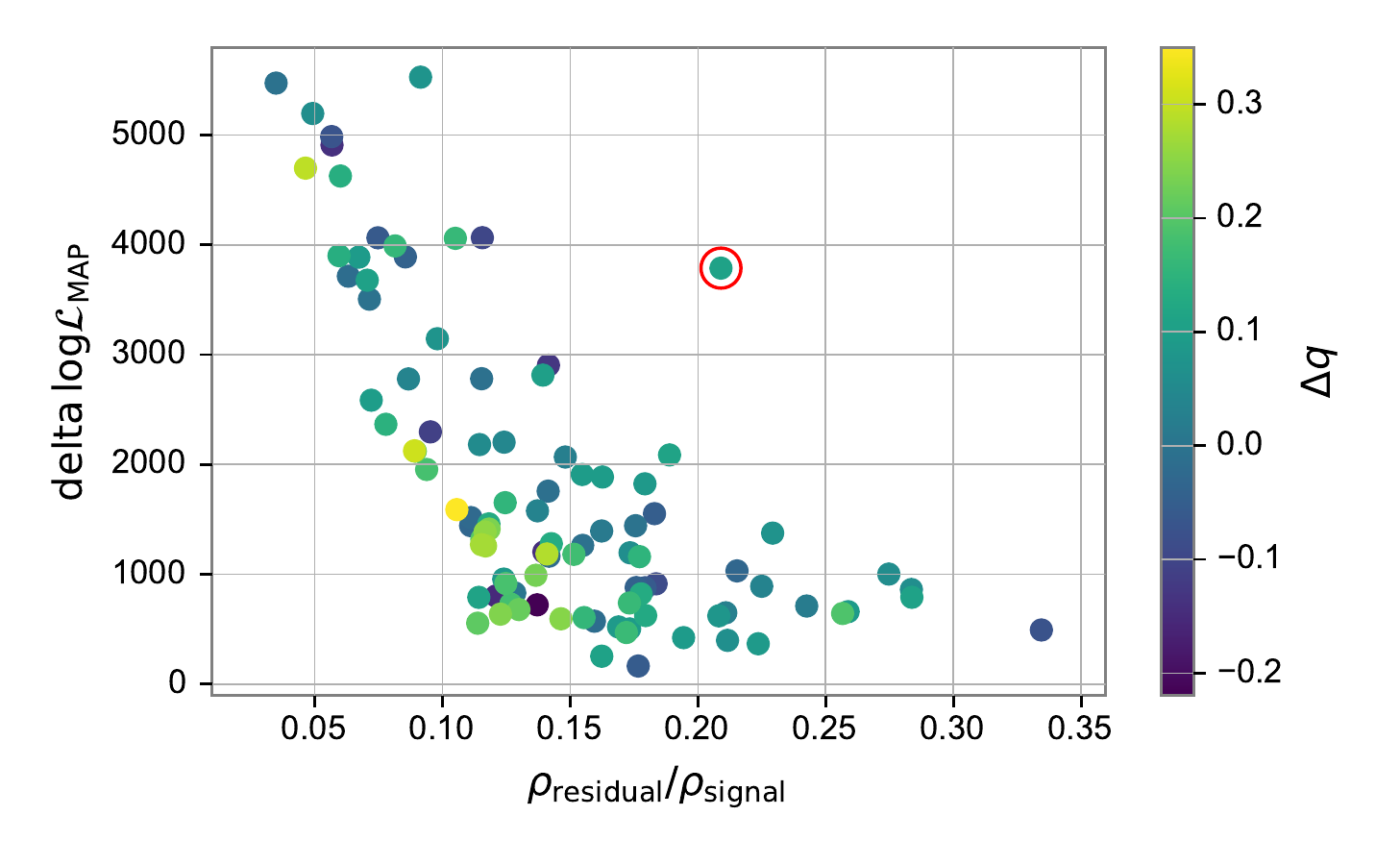}
  \caption{Fractional network \ac{SNR} in strain residuals $h_\mathrm{s}(t; \theta_s) - h_\mathrm{m}(t; \theta_\mathrm{MAP})$ between the \texttt{NRSur7dq2} signal strains $h_\mathrm{s}(t; \theta_s)$ and \texttt{IMRPhenomPv2} \ac{MAP} template strain $h_\mathrm{m}(t; \theta_\mathrm{MAP})$ for each detector in the HLVK network compared to the \ac{SNR} of the respective signal strain.
  \emph{Left:} Residual \ac{SNR} as a function of signal (injection) \ac{SNR} with fractional \ac{SNR} indicated by color for the events in the population study.
  \emph{Right:} Delta Log-likelihood $\langle d|h \rangle - 0.5\langle h|h \rangle$ at \ac{MAP} against fractional \ac{SNR} colored by the mass-ratio bias. The bias in mass-ratio is indicated by color. 
  The event with the highest residual \ac{SNR} is indicated by a red circle around the marker. The parameters for this event are given in Table~\ref{tab:parameters-loudest-residual}.
  }
  \label{fig:Strain_residual_SNR}
\end{figure*}

In Fig.~\ref{fig:Strain_residual_SNR} we show the network \acp{SNR} found in the signal strain $h_\mathrm{s}(t; \theta_s)$ and in the residual strain $h_\mathrm{s}(t; \theta_s) - h_\mathrm{m}(t; \theta_\mathrm{MAP})$. 
In each detector of the network we compute the strain by projecting the waveform polarizations on the detector as defined in Eq.~\eqref{eq:detector-response}. 
We observe that residuals reach \acp{SNR} of about 12, expect for one event with residual \ac{SNR} $\sim 18.37$. 
Parameters for this event are shown in Table~\ref{tab:parameters-loudest-residual}. 
We find residual \acp{SNR} up to 30\% of the signal \ac{SNR}. 
The log-likelihood at \ac{MAP} is highest for events where the agreement between the signal and template waveforms is good and thus the residual is small, and it drops substantially for events where the residual contains a sizable fraction of the signal \ac{SNR}. 
For the event with the highest residual \ac{SNR} the biases are only moderate $\Delta\mathcal{M} = -0.27 M_\odot$, $\Delta q = 0.11$, $\Delta\chi_\mathrm{eff} = -0.02$, and $\Delta\chi_\mathrm{p} = -0.04$, but it has a high signal \ac{SNR} $87.91$.

\begin{figure*}[h]
  \centering
  \includegraphics[width=.45\textwidth]{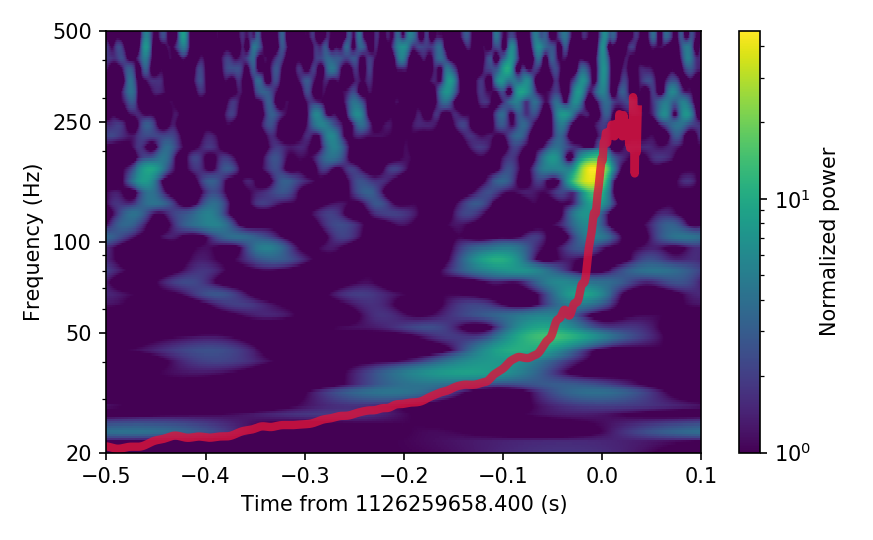}
  \includegraphics[width=.45\textwidth]{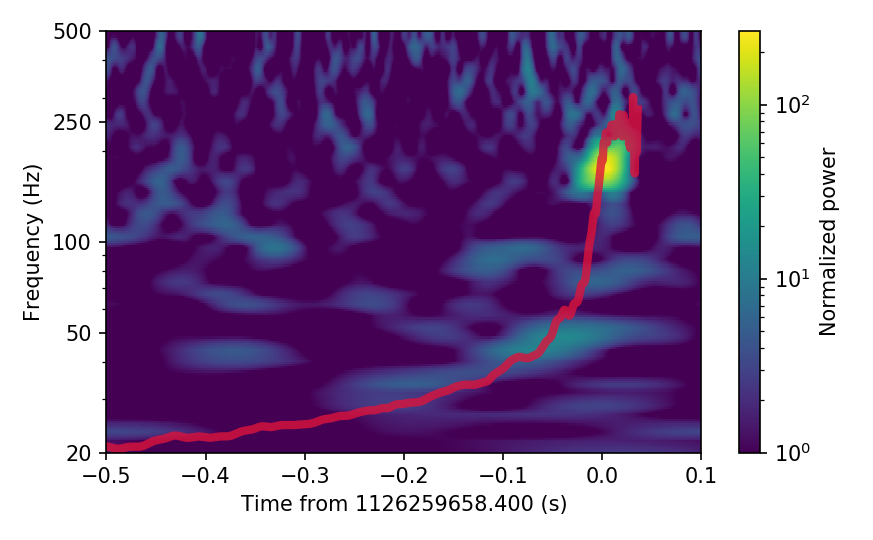}
  \caption{Q-transform of strain residuals for the event with the loudest residual (optimal network SNR 18.372). The parameters for this event are given in Table~\ref{tab:parameters-loudest-residual}.
  The strain residuals were computed by subtracting the \texttt{IMRPhenomPv2} \ac{MAP} template from the \texttt{NRSur7dq2} signal for each detector. Gaussian colored noise was added to the residual before computing the Q-transform.
  \emph{Left:} Residual in the interferometer where the residual is loudest: LIGO Hanford, \ac{SNR} $11.5$.
  \emph{Right:} Coherent sum of the time-shifted residuals in LIGO Hanford, LIGO Livingston, Virgo and KAGRA.
  The normalized power is shown in color.
  The chirp-trace of the injected \texttt{NRSur7dq2} signal is shown in crimson.
  }
  \label{fig:qscan-residuals}
\end{figure*}

Next we take a look at the power in the time frequency plane and compare the loudest residual against a chirp signal. 
In Fig.~\ref{fig:qscan-residuals} we plot the Q-transform~\cite{Brown91Q} using PyCBC~\cite{pycbc_release_zenodo} of the residual with the highest \ac{SNR}. 
As shown in the left panel, the power in the residual in LIGO Hanford traces out a chirp signal and agrees well with the overlaid time frequency evolution of the waveform emitted by the source. 
The right panel shows that the coherent power in the residual (taking into account time-shifts for each detector) is about a factor 5 larger than the power of the loudest single detector residual. 
Most of the coherent residual power is concentrated near merger where the GW signal is most non-linear.

\begin{figure*}[t]
  \centering
  \includegraphics[width=.45\textwidth]{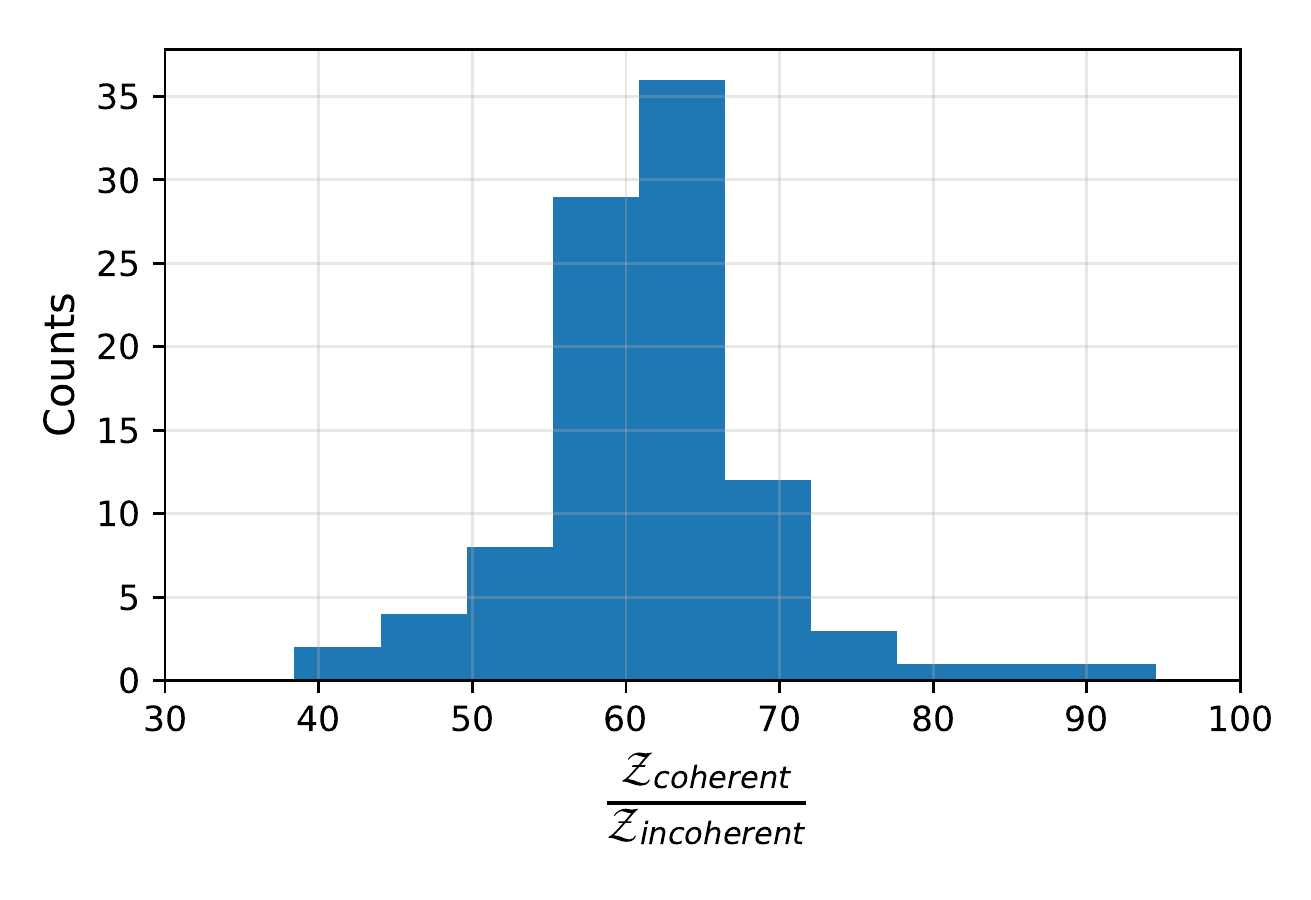}

  \caption{
Histogram of the Bayes factors, for all residuals (except the event highlighted in red in Fig~\ref{fig:Strain_residual_SNR} and Table~\ref{tab:parameters-loudest-residual}), between a coherent (assuming the same incoming signal in all participating detectors) and an incoherent (where the signal in each detector is assumed independent) model \cite{Littenberg:2015kpb}.
Both the coherent and incoherent models are constructed from a superposition of Morlet-Gabor wavelets, where the number of wavelets is in itself a variable, as implemented in \bayeswave~\cite{Cornish:2014kda}.
This shows unequivocaly that although the modelled analysis, where a known GR waveform approximants attempts to match the signal that best matches what is observed across the detector network, there is a significant fraction of observable coherent signal left.
The properties of this left-over signal are not strongly constrained by this analysis however, as expected by the typical SNR $\sim 12$ for these residual signals.
The excluded event, with properties listed in Table~\ref{tab:parameters-loudest-residual}, has a Bayes factor of $\sim 4\times 10^{11}$.
  }
  \label{fig:BW_BF}
\end{figure*}

Finally, we analyse the residual strain across the detector network using the \bayeswave~\cite{Cornish:2014kda} code, assuming no pre-defined signal model apart from constraining signals coherent across the detector network to an elliptical polarization.
Here, the waveform is reconstructed directly, through a superposition of Morlet-Gabor, or sine-Gaussian, wavelets~\cite{Cornish:2014kda,Gabor1946}, where the number, placement and properties of the wavelets are themselves variables in the analysis.
For this study, we compare two competing models for the observed residual data~\cite{Littenberg:2015kpb}.
The coherent model assumes a common waveform across the entire network, as originating from a point in the sky and projected onto each detector assuming standard antenna pattern functions for the two tensorial polarization modes as defined in Sec.~\ref{sub:bayesian_parameter_estimation}\footnote{This is also the model generally assumed in \lalinference and \bilby}.
The incoherent model assumes complete independence between the observed signals across the network. 
Instead of the data being represented through a common set of wavelets projected onto the detectors this model constructs a separate waveform for each detector where the placement and structure of the wavelets is independent from other detectors and no phase and time coherence across the network is required.
The two models\footnote{The nomenclature used internally by  \bayeswave is to call the coherent model \textit{signal} and the incoherent model \textit{glitch}.} can then be directly compared through a Bayes factor for each set of residual strains as shown in Fig.~\ref{fig:BW_BF}.
As \bayeswave is constructed, it has a strong dependance on signal complexity, as opposed to simply depending on signal strength only, in order to make observational claims such as for example preferring a coherent description of the signal over an incoherent one~\cite{Kanner:2015xua}.
This means that the Bayes' factors inherently incorporate the Occam factor between the two models, where the incoherent model can require a larger number of wavelets (and hence a larger number of signal parameters) to reconstruct the data across the network as it does not need to consider extrinsic parameters (sky position and two angles describing the polarization and ellipticity of the gravitational wave).
For the set of residual strains in this study, we often find the incoherent model incapable of capturing the signal in an individual detector, with a median number of 0 wavelets per detector.
The coherent signal on the other hand always captures the common signal, but even here the median number of wavelets is ``only'' 1.
We interpret this as \bayeswave being consistently able to determine that there is \textit{something} originating from a common coherent source in the data, but due to the relatively low \ac{SNR} we are not generally able to make strong inference on the physical description of what this coherent signal would be.
Even so, we argue that this type of analysis will be a valuable tool in determining the power and accuracy of future modelled inference~\cite{TheLIGOScientific:2016src,LIGOScientific:2019fpa}, and can ensure that all of the observable signal can be captured and characterized.
Note that the analysis here is performed in a noise-free set of data, assuming a known and fixed set of detector sensitivities shown in Fig.~\ref{fig:plots_ASDs_evolution}.
For ``real'' data, the presence of time-varying random Gaussian noise~\cite{LIGOScientific:2019hgc}, as well as actual detector glitches~\cite{Zevin:2016qwy}, is expected to reduce the fidelity of this category of tests, however \bayeswave is already capable of accounting for such variance~\cite{Littenberg:2014oda, Kanner:2015xua}.
The level to which variations in data will affect a study of residual recovery will be left for future investigation.


\subsection{Population inference} 
\label{sub:population_inference}

\begin{figure*}[h]
  \centering \includegraphics[width=.45\textwidth]{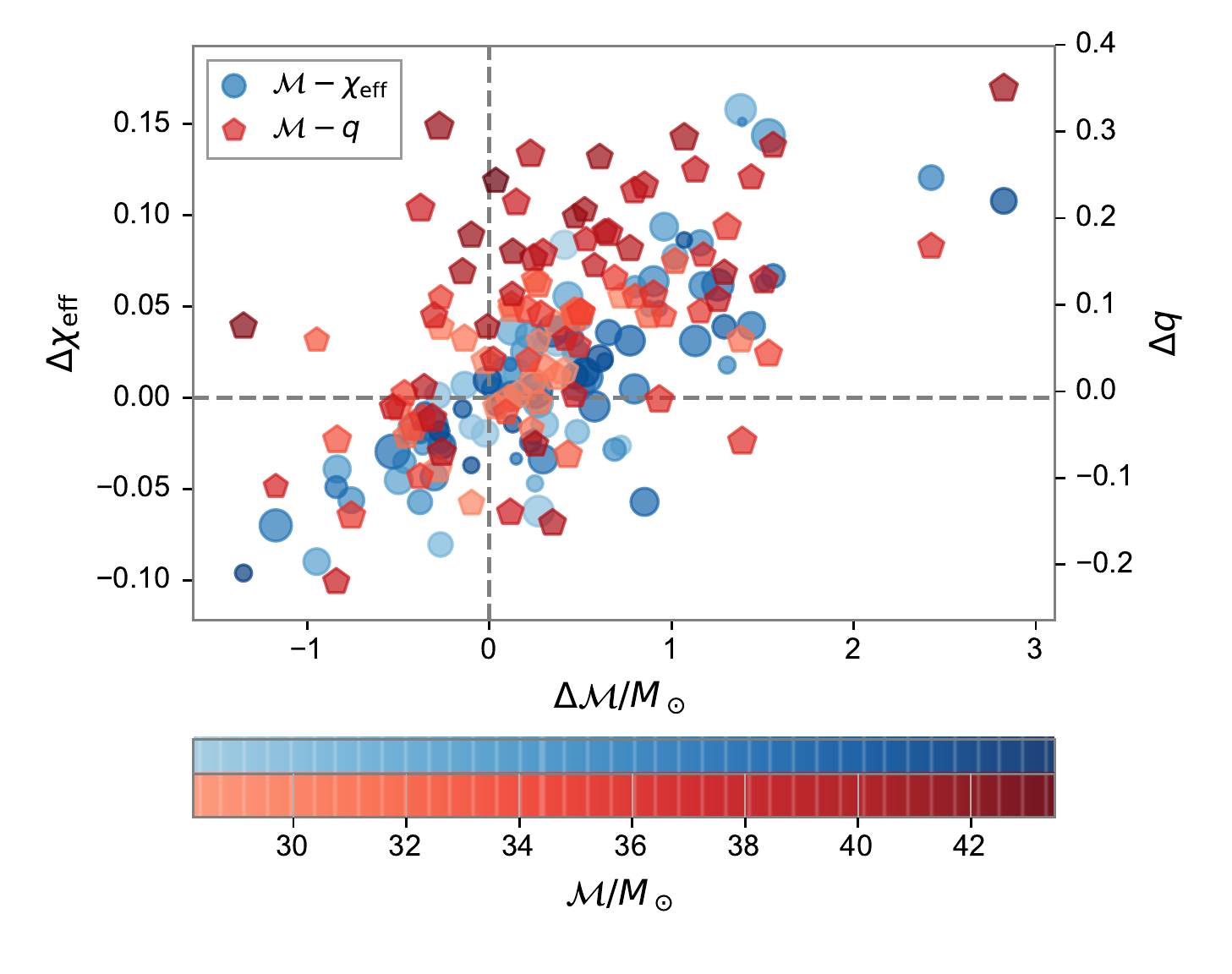} \includegraphics[width=.45\textwidth]{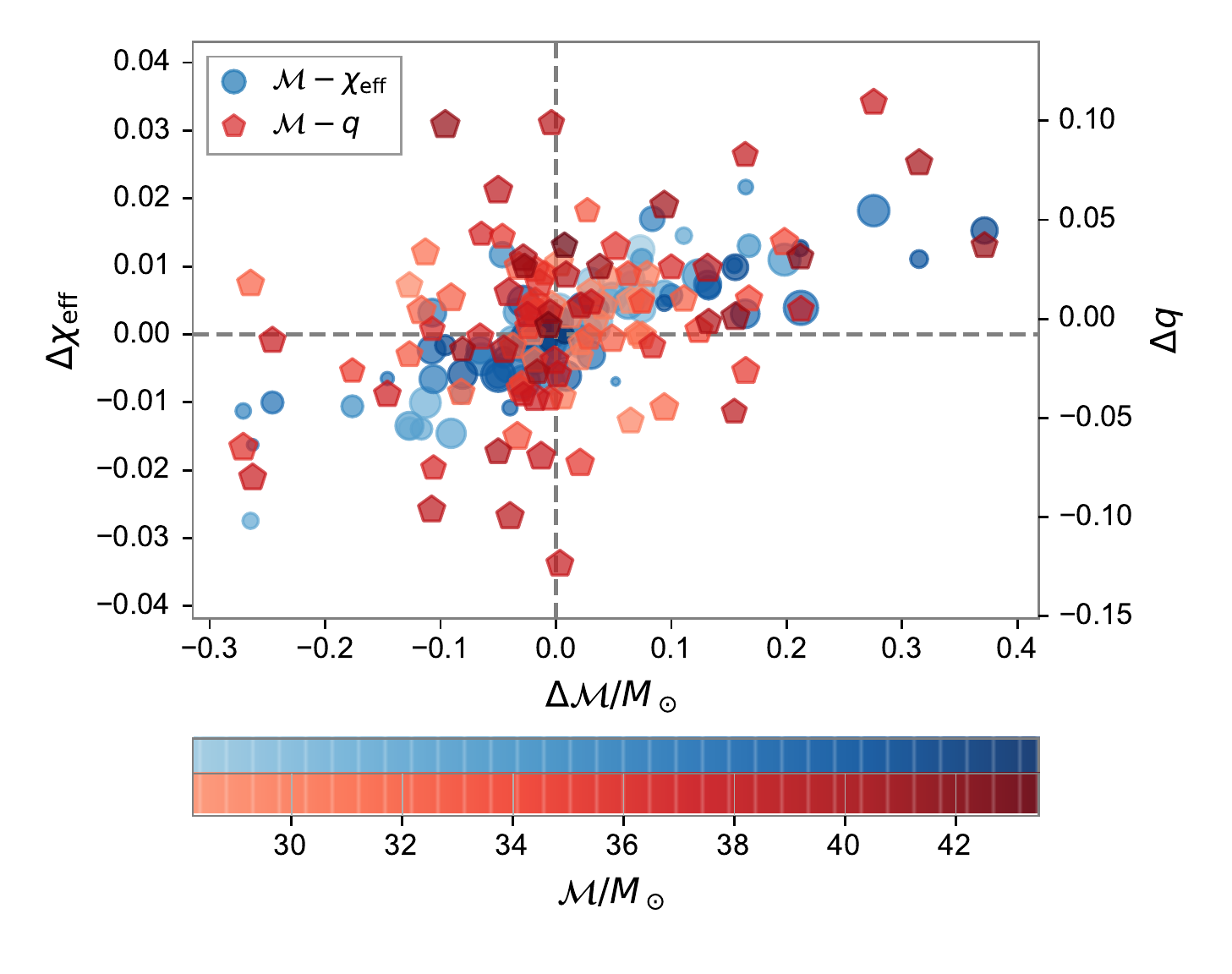}
  \caption{Absolute biases $\theta_{s} - \theta_\mathrm{MAP}$ between chirp-mass $\mathcal{M}/M_\odot$ and effective aligned-spin $\chi_\mathrm{eff}$ (blue circles) and chirp-mass and mass-ratio $q$ (red pentagons). 
  Left: \texttt{NRSur7dq2} signals recovered with \texttt{IMRPhenomPv2}. The Pearson correlation coefficients are $R_{\mathcal{M}, \chi_\mathrm{eff}} \sim 0.8$ and $R_{\mathcal{M}, q} \sim 0.5$.
  Right: \texttt{IMRPhenomPv2} signals recovered with \texttt{IMRPhenomPv2}. The Pearson correlation coefficients are $R_{\mathcal{M}, \chi_\mathrm{eff}} \sim 0.8$ and $R_{\mathcal{M}, q} \sim 0.3$.
  We indicate the signal chirp mass by the luminosity of the symbol colors and the absolute value of the effective aligned-spin and mass-ratio of the signal by the symbol area to give a sense of how the biases are distributed over the parameter space.
  The symbol area was calculated as $200 \sqrt{|\chi_\mathrm{eff}|}$ and $100 \sqrt{q}$.
  }
  \label{fig:absolute_biases_pop}
\end{figure*}

\begin{figure*}[h]
  \centering \includegraphics[width=.45\textwidth]{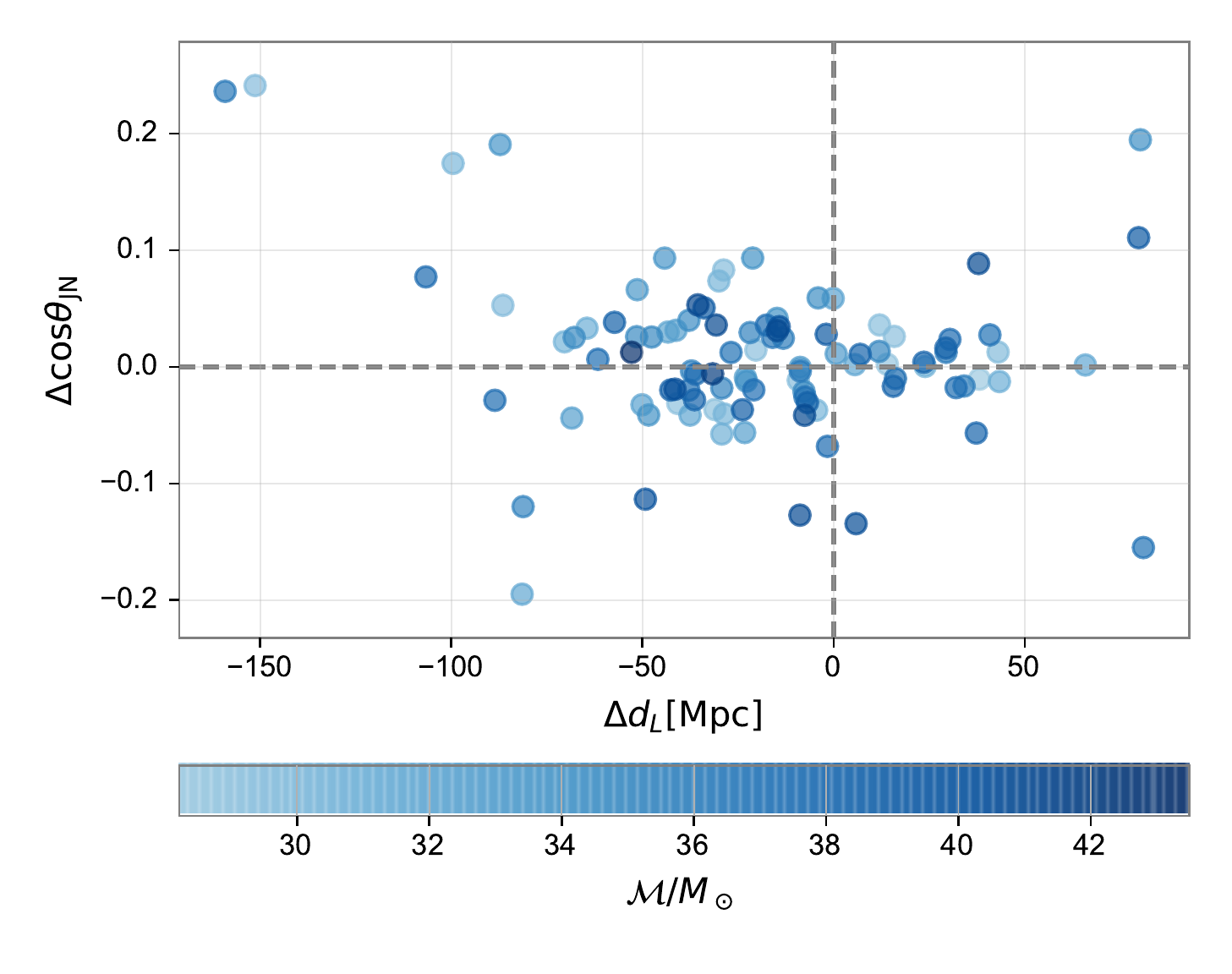} \includegraphics[width=.45\textwidth]{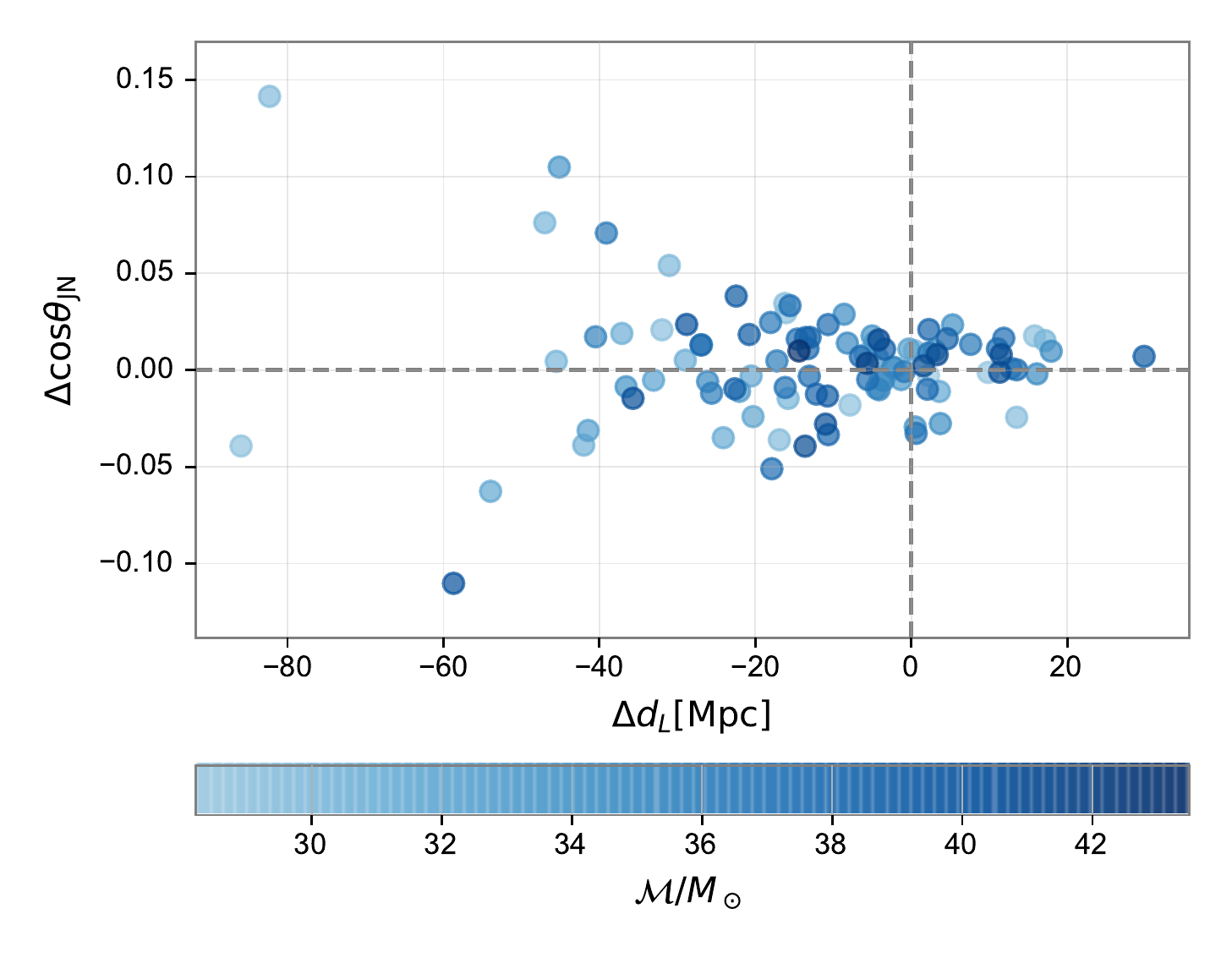}
  \caption{Absolute biases $\theta_{s} - \theta_\mathrm{MAP}$ between luminosity distance and 
  the cosine of the inclination angle
  Left: \texttt{NRSur7dq2} signals recovered with \texttt{IMRPhenomPv2}.
  Right: \texttt{IMRPhenomPv2} signals recovered with \texttt{IMRPhenomPv2}.
  As in Fig. \ref{fig:absolute_biases_pop} we indicate the signal chirp mass by the luminosity of the symbol colors.
  }
  \label{fig:absolute_biases_pop_distance_cos_inclination}
\end{figure*}

\begin{figure*}[]
  \centering
  \includegraphics[width=.45\textwidth]{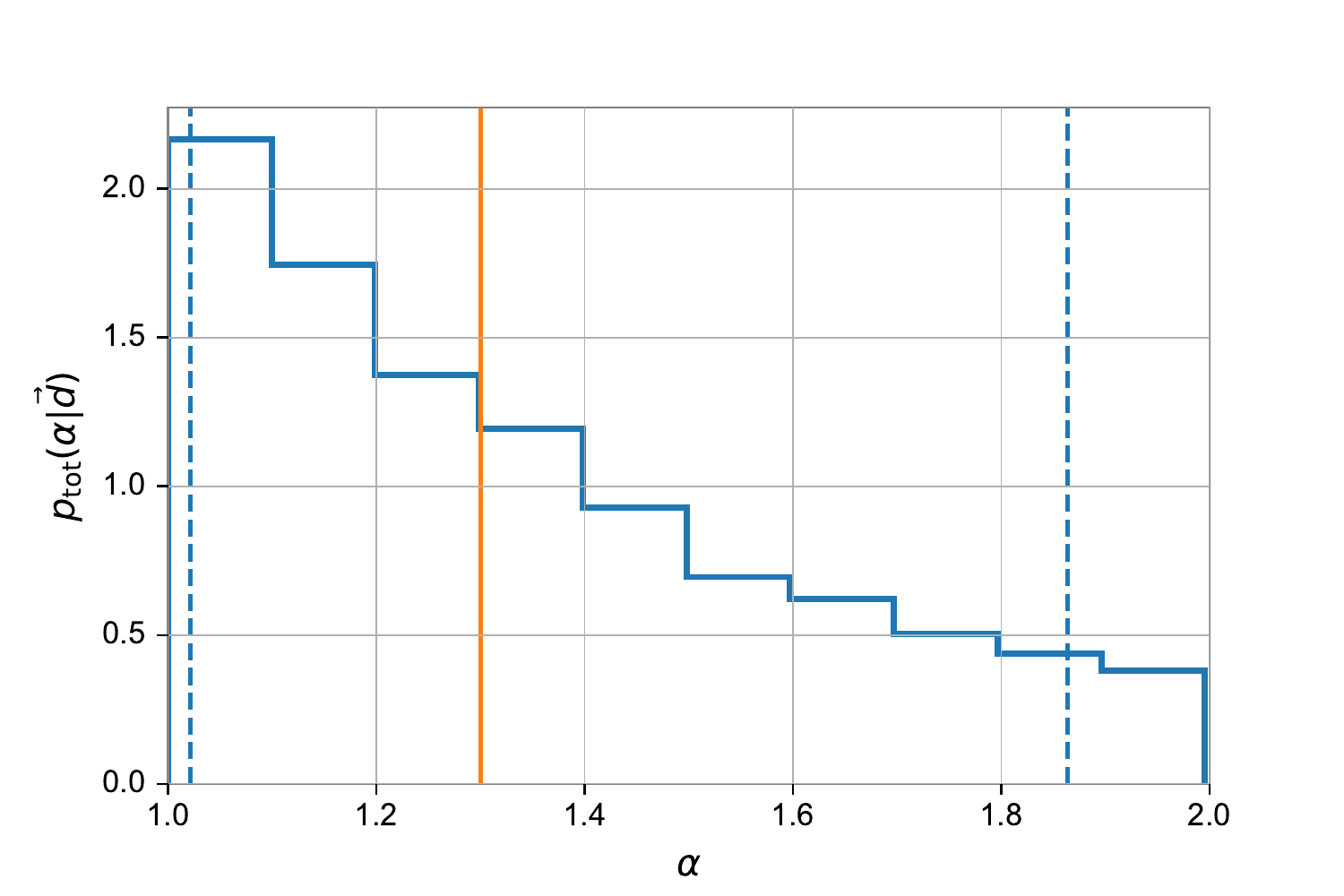}
  \includegraphics[width=.45\textwidth]{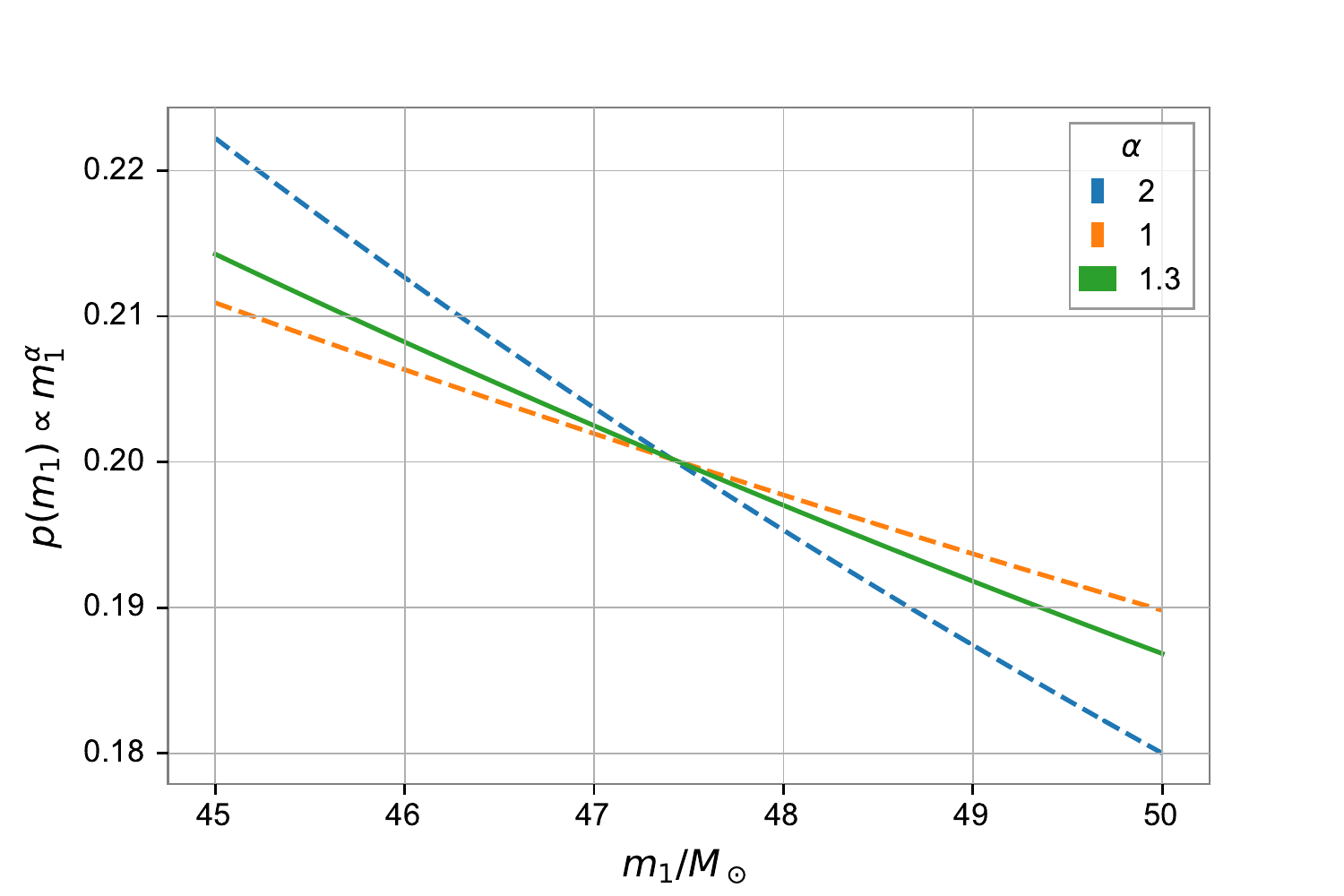}
  \caption{Left: hyper-posterior $p_\mathrm{tot}(\alpha| \vec d|)$ for power-law index $\alpha$ for $m_1$. Right: power-law PDFs for $m_1$ for $\alpha = 1.3$ (true value), and the bounds of the uniform hyper-prior on $\alpha$: $\alpha = 1, 2$.
  }
  \label{fig:hyper_posterior_alpha}
\end{figure*}

We follow the hierarchical Bayesian inference method described in Sec.~\ref{ssub:hierarchical_inference}. 
We show the hyper-posterior for $\alpha$, the power-law index of the primary \ac{BH} mass in Fig.~\ref{fig:hyper_posterior_alpha}, where we assumed a hyper-prior $\pi(\alpha) \sim U(1, 2)$. 
Unfortunately, the PDFs of the power-law distribution for the true value and the boundary values of $\alpha$ are rather similar over the narrow mass interval considered here. 
This is probably due to the rather tight lower mass bound which is set by the finite length of the \texttt{NRSur7dq2} waveform model. 
Given that there is not much information in the hyper-posterior we ask the question whether we prefer $\alpha=1$ or $\alpha=2$. Clearly, $\alpha=1$ is preferred by the hyper-posterior. 
This agrees with the observation that in the single event posterior PDFs we overestimate the masses (see Fig.~\ref{fig:absolute_biases_pop}).



\begin{figure*}[h]
  \centering
    \includegraphics[width=.45\textwidth]{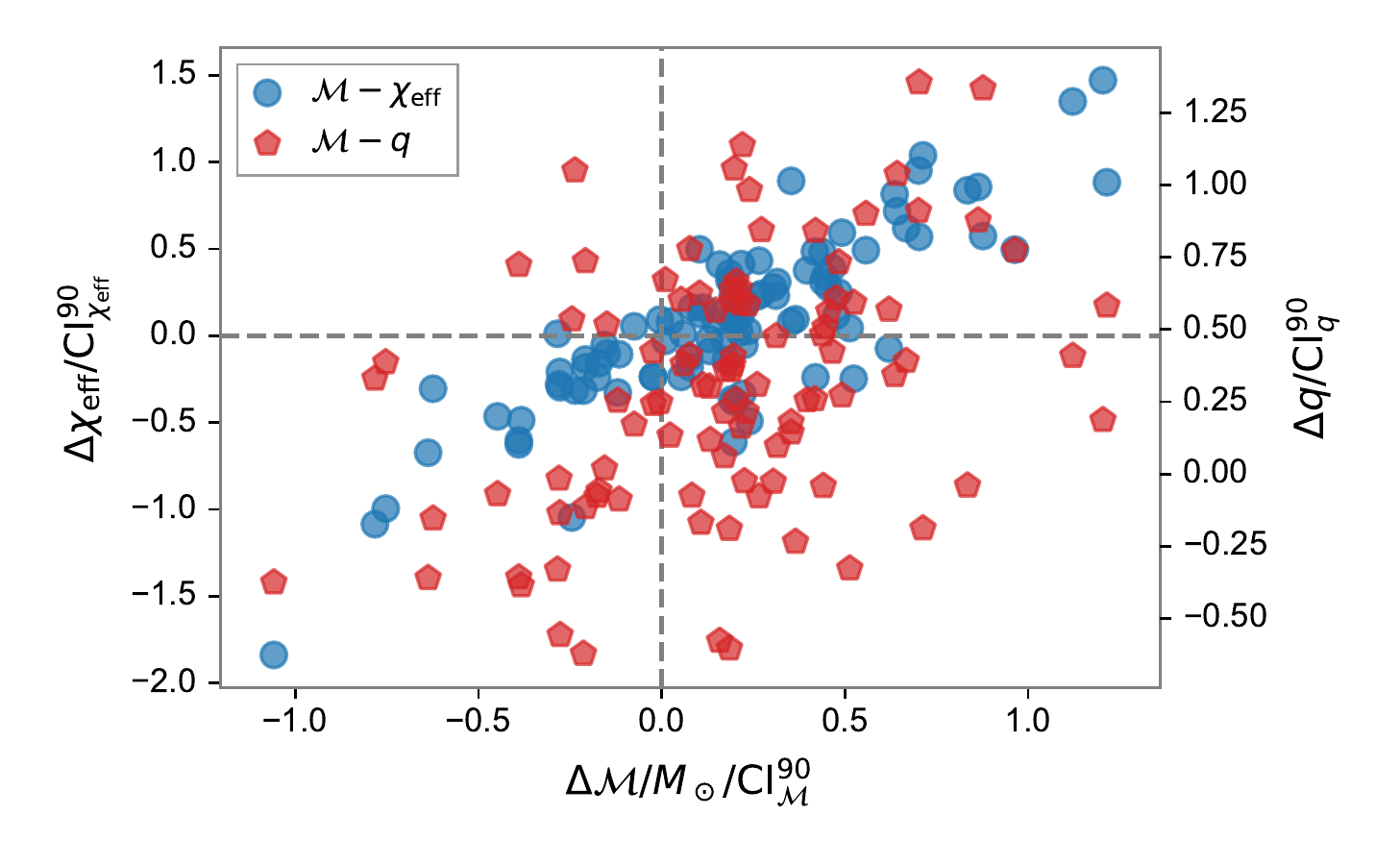}
    \includegraphics[width=.45\textwidth]{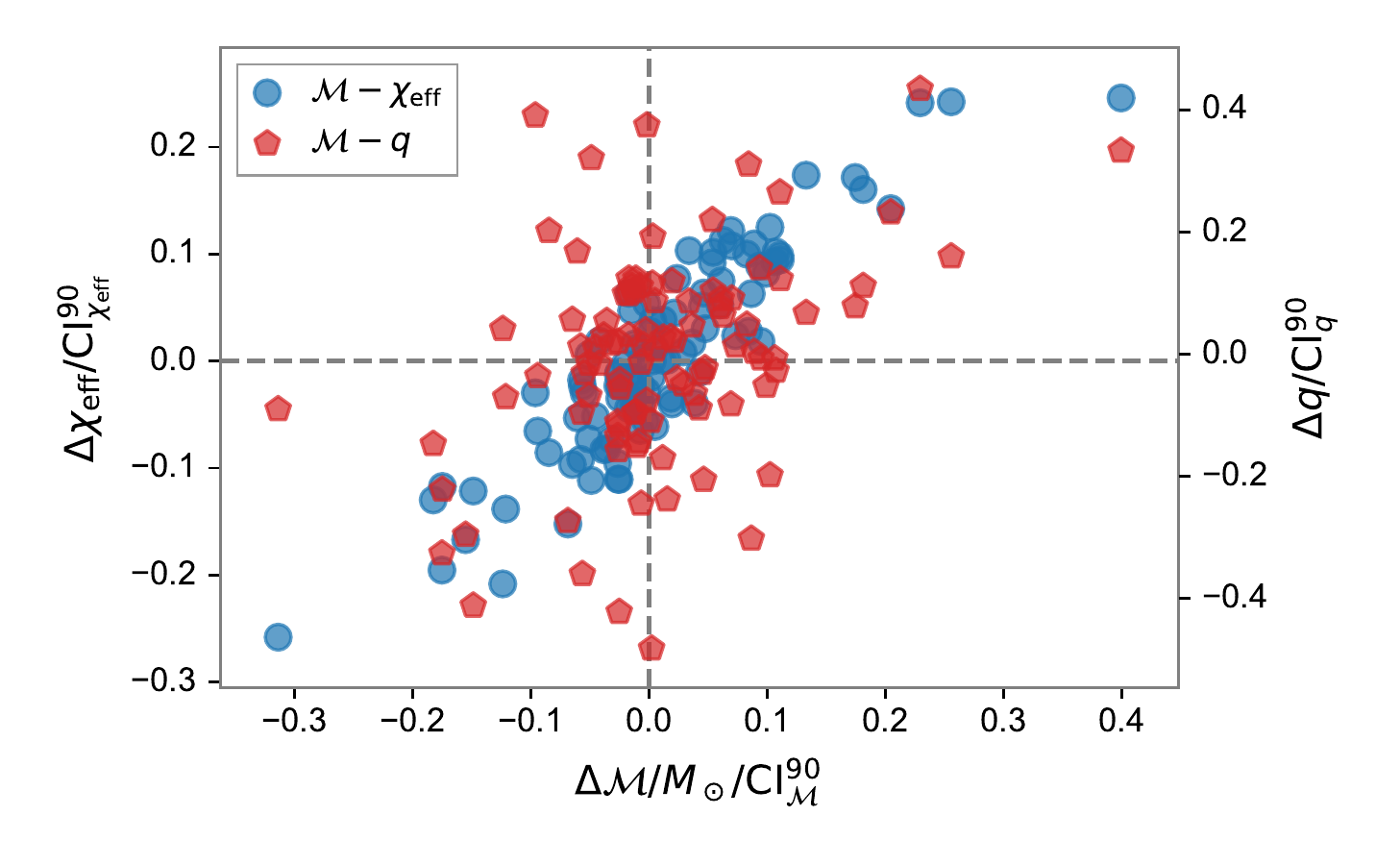}
  \caption{Ratio of absolute biases and 90\% credible intervals $\mathcal{R}(\theta, d)$ as defined in Eq.~\eqref{eq:bias-ratio}
  between chirp-mass $\mathcal{M}/M_\odot$ and effective aligned-spin $\chi_\mathrm{eff}$ and chirp-mass and mass-ratio $q$. 
  $\mathcal{R} = \pm 1/2$ when the true value is found at the boundaries of the 90\% interval.
  Left: \texttt{NRSur7dq2} signals recovered with \texttt{IMRPhenomPv2}.
  Right: \texttt{IMRPhenomPv2} signals recovered with \texttt{IMRPhenomPv2}.
  }
  \label{fig:absolute_biases_over_CI90}
\end{figure*}

\begin{table*}
\begin{tabular}{l l l d{2.2} d{2.2} d{2.2} d{2.2}}
  \hline
  \hline
 Quantity & Signal waveform & Point estimate & \multicolumn{1}{c}{$\mathcal{M}^\mathrm{det} / M_\odot$} & \multicolumn{1}{c}{$q$} & \multicolumn{1}{c}{$\chi_\mathrm{eff}$} & \multicolumn{1}{c}{$\chi_\mathrm{p}$}\\
  \hline
  \hline
  \multirow{3}{*}{$\sum_i \mathcal{B}_i$} & \multirow{3}{*}{\texttt{NRSur7dq2}} & Mean   & 17.22 & 4.36 & 0.49 & 5.49\\
                        &                               & Median & 16.45 & 4.67 & 0.41 & 6.20\\
                        &                               & MAP    & 34.01 & 7.78 & 1.12 & 5.37\\
  \hline
  \multirow{3}{*}{$\sum_i \mathcal{B}_i$} & \multirow{3}{*}{\texttt{IMRPhenomPv2}} & Mean   & -4.68 & -0.44 & -0.26 & -0.38\\
                        &                               & Median & -3.87 & -0.25 & -0.25 & 0.02\\
                        &                               & MAP    &  0.90 & -0.13 & 0.04  & -1.27\\
  \hline
  \hline
  \multirow{3}{*}{$\sum_i \mathcal{R}_i$} & \multirow{3}{*}{\texttt{NRSur7dq2}} & Mean   & 9.92  & 18.94 & 3.36 & 27.73\\
                        &                               & Median & 9.58  & 20.18 & 2.79 & 30.19\\
                        &                               & MAP    & 17.89 & 30.82 & 7.39 & 28.78\\
  \hline
  \multirow{3}{*}{$\sum_i \mathcal{R}_i$} & \multirow{3}{*}{\texttt{IMRPhenomPv2}} & Mean   & -2.39 & -1.80 & -2.56 & -1.33\\
                        &                               & Median & -2.30 & -1.23 & -2.78 & 0.20\\
                        &                               & MAP    & 0.86 & -0.24 & 0.31  & -5.10\\
  \hline
  \hline
\end{tabular}
\caption{
    \label{tab:pop-biases}
    Population biases $\sum_i \mathcal{B}_i$, and $\sum_i \mathcal{R}_i$ (see Eqs.~\eqref{eq:abs_bias} and~\eqref{eq:bias-ratio}) for chirp mass, mass-ratio, effective aligned spin and effective precession spin. The point estimate $\theta_\mathrm{p}$ is either the mean, median, or \ac{MAP}. We show biases for \texttt{NRSur7dq2} and \texttt{IMRPhenomPv2} signals.
    }
\end{table*}

\begin{table*}
\begin{tabular}{l|l|l|l|l|l|l|l}
  \hline
  \hline
   $M_\mathrm{tot}^\mathrm{src} / M_\odot$ & $\mathcal{M}^\mathrm{src} / M_\odot$ & $q$ & $\vec\chi_1$ & $\vec\chi_2$ & $\chi_\mathrm{eff}$ & $\chi_\mathrm{p}$ & $\theta_\mathrm{JN}$\\
  \hline
  \hline
  81.35 & 34.64 & 0.68 & (-0.23, -0.64, 0.31) & (0.21, 0.53, 0.30) & 0.31 & 0.68 & 2.63\\
  \hline
  \hline
\end{tabular}
\caption{
    \label{tab:parameters-loudest-residual}
    Source parameters for \ac{BBH} with the loudest residual, shown in Fig.~\ref{fig:qscan-residuals}.
    }
\end{table*}

\section{Discussion}
\label{sec:discussion}

\subsection{Summary of Results} 
\label{sub:summary_of_results}

In this study we have looked at the impact of inaccuracies in models of the \ac{GW} waveform on inferring parameters for single loud events and for populations of \acfp{BBH}.
In Sec.~\ref{sec:results-golden-binaries} we presented results from parameter estimation analyses with current waveform models (\texttt{IMRPhenomPv2}, \texttt{SEOBNRv4\_ROM}) for two simulated PN-NR signals at fixed luminosity distance for a series of detector networks. These ``golden binaries'' are therefore observed with increasingly high \ac{SNR} as we look towards future detectors which are about a hundred times more sensitive than the current ones.
From the posterior distributions we calculated systematic and statistical errors and produced a tuned version of the indistinguishability criterion (see Eqns.~\eqref{eq:indistinguishability-criterion} and~\eqref{eq:D_equation}). 
In Fig.~\ref{fig:plots_indistinguishability} we show the resulting ``acceptable error'' as a function of \ac{SNR}.
The main result of this paper shows that current waveform models used as templates in our \ac{PE} analyses need to be improved for aLIGO design sensitivity and beyond: 
For 3G detectors such as Cosmic Explorer and the Einstein Telescope, the mismatch error for semi-analytical models needs to be reduced by \emph{three orders of magnitude} and by \emph{one order of magnitude} for \ac{NR} waveforms.

In Sec.\ref{ssub:Discussion} we saw that waveform inaccuracies can come from a combination of factors: errors in the dominant  $(2, \pm 2)$ modes in the co-precessing frame, approximate 
modeling of the precessing reference frame of the binary, and from missing higher harmonics 
in the waveform. Better semi-analytical models that include more physics are becoming available~\cite{London:2017bcn,Cotesta:2018fcv,Cotesta:2019inprep,Khan:2018fmp,Khan:2019kot,SEOBNRv4P,SEOBNRv4PSur}.

It stands to reason that if inferred binary parameters for single events are affected by inaccuracies in waveform models that these deficiencies will also impact the analysis of populations of compact binaries. 
In populations, many events will be significantly weaker than the loud ``golden binaries'' we have considered before. 
Still, many small errors may sum up to give a sizable effect that can impact analyses. 
Therefore, in Sec.~\ref{sec:population} we presented a study for one hundred high mass \acp{BBH}  mock signals (either \texttt{NRSur7dq2} or \texttt{IMRPhenomPv2}) drawn from an astrophysically motivated distribution in the intrinsic parameters. 
We again performed \ac{PE} with the semi-analytical \texttt{IMRPhenomPv2} model for the aLIGO-Virgo-KAGRA design sensitivity network.

In Fig.~\ref{fig:absolute_biases_over_CI90} we find that parameter biases between key parameters such as chirp mass, mass-ratio, and effective spin are strongly correlated, the population sum of these biases is nonzero for the \texttt{NRSur7dq2} signals, and the largest parameter biases lie outside 90\% credible intervals. 
Posteriors for \texttt{IMRPhenomPv2} signals still show the correlations, but as shown in Table~\ref{tab:pop-biases} the population sum of their biases is close to zero.

The residual between the GW data recorded by a detector and the best fit template waveform obtained from \ac{PE} can be analysed further. 
If the waveform template cannot capture all the features in the signal then the residuals (for the detectors in a network) will contain some coherent power (i.e. the residuals are not just due to random noise fluctuations in each detector). 
We show that this is the case for a \texttt{NRSur7dq2} event in our population study which has significant \ac{SNR} in its residual (see Fig.~\ref{fig:Strain_residual_SNR}), and significant power in the time frequency plane (see Fig.~\ref{fig:qscan-residuals}). 
We have also carried out a \bayeswave analysis and show in Fig.~\ref{fig:BW_BF} the Bayes factors between a coherent and an incoherent wavelet model for the population events.
Most events have a residual \ac{SNR} of about 12 and Bayes factors of about 60 in favor of the coherent model while the event with the loudest residual has a residual \ac{SNR} of 18 and BF of $4 \times 10^{11}$.

We have computed the population hyper-posterior of the power-law index of the larger \ac{BH}, the only free parameter in the distribution of source parameters.
Due to the shortness of the signals in time, the hyper-posterior is not very informative, but it shows preference for the lower bound of the prior $\alpha = 1$ over then upper bound $\alpha = 2$, and is thus closer to the true value $\alpha = 1.3$.


\subsection{Outlook} 
\label{sub:outlook}

Let us discuss several further implications of systematic errors in measured binary parameters caused by inaccuracies in waveform models. They concern the astrophysical relevance of biases.
the future of waveform modeling and \ac{NR} simulations, and how tests of \ac{GR} will be affected.

In this study we have reported extensively about biases in inferred binary parameters. 
How much should we care about these biases? 
Beyond the simple statement that parameter biases will matter more when they are large, we would like to point out particular situations when biases are especially important and can severely impact the interpretation of GW observations.
Severe biases could cause a misidentification of the class of a compact binary, e.g. confusing BNS, NSBH, BBH sources near the lower mass gap~\cite{Mandel:2015spa,Littenberg:2015tpa}.
Large biases in spin parameters such as the effective precession spin $\chi_\mathrm{p}$ could lead to a misidentification of formation channel of a binary. 
This could also happen if the effective aligned spin parameter $\chi_\mathrm{eff}$ was heavily biased, but in general $\chi_\mathrm{eff}$ measurements are a lot more robust since this parameter is connected with the length of the inspiral signal~\cite{Ng:2018neg}. 
We have seen in Fig.~\ref{fig:0308_posterior_PDFs} that $\chi_\mathrm{p}$ can indeed be significantly underestimated, especially if the precession modulations are suppressed when the binary is viewed nearly face-on or face-off.

Extrinsic parameters are in general less affected by waveform systematics and we do not expect their measurement errors to have a big impact. 
Sky location parameters enter in the detector pattern functions and should not be affected. 
We expect luminosity distance measurements to be affected mainly through their correlation with the binary's inclination angle. 
The latter can be better measured~\cite{Graff:2015bba,London:2017bcn,Cotesta:2018fcv,Kalaghatgi:2019log,Shaik:2019dym} when the waveform includes higher harmonics beyond the dominant $(2, \pm 2)$ modes. Amplitude errors should play a lesser role than phase errors, which could lead to us to misestimate $\mathcal{M}$
and thus bias the recovered distance.
If we misestimate $\mathcal{M}$ due to phase errors, that will also bias the recovered distance.
Through this correlation mis-estimation of distance can lead to additional bias on the source-frame masses. 
This can be significant for very distant binaries.
Finally, based on the discussion in~\ref{sub:bias_results} we expect that for population analyses parameters characterizing the mass and spin distributions will be affected to some degree since the events making up the population will suffer some amount of parameter biases.

How can waveform models be improved and made ready for the planned future 3G detectors, such as Cosmic Explorer and Einstein telescope? 
On the one hand the accuracy in the inspiral regime needs to be improved. 
This requires a higher order and more complete \ac{PN} description and further work on re-summation and effective-one-body theory to extend the validity of the inspiral to higher frequencies. The inclusion of self force terms into \ac{EOB} could help accuracy for large mass ratios~\cite{Antonelli:2019fmq}. Post-Minkowskian results obtained with modern scattering amplitude methods could be useful to improve the accuracy, if pushed to higher order~\cite{Antonelli:2019ytb,Bern:2019nnu}. PN calculations have being made at 4PN order for non-spinning \acp{BBH}~\cite{Damour:2014jta,Marchand:2017pir,Foffa:2016rgu}.
Practical semi-analytic inspiral-merger-ringdown waveform models for \acp{BBH}, whether they are phenomenological or \ac{EOB} models, require more \ac{NR} waveforms covering larger parts of the binary parameter space and ultimately higher \ac{NR} accuracy. 
Especially for unequal mass-ratios we will also require longer \ac{NR} simulations in time in order to be able to combine \ac{NR} waveforms with \ac{PN} or \ac{EOB} inspirals to form highly accurate hybrid waveforms~\cite{Hannam:2010ky,Boyle:2011dy,MacDonald:2011ne}, and to better determine inspiral coefficients in the construction of \ac{EOB} models~\cite{Bohe:2016gbl}. 
So far semi-analytic models have been tuned only in the non-precessing sector. 
Extending calibration as more precessing \ac{NR} simulations are becoming available will be essential to improve their accuracy. 
In addition, a novel NR-independent, analytical approach for modeling the merger has been put forward~\cite{McWilliams:2018ztb}. The accuracy of this approach beyond NR accuracy could be assessed with constraints on waveforms obtained from balance laws at future null infinity~\cite{Ashtekar:2019viz}.

Surrogate and reduced order models of \ac{NR} waveforms~\cite{Blackman:2015pia,Blackman:2017dfb,Blackman:2017pcm,Varma:2018mmi,Varma:2019csw} and of \ac{EOB} waveform models~\cite{Field:2013cfa,Purrer:2014fza,Purrer:2015tud,Bohe:2016gbl,Lackey:2016krb} have come to prominence in the past several years. 
They preserve the accuracy of the training set waveforms they are constructed from and are orders of magnitude faster to evaluate making them crucial for data analysis applications. 
They depend on their input data and so their accuracy is limited by the accuracy of the training set waveforms, and the requirement that the training data is sufficiently dense in the parameter space, since they need to fit or interpolate waveform coefficients over parameter space.

Waveform models should also include all physical effects that will leave a measurable trace in the emitted \ac{GW} signal. 
This includes spin effects (aligned and precessing spins), higher harmonics beyond the dominant $(2, \pm 2)$ modes in the waveform, imprints of eccentricity, and tidal effects if the binary contains at least one neutron star. 
As the number of waveform parameters increases it becomes harder to carry out enough \ac{NR} simulations to accurately tune models.
A further desirable improvement for waveform models is to also model internal errors in waveform models and marginalize over these parameters in \ac{PE}, which can be achieved with \ac{GPR}~\cite{Moore:2014pda,Moore:2015sza,Doctor:2017csx,Huerta:2017kez,Lackey:2018zvw,Williams:2019vub}. 
Posterior distributions obtained with such models should be more accurate (reduced bias) but somewhat less precise.

We have seen that \ac{NR} waveforms are central for \ac{IMR} waveform modeling. 
According to our results, \ac{NR} waveforms will have to be improved in the future along three different dimensions: 
First, accuracy; second, length; third improved parameter space coverage.
It turns out that each of these aspects will make simulations more expensive.
Regarding length, the cost of an \ac{NR} simulation is at least proportional to the time-to-merger; hence the cost will increase as $1 / \eta \; (M \Omega_i)^{-8/3}$, 
so that starting at one half the initial (orbital) frequency $M \Omega_i$ increases cost by at least a factor of $5$. 
This scaling of the time-to-merger already indicates that making the mass-ratio more unequal will also increase the computational cost: 
The time-to-merger (and this computational cost) will increase at least as $1/\eta$. In addition, current \ac{NR} codes use explicit time integration and are therefore limited by the Courant condition, so that each time-step can cover at most a time-interval $\propto q$ (for $q \leq 1$), giving a second power of the mass-ratio. 
Regarding accuracy, it is difficult to predict how the achieved accuracy scales with computational cost; one estimate for SpEC is that the cost goes as $\epsilon^{-1/3}$~\cite{HP-pcomm}, where $\epsilon$ is the \ac{NR} error. 
Therefore, reducing the mismatch-error by a factor of 10 -- at same parameters and length of 
the simulation -- increases computational cost by about 50\% for SpEC since the mismatch error 
goes as the square of the NR error~\cite{Flanagan:1997kp, McWilliams:2010eq, Ohme:2011rm}.
Finally, both higher spins and higher mass-ratio make \ac{NR} simulations more expensive, with the mass-ratio dependence most pronounced.

The accuracy and number of \ac{NR} simulations have improved dramatically since the breakthrough in 2005~\cite{Pretorius:2005gq,Campanelli:2005dd,Baker:2005vv}. 
What improvements can we expect for the future that can deliver the simulations needed to be ready for 3G science? 
We can no longer rely on Moore's law to deliver massive improvements of CPU clock speeds. Instead advances in CPU development have shifted to increasing the number of cores and to exploit that \ac{NR} codes need better parallelization and scaling.
New codes are being developed to address these accuracy and performance issues. 
The SpECTRE code~\cite{Kidder:2016hev,SpECTRE} from the SXS Collaboration uses task-based parallelism combined with the discontinuous Galerkin method to significantly increase the efficiency and scalability of relativistic astrophysics simulations. 
Work to significantly reduce computational cost for \ac{NR} simulations is also under way for finite difference codes~\cite{BlackHolesAtHome}. 
These approaches could lead to a two order of magnitude improvement in efficiency and bring us closer to solving the problems we have pointed out here.
In addition to the truncation error which results from the finite degree polynomial approximations to continuum derivatives in Einstein's equations, errors are made when extracting the \ac{GW} waveform on computer grids extending finite distances away from the merging binary. 
Traditionally, the waves are extracted (ideally on spherical shells) at several radii as far away from the origin as possible and the ideal waveform at future null infinity is extrapolated from that data. 
The Cauchy characteristic extraction method~\cite{Babiuc:2010ze,Taylor:2013zia,Handmer:2015dsa,Handmer:2016mls} can compute the emitted \acp{GW} with higher accuracy and should be available for future \ac{NR} simulations.
Combing waveforms from SpEC and finite difference codes by hybridization is a promising technique for especially challenging configurations~\cite{Hinder:2018fsy}.

Our study on the impact of waveform inaccuracies should be extended to tests of \ac{GR} which we expect to be especially susceptible to systematic effects which could be misinterpreted as genuine deviations from \ac{GR}. 
All of the current tests of \ac{GR}~\cite{LIGOScientific:2019fpa} should be scrutinized. 
This includes tests on the distribution of the \ac{SNR} of residuals in detector noise, testing whether the final mass and spin inferred from the low and high frequency parts of the \ac{GW} signal are consistent, computing posterior distributions of deviations in e.g. \ac{PN} waveform coefficients, computing posteriors on parameters in phenomenological dispersion relations and tests that put constraints on alternative \ac{GW} polarizations. 
Ultimately, tests of \ac{GR} should be done by estimating parameters of waveform models for alternative theories of gravity, along with Bayesian model comparisons.
Work is under way to identify well-posed alternative theories of gravity~\cite{Healy:2011ef,Berti:2013gfa,Berti:2015itd,Yunes:2016jcc,Witek:2018dmd} and to numerically compute what the emitted \ac{GW} will look like in the strong field regime~\cite{Okounkova:2019zjf}.

Finally, we expect that LISA analyses of massive \acp{BBH}, which are should have \acp{SNR} of hundreds to thousands, will be affected in similar ways as demonstrated here for 3G ground-based detectors~\cite{Cutler:2007mi}. Updated estimates for current \ac{IMR} waveform models will need to be explored in future studies.


\begin{acknowledgments}
This work was stimulated by the Gravitational Wave International Committee (GWIC) 3G science-case study~\cite{GWIC-3G-report}.
The authors would like to thank Harald Pfeiffer, Katerina Chatziioannou, Will Farr, Sergei Ossokine, John Veitch, Alessandra Buonanno, Mark Hannam, Salvatore Vitale, Frank Ohme, Sascha Husa, Badri Krishnan, Bruce Allen for useful discussions.
We thank Patricia Schmidt for help with the LVC \ac{NR} injection infrastructure, Tito dal Canton for help with PyCBCs's qscans, Sergei Ossokine and Stas Babak for the code to compute the max-max overlap.

C.-~J.~H.~acknowledges support of the MIT physics department through the Solomon Buchsbaum Research Fund, the National Science Foundation, and the LIGO Laboratory.
The authors acknowledge usage of LIGO Data Grid clusters and AEI's Slarti computer.
LIGO was constructed by the California Institute of Technology and Massachusetts Institute of Technology with funding from the National Science Foundation and operates under cooperative agreement PHY-0757058. 
This is LIGO Document Number DCC-1900377.

\end{acknowledgments}

\appendix

\section{Hybridization procedure}
\label{app:hybrids}

We construct hybrid waveforms by combining multi-modal precessing \ac{PN} and \ac{NR} waveforms using the \texttt{GWFrames}~\cite{GWFrames} code.
The code first reads the \ac{NR} waveform data and transform it to the co-rotating frame~\cite{Boyle:2013nka} and shifts it in time so that the merger lies at $t=0$. 
Data for the evolution of the positions, masses and spin vectors of the \acp{BH} as determined by locating their apparent horizons in the \ac{NR} code is read in as well. 
Next, the separation vector between the two \acp{BH} and the orbital frequency are computed, along with the rotor of the reference frame at the relaxed time (after the junk radiation has passed).

We compute a \ac{PN} waveform from the \texttt{PNWaveform} package included in the \texttt{GWFrames} code.
The PN implementation includes nonspinning orbital binding up to 4pN~\cite{BlanchetLRR}. The 5pN term is set to zero. Spin-orbit terms in the angular momentum are included up to 3.5pN~\cite{Bohe:2012mr}. Nonspinning flux terms are included up to 3.5PN~\cite{BlanchetLRR}, and higher-order terms from~\cite{Fujita:2012cm} up to 6 PN along with absorption terms from~\cite{Alvi2001}. Spin-spin and spin-orbit squared terms at 2PN order are included~\cite{Kidder95,Will96,Arun:2008kb} and spin-orbit terms in the flux are included up to 4.0PN~\cite{Marsat:2013caa}. Precession of the orbital angular velocity and spins follows~\cite{Kidder95,Racine08,Bohe:2012mr}. Expressions for waveform modes are taken from~\cite{Blanchet:2008je,Faye:2012we,Faye:2014fra,Buonanno:2012rv}.
We use the \texttt{SpinTaylorT1} and \texttt{SpinTaylorT4} implemented in this code which are simply called \texttt{TaylorT1} and \texttt{TaylorT4} there, but we add the prefix \texttt{Spin} to make it clear that they support precession.

Initial data for the \ac{PN} integration is set at the \ac{NR} relaxed time and the \ac{PN} equations are also evolved backwards in time to the desired starting orbital frequency $M\Omega_i$. 
The PN waveform is then transformed to the co-rotating frame.
To prepare for hybridization, the \ac{PN} and \ac{NR} waveforms are aligned by minimizing the distance between their rotors in their co-rotating frames.
The aligned waveforms are then blended and hybridized.

In this study we choose $M\Omega_i = 0.002$ due to computational restrictions.
This corresponds to $f_{GW} \approx 3.5 \, \mathrm{Hz}$ for the $(2,2)$ mode. Higher $(\ell, m)$ modes in the waveform enter the frequency band at $m/2$ the frequency at which the $(2,2)$ mode enters.
Therefore, some of the higher harmonics are truncated at low frequencies but this effect is minor because they are very small compared to the dominant modes.

To use the LVC NR-injection infrastructure~\cite{Schmidt:2017btt} we also hybridize dynamics quantities, namely the spin vectors, orbital frequency, the Newtonian orbital angular momentum vector, the vector $\hat n$ pointing from one \ac{BH} to the other, and the position vectors of the \acp{BH}. 
This allows us to define the spin vectors at a particular reference frequency and to output the result in ``LVC NR'' format.

\begin{figure*}[t]
  \centering
  \includegraphics[width=.45\textwidth]{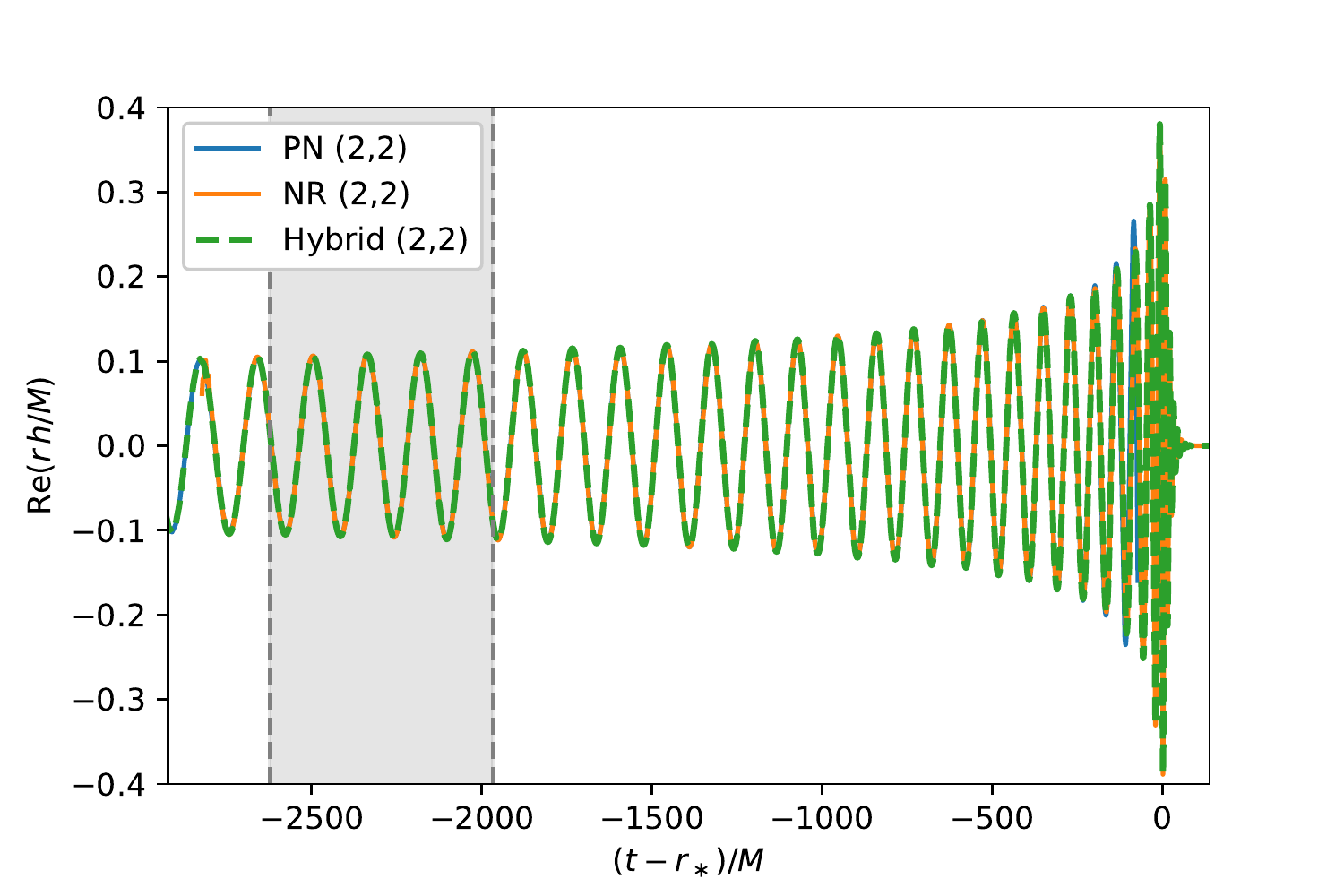}
  \includegraphics[width=.45\textwidth]{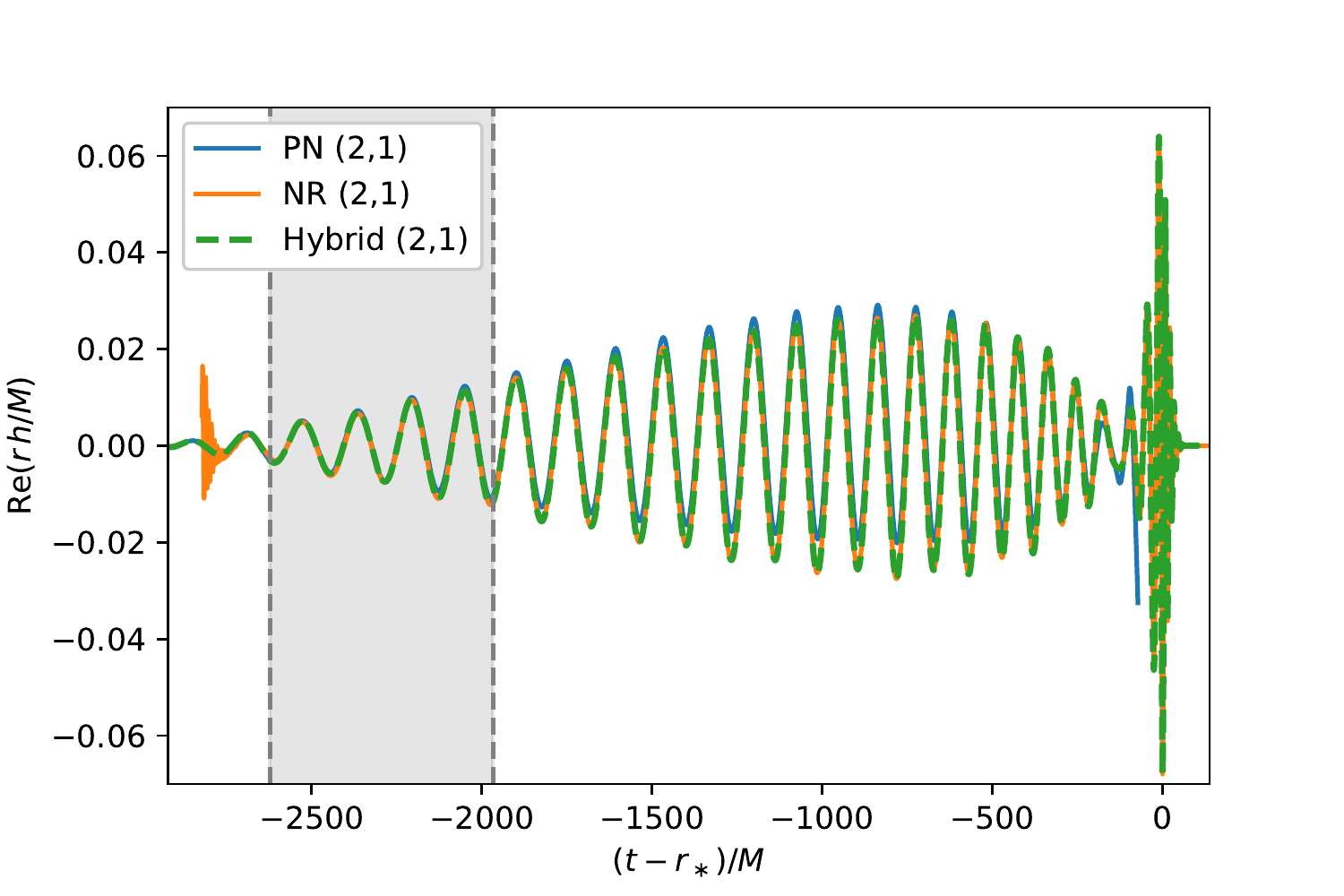}
  \includegraphics[width=.45\textwidth]{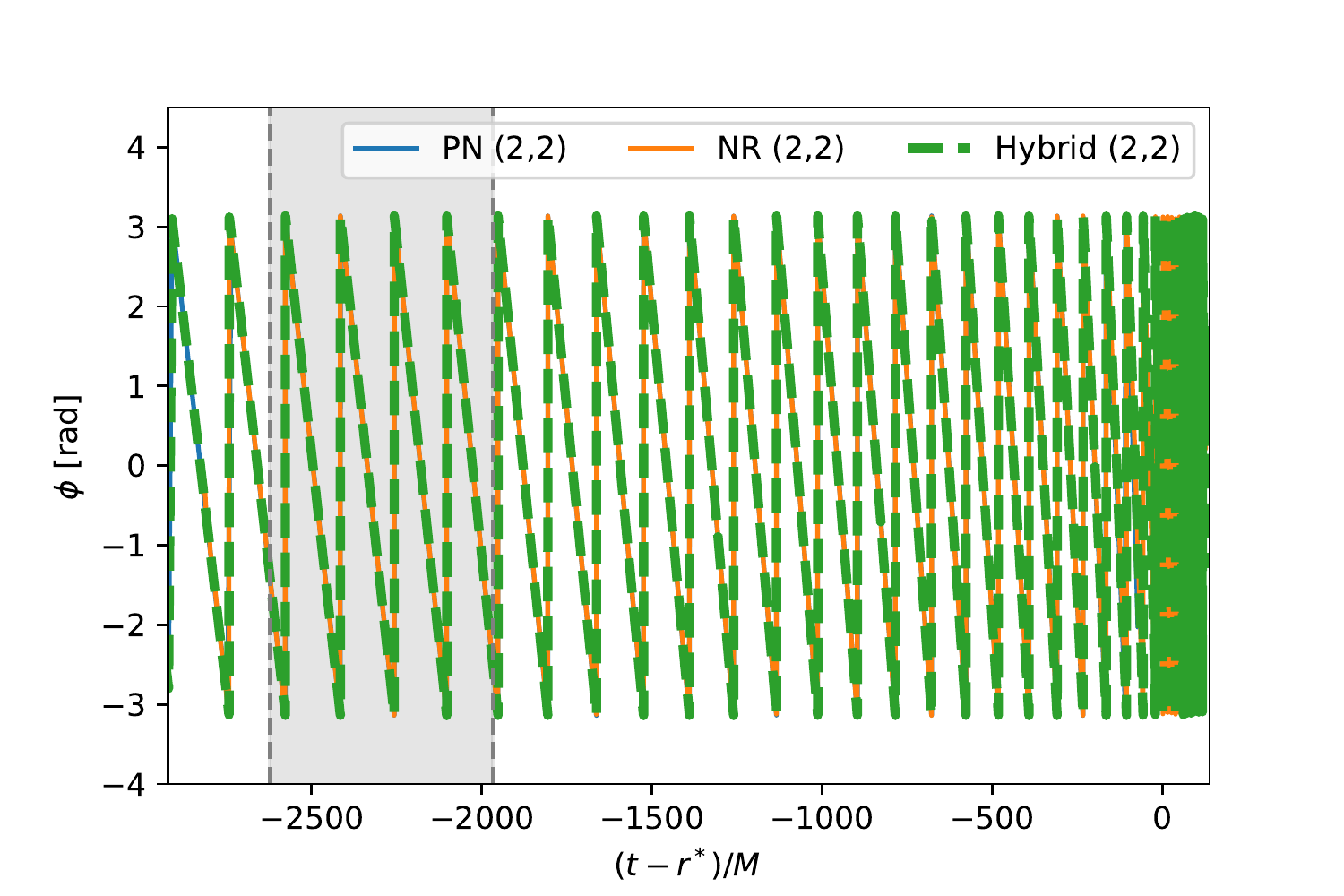}
  \includegraphics[width=.45\textwidth]{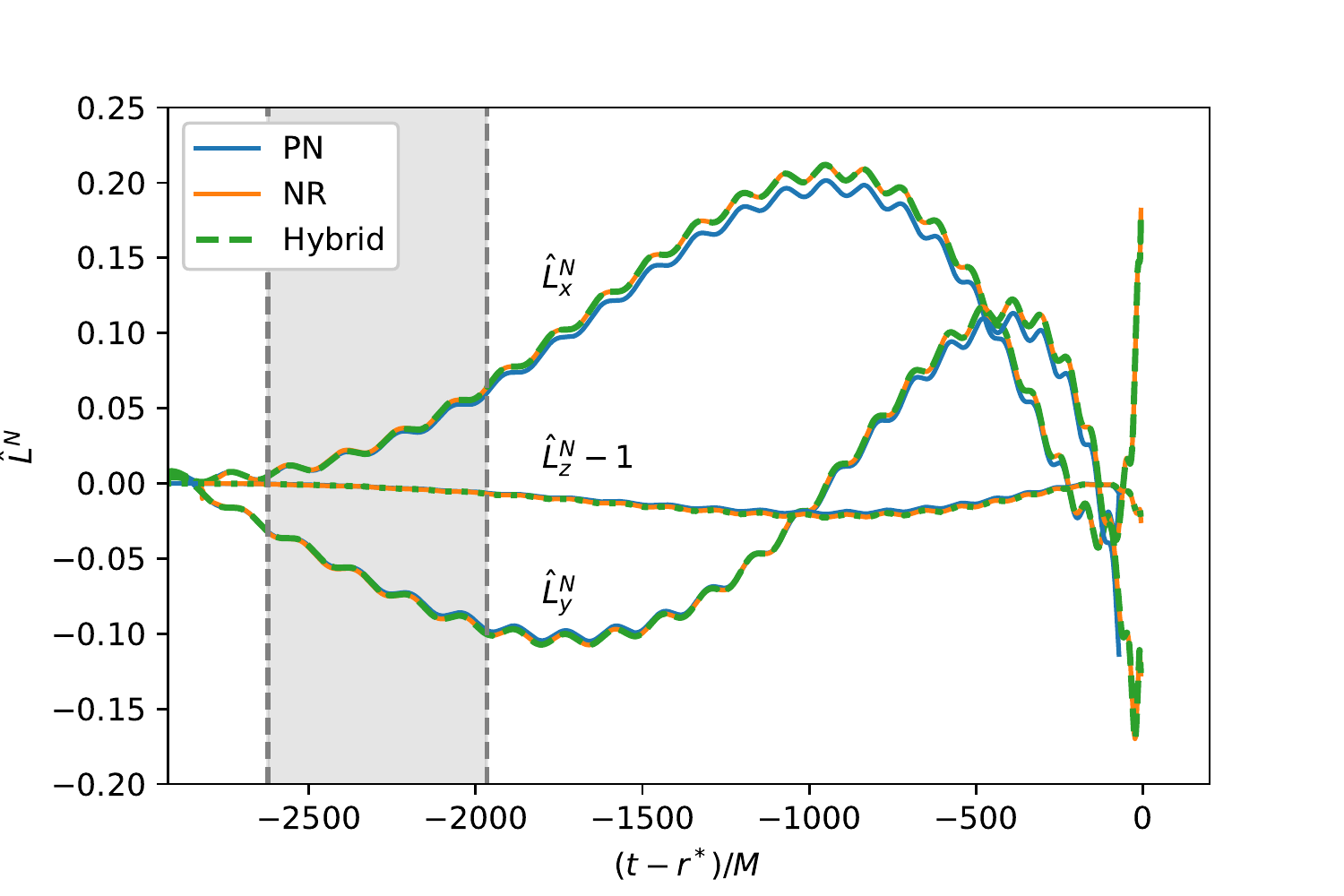}
  \includegraphics[width=.45\textwidth]{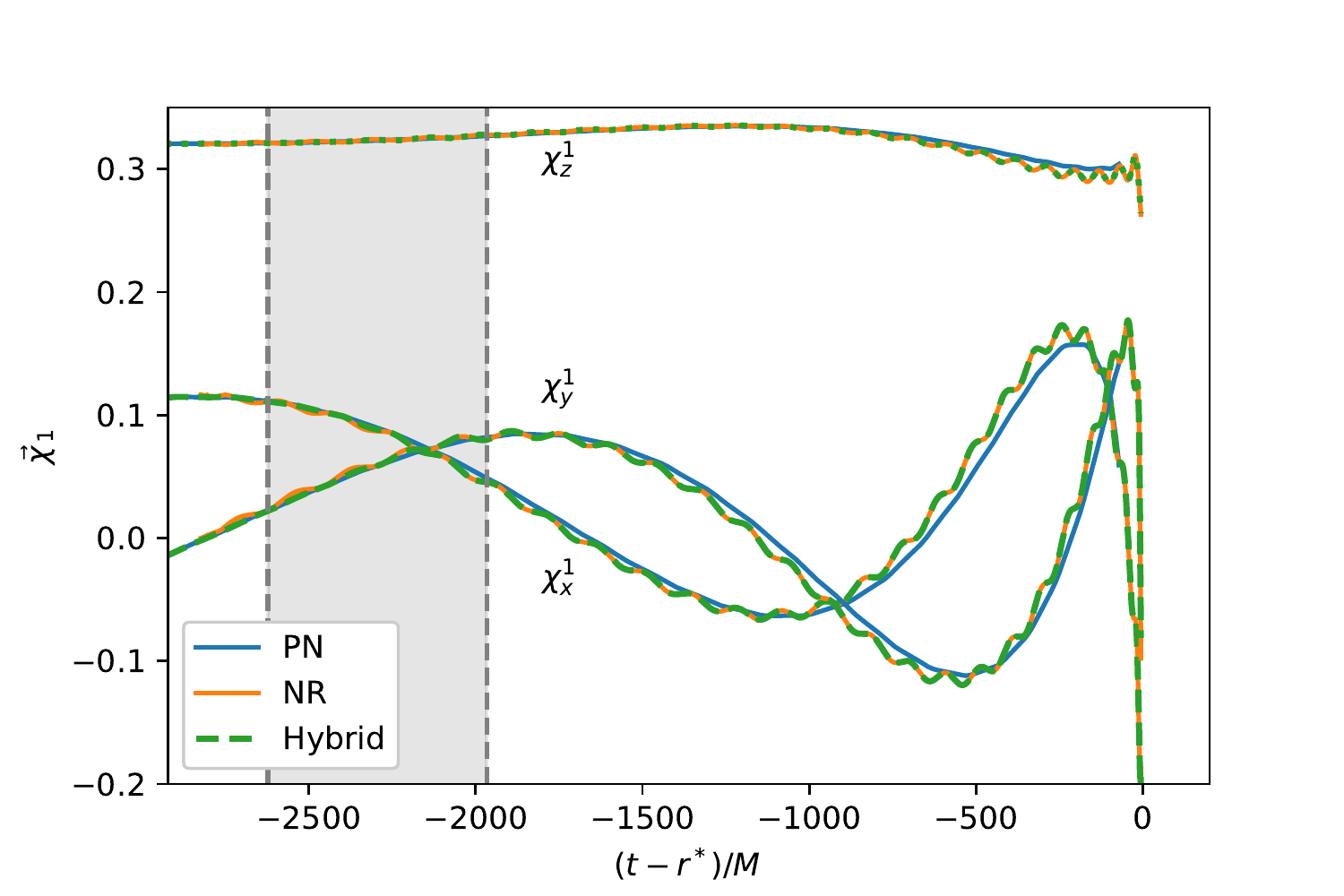}
  \includegraphics[width=.45\textwidth]{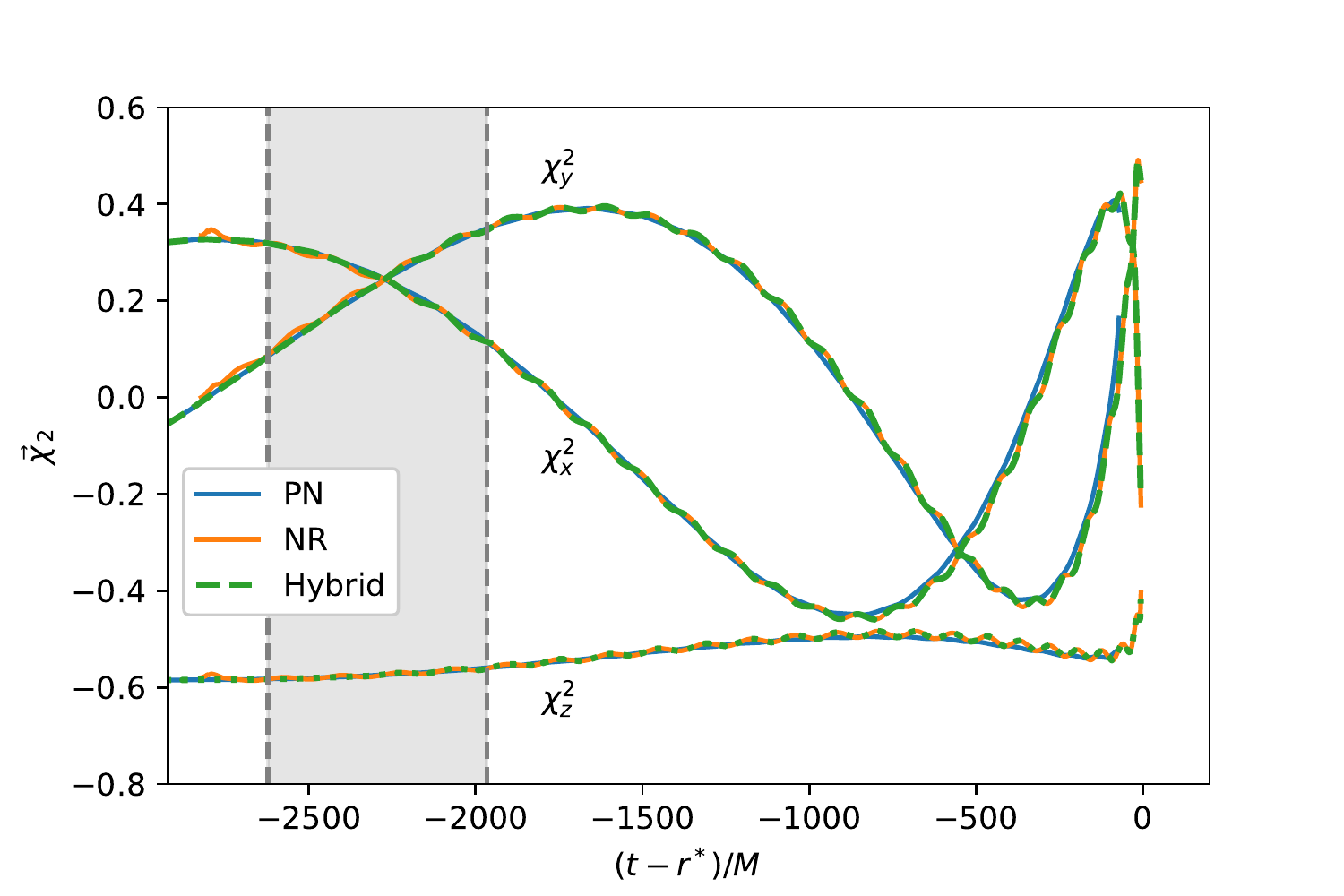}
  \caption{PN-NR hybrid for SXS\_BBH\_0308 with the \texttt{SpinTaylorT1} approximant.
  For each quantity we show the \ac{PN}, \ac{NR} and hybrid data as a function of retarded time before merger. The waveforms have been blended together in the gray shaded hybridization region.
  \emph{Top:} real part of the $(2, 2)$ and $(2, 1)$ modes in the inertial frame
  \emph{Middle:} 
    \emph{left:} (wrapped) phase of the $(2, 2)$ mode in the inertial frame
    \emph{right:} Cartesian components of the Newtonian orbital angular momentum unit vector in the inertial frame
  \emph{Bottom:} Dimensionless spin vectors of the \acp{BH}.
  }
  \label{fig:0308_hybrid}
\end{figure*}

\begin{figure*}[t]
  \centering
  \includegraphics[width=.45\textwidth]{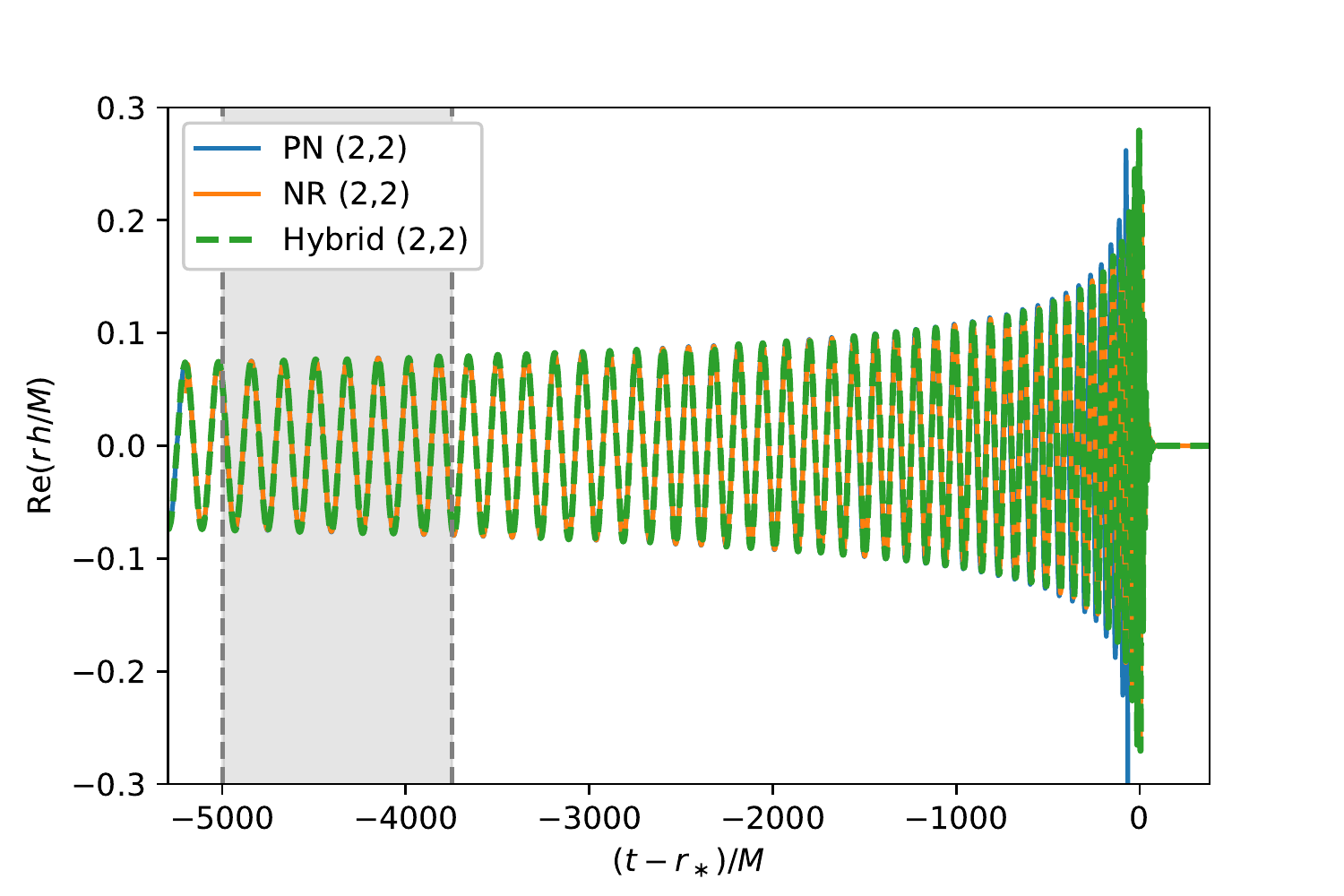}
  \includegraphics[width=.45\textwidth]{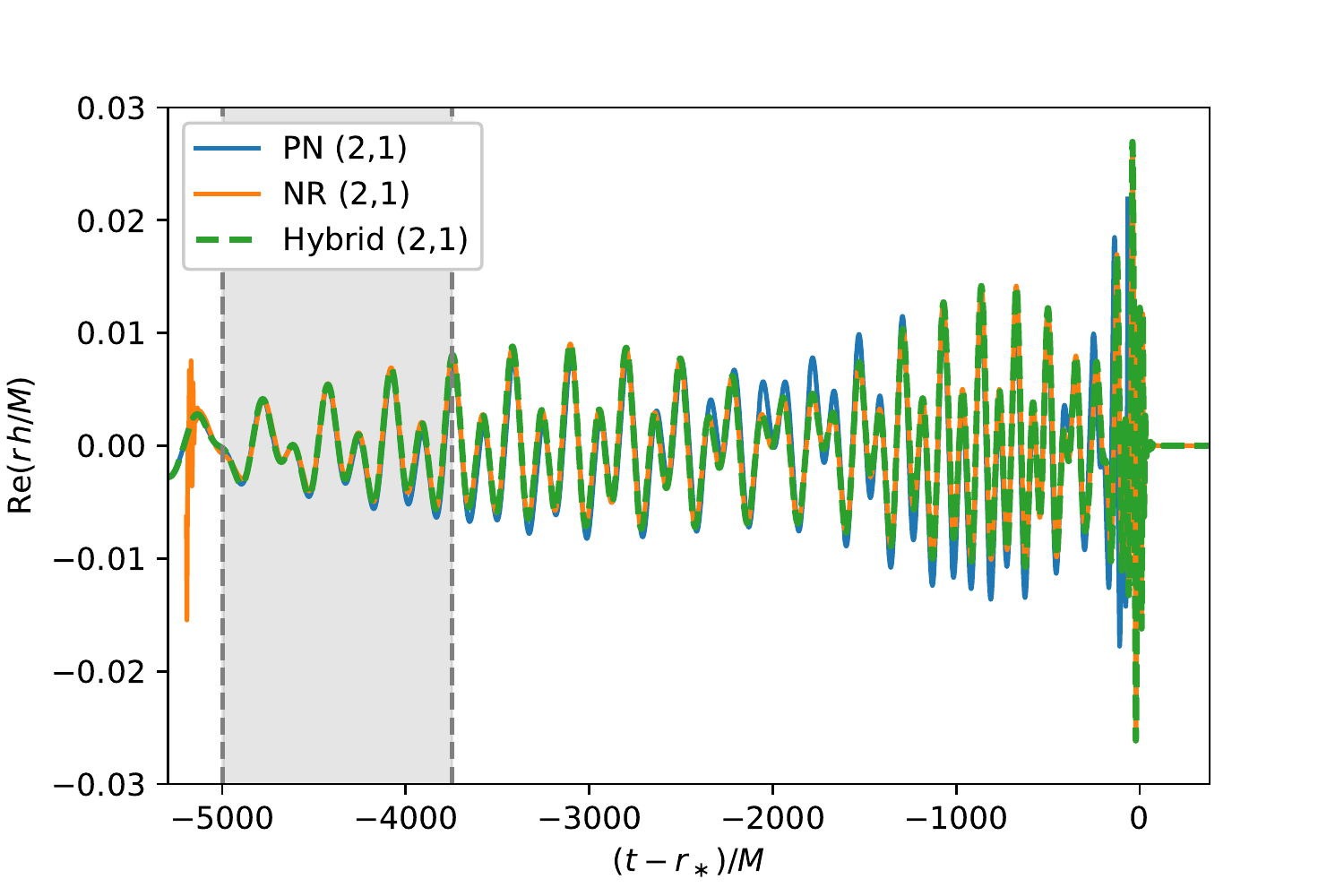}
  \includegraphics[width=.45\textwidth]{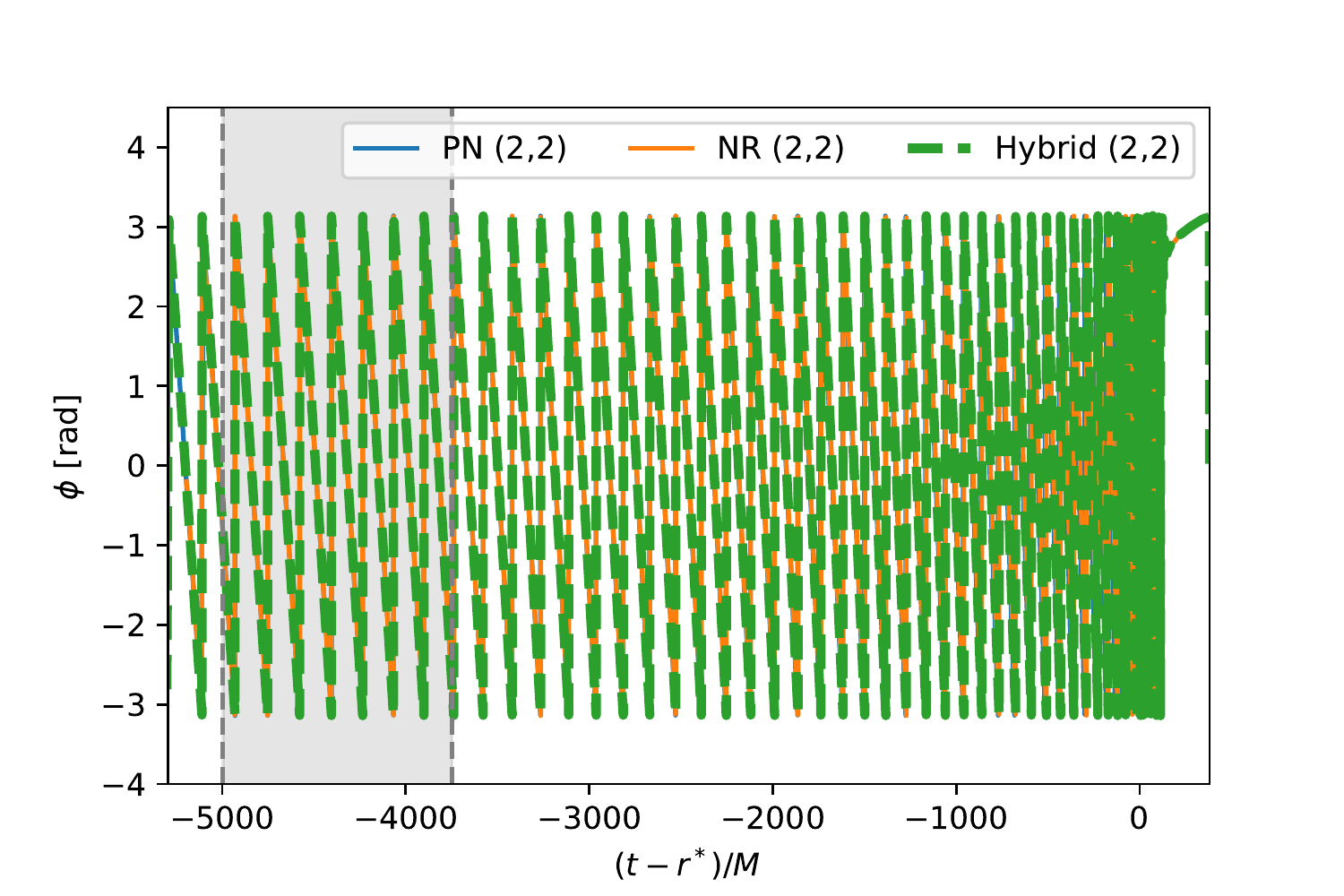}
  \includegraphics[width=.45\textwidth]{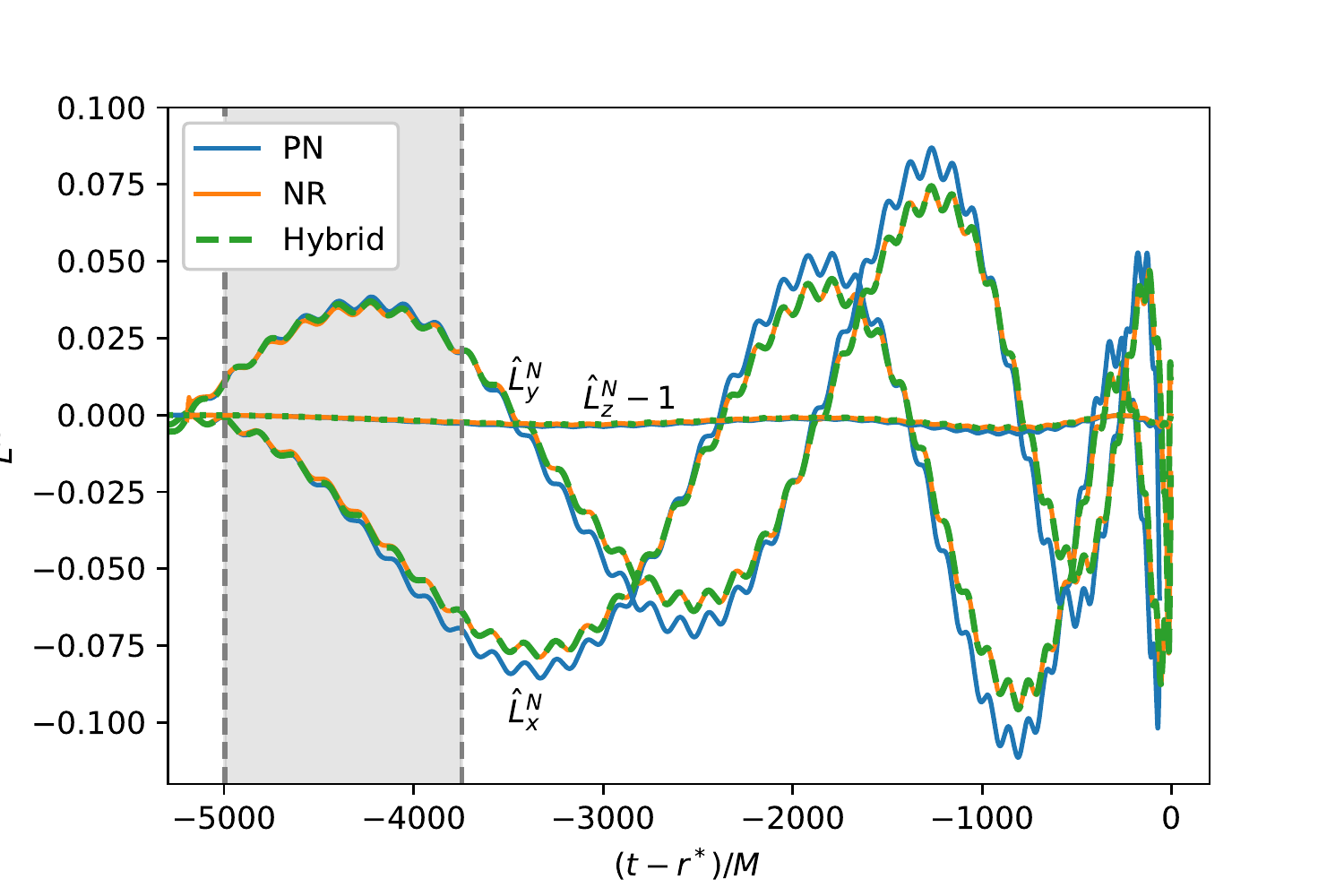}
  \includegraphics[width=.45\textwidth]{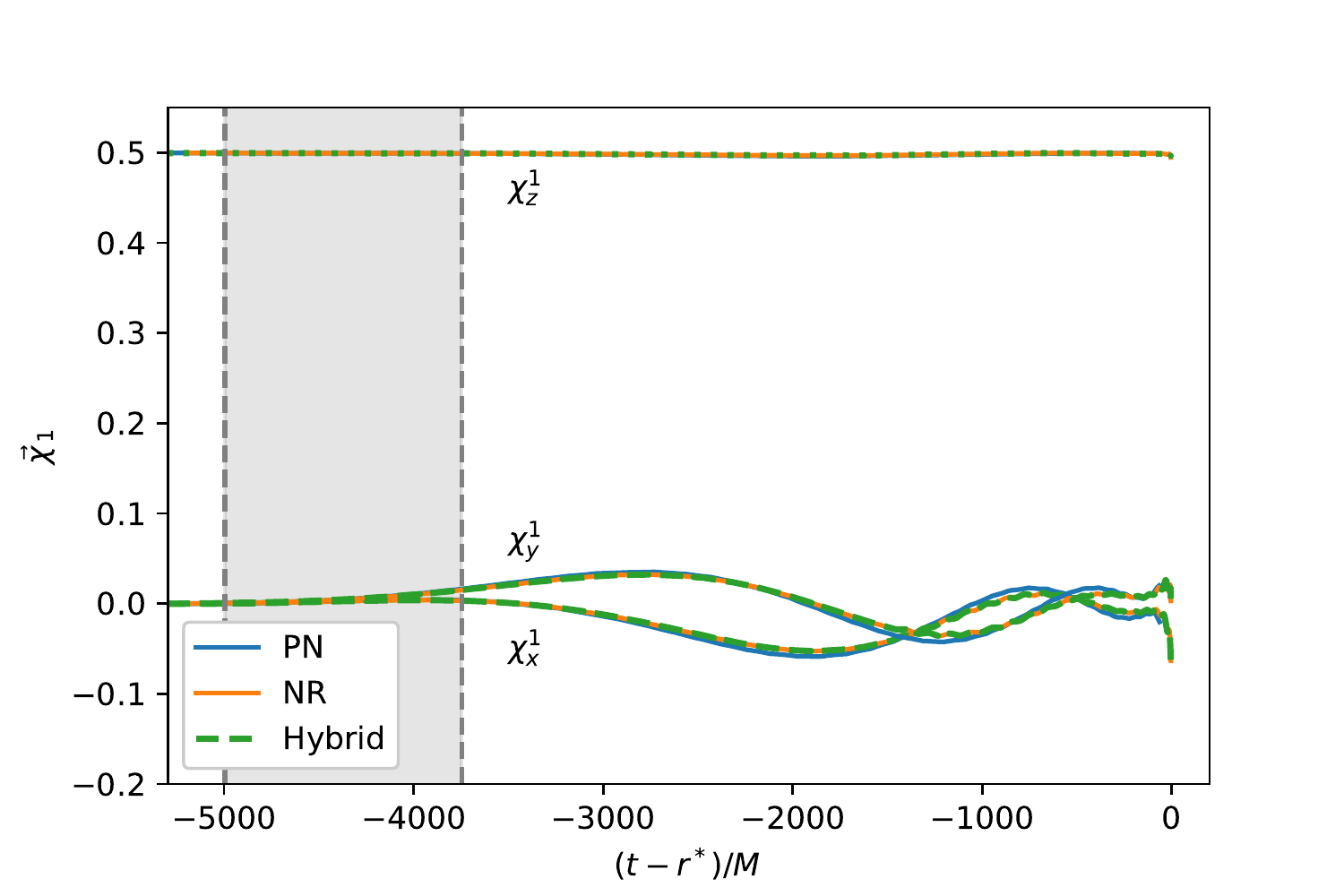}
  \includegraphics[width=.45\textwidth]{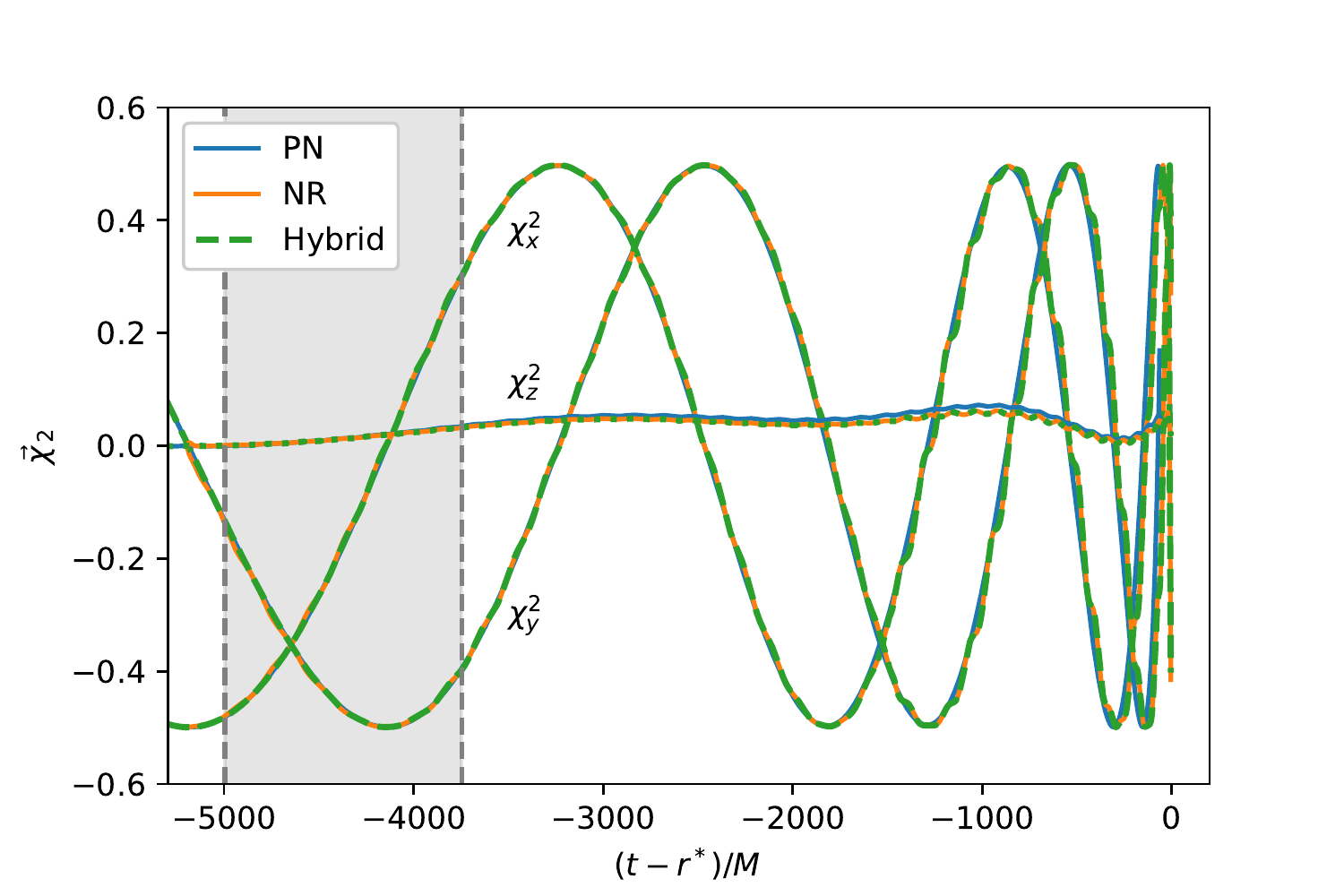}
  \caption{PN-NR hybrid for SXS\_BBH\_0104 with the \texttt{SpinTaylorT1} approximant.
  For each quantity we show the \ac{PN}, \ac{NR} and hybrid data as a function of retarded time before merger. The waveforms have been blended together in the gray shaded hybridization region.
  \emph{Top:} real part of the $(2, 2)$ and $(2, 1)$ modes in the inertial frame
  \emph{Middle:} 
    \emph{left:} (wrapped) phase of the $(2, 2)$ mode in the inertial frame
    \emph{right:} Cartesian components of the Newtonian orbital angular momentum unit vector in the inertial frame
  \emph{Bottom:} Dimensionless spin vectors of the \acp{BH}.
  }
  \label{fig:0104_hybrid}
\end{figure*}

\begin{figure*}[t]
  \centering
  \includegraphics[width=.45\textwidth]{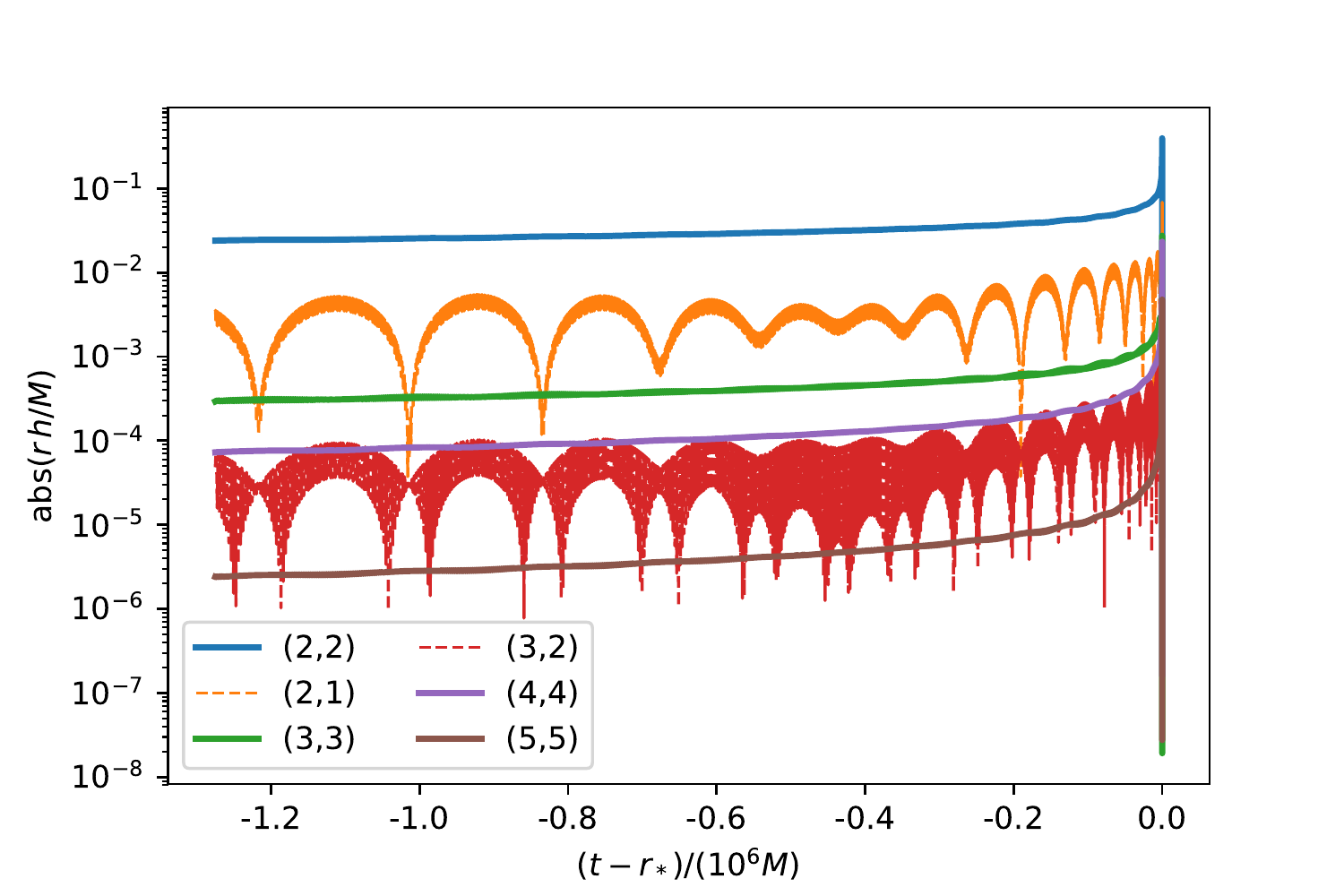}
  \includegraphics[width=.45\textwidth]{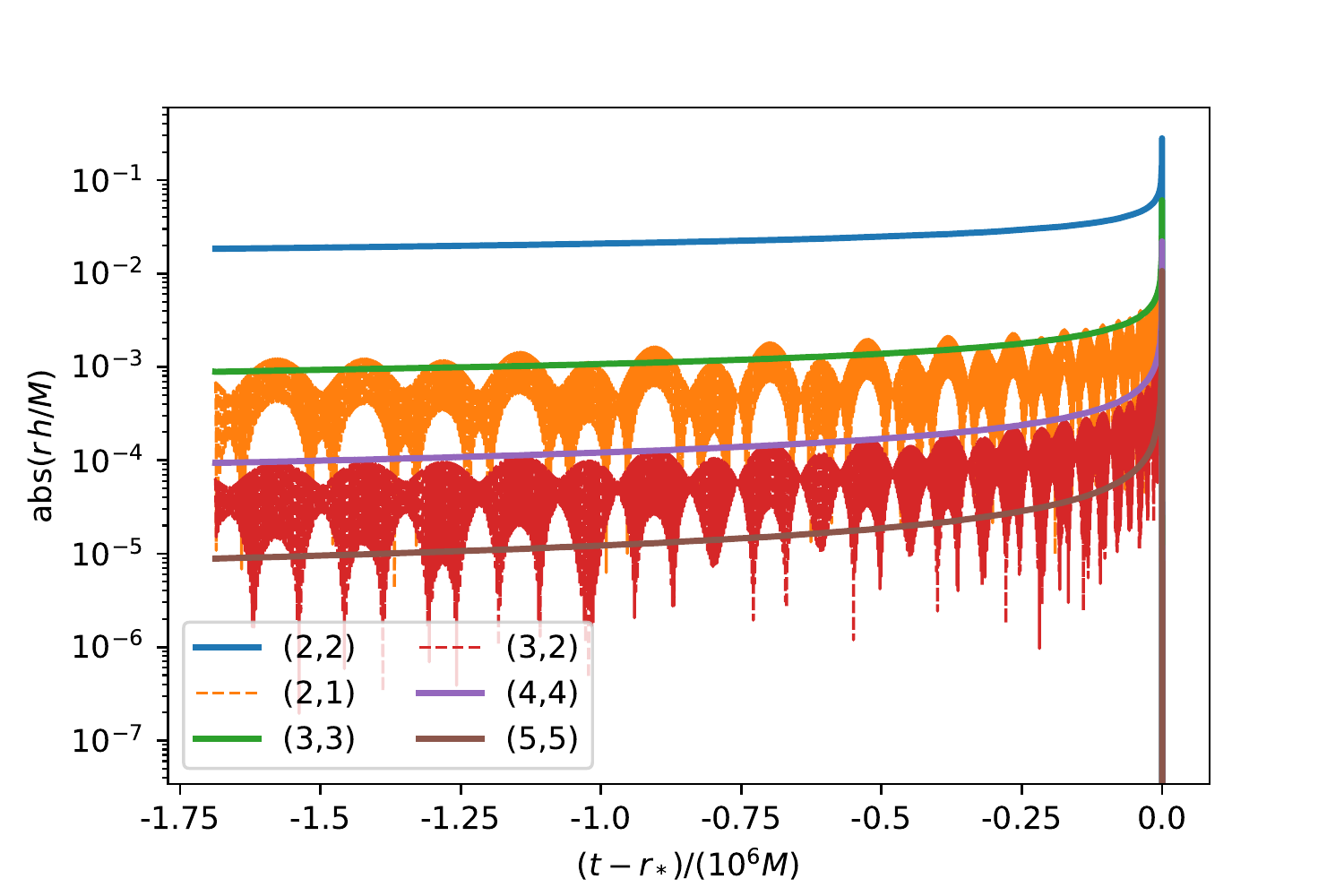}
  \includegraphics[width=.45\textwidth]{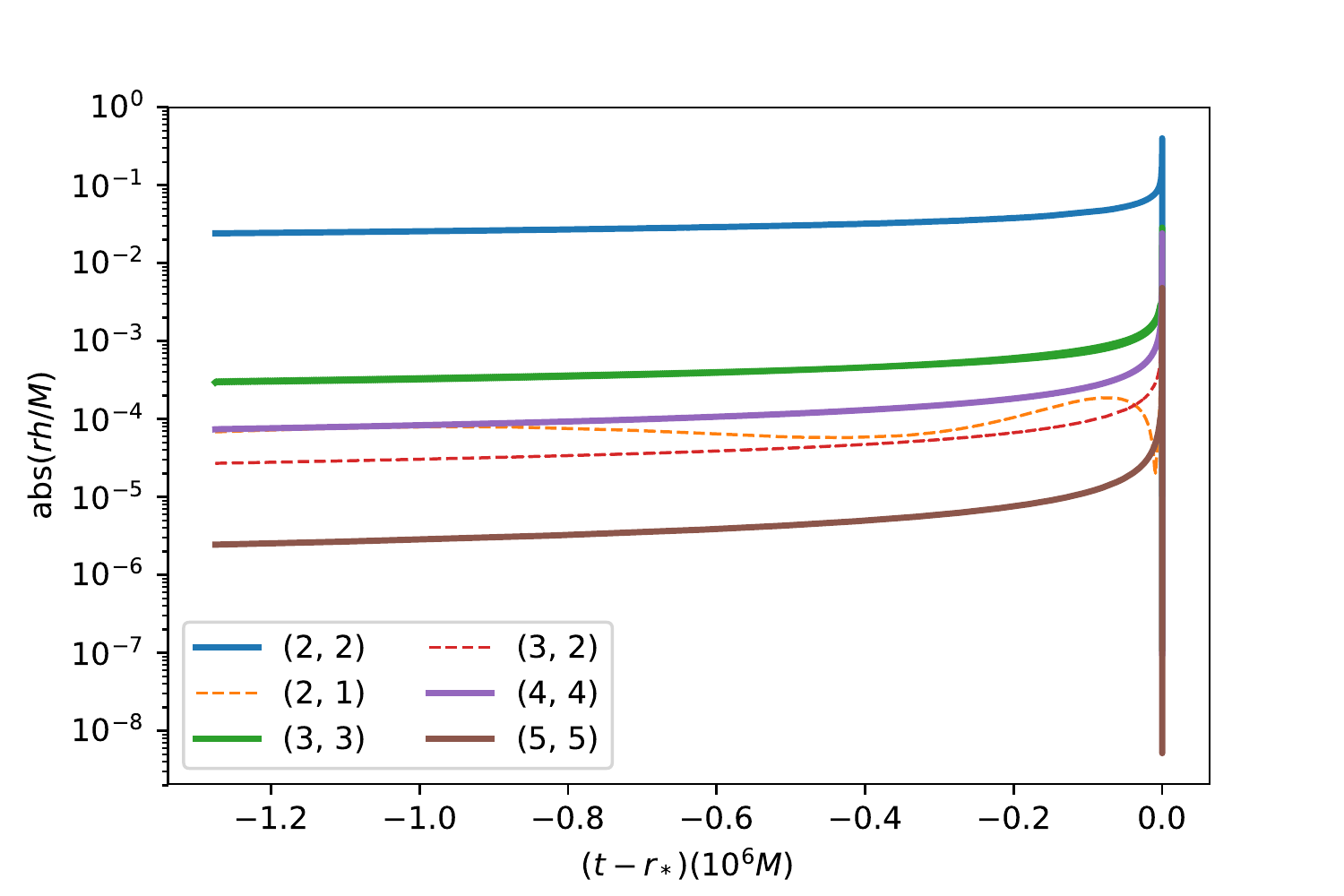}
  \includegraphics[width=.45\textwidth]{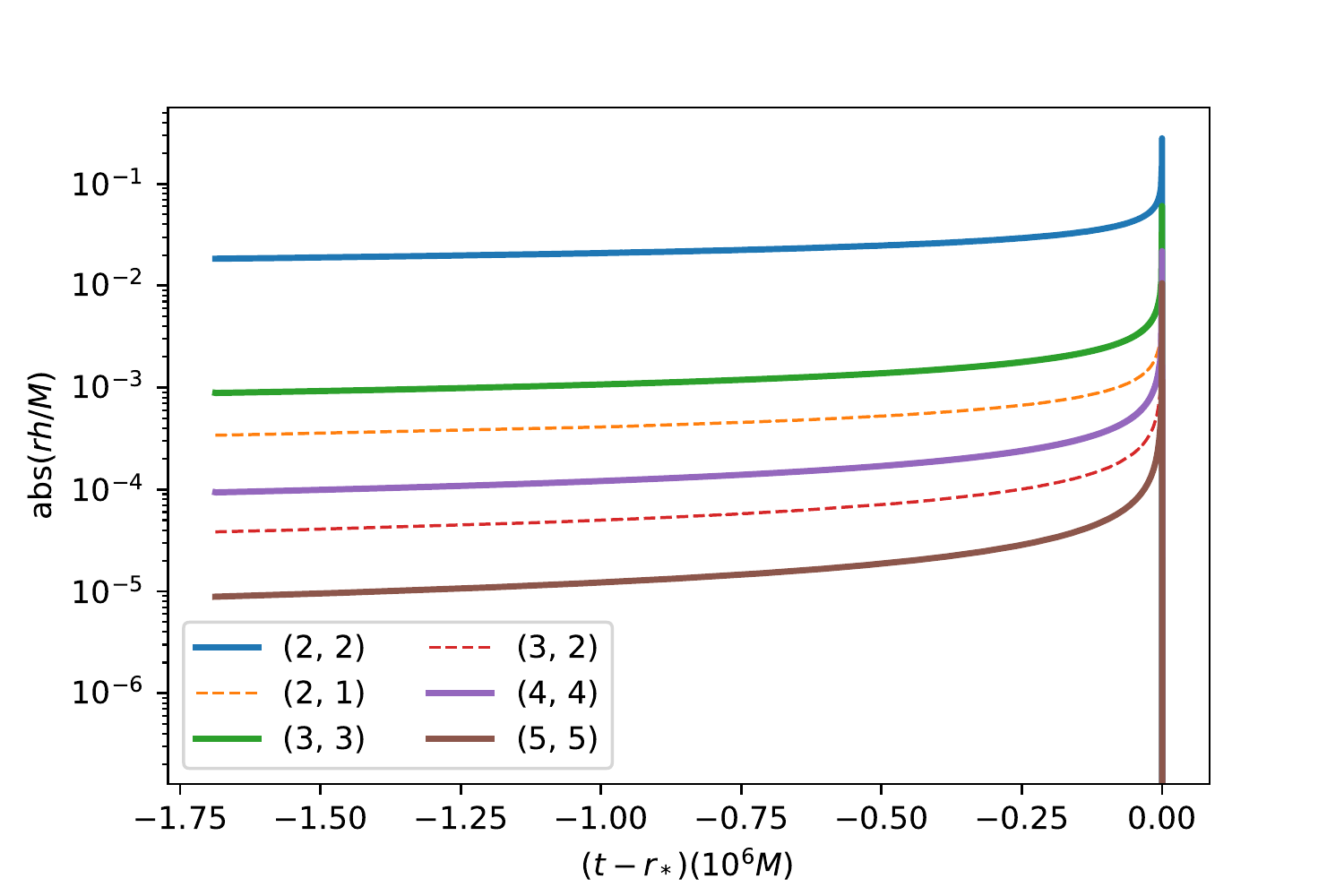}
  \caption{Selected modes for PN-NR hybrids with the \texttt{SpinTaylorT1} approximant.
  Waveform modes are shown in the inertial frame (top panel) and co-precessing frame
  (bottom panel).
  The hybrids start at an orbital frequency of $M\Omega = 0.002$.
   \emph{Left:} SXS\_BBH\_0308 \emph{Right:} SXS\_BBH\_0104
   Modes in the co-precessing frame are close to non-precessing waveforms, while inertial
   modes are modulated by the precession of the orbital plane.
  }
  \label{fig:hybrid_modes}
\end{figure*}

Figs.~\ref{fig:0308_hybrid} and~\ref{fig:0104_hybrid} show
selected waveform modes, the phase of the $(2,2)$ mode, the orbital angular momentum vector and the spin vectors for the two configurations used in this study. 
These plots demonstrate the good blending between the \ac{PN} and \ac{NR} data in the hybridization time region (gray shaded).
The absolute value of inertial frame modes for the hybrids are shown in~\ref{fig:hybrid_modes}.
Higher harmonics are stronger for the more unequal mass SXS\_BBH\_0104 configuration, whereas precession effects that give rise to modes like the $(2,1)$ and $(3,2)$ mode are stronger for SXS\_BBH\_0308.

For SXS\_BBH\_0308 there is a disagreement in the co-rotating frame (2,1) mode between PN and NR. 
This mode is weak since the system is almost equal mass which likely exacerbates the disagreement. In contrast, we find excellent agreement in the same mode for SXS\_BBH\_0104. 
The effect of this discrepancy for SXS\_BBH\_0308 is very small. 
The mismatch between the hybrid with and without the co-rotating frame $(2, \pm 1)$ modes is on the order of hybridization error and \ac{NR} error, about $\sim 10^{-5}$.

\begin{figure*}[t]
  \centering
  \includegraphics[width=.45\textwidth]{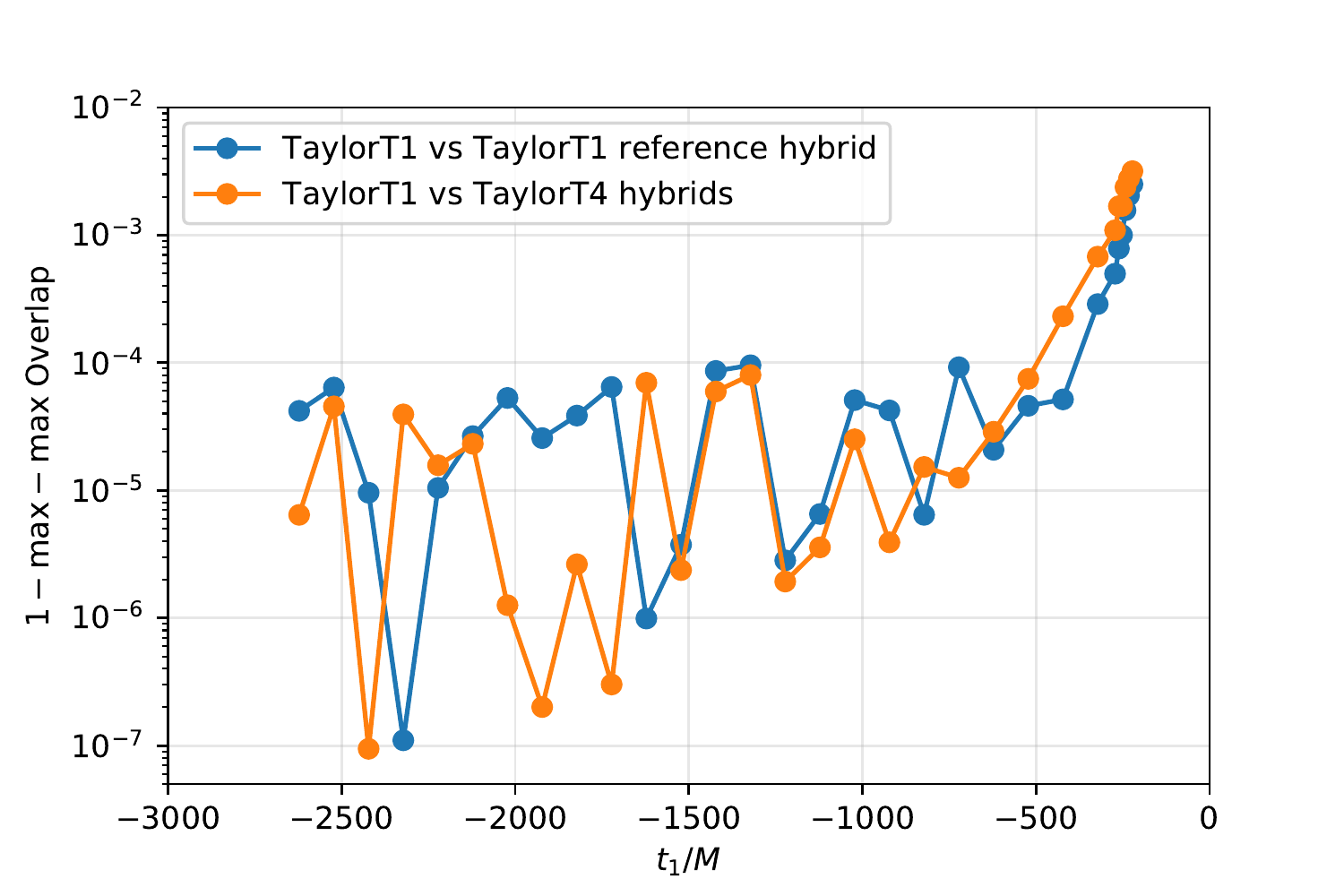}
  \includegraphics[width=.45\textwidth]{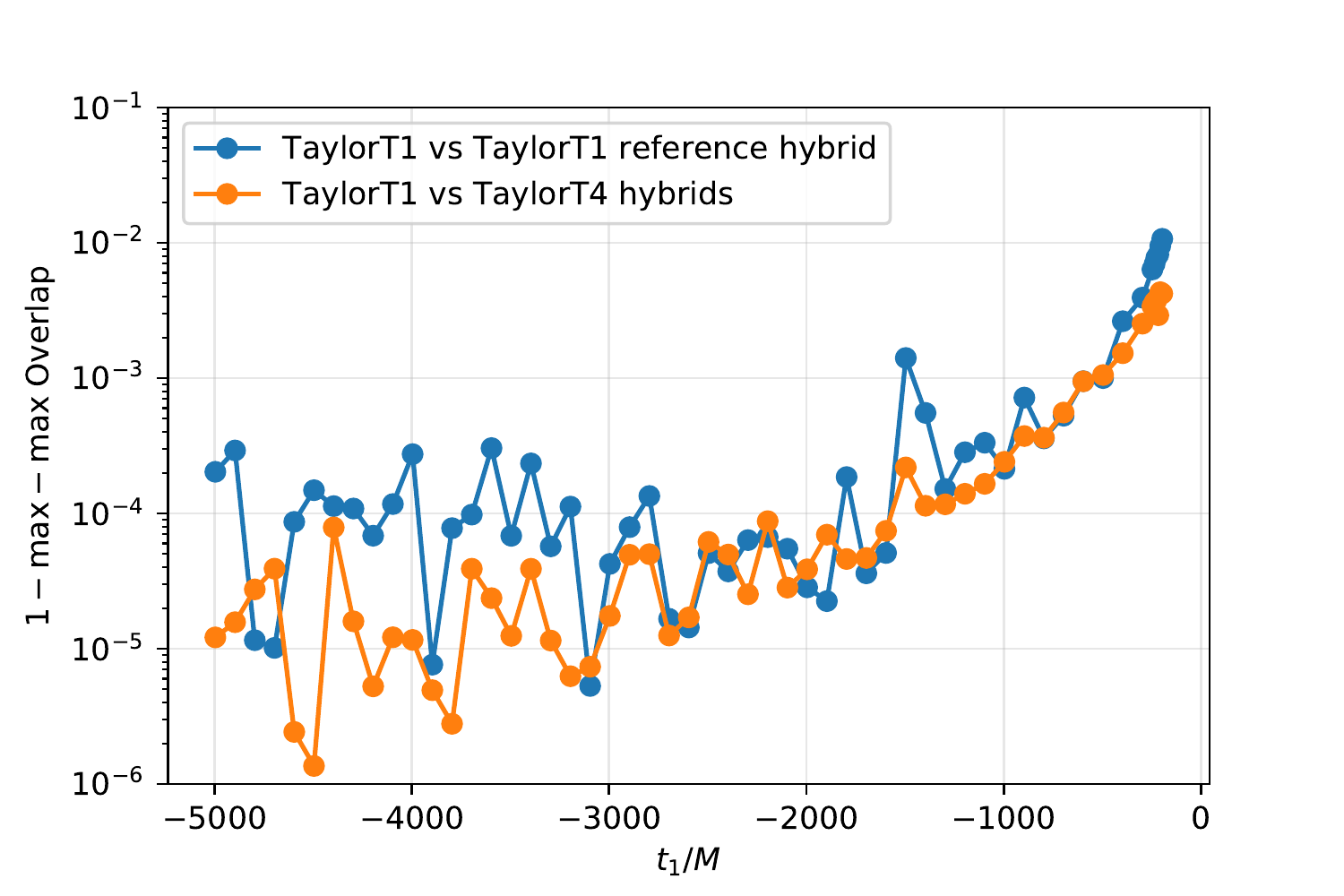}
  \caption{Mismatch between PN-NR hybrid waveforms.
  \emph{Left:} SXS\_BBH\_0308.
  \emph{Right:} SXS\_BBH\_0104.
  The mismatch is computed as one minus the $\max-\max$ overlap including higher modes up to $\ell = 8$ in the \ac{NR} waveforms. 
  Here, hybrids are constructed from an orbital frequency of $M\Omega = 0.01$ and the overlap integral starts at $10$ Hz and uses the aLIGO design \ac{PSD}. (Therefore, higher harmonics will be incomplete in the frequency band, but in a consistent manner. Degradation of the mismatch will come from high frequencies close to merger.)
  The blue curves show mismatches between \texttt{SpinTaylorT1}-NR hybrids constructed with a $100 M$ hybridization window as a function of the start time $t_1$ of this window before merger against a reference hybrid with a broader window at $t = [200, 800] M$ measured from the beginning of the NR waveform, and after the relaxation time. 
The binaries merge at $2822.24 M$ and $5196.2 M$ from the beginning of the \ac{NR} simulations, respectively for SXS\_BBH\_0308 and SXS\_BBH\_0104.
  The orange curves show mismatches between \texttt{SpinTaylorT1}-NR and \texttt{SpinTaylorT4}-NR hybrids constructed with the same $100 M$ hybridization window as a function of the start time $t_1$.
  }
  \label{fig:hybridization_error}
\end{figure*}

To study the error introduced by hybridizing \ac{PN} and \ac{NR} waveforms the optimal test would be to compute the mismatch of a hybrid against a very long high accuracy \ac{NR} waveform that fills the detector band. 
Since this is in practise not possible we perform the following experiments to make sure that hybridization errors are subdominant.
We compute overlaps between (a) a reference hybrid in the time window $t = [200, 800] M$, measured from the beginning of the NR waveform, and ``sliding hybrids'', a series of hybrids blended with $100M$ long time windows that approach the merger in discrete steps. 
We also compute (b) overlaps between sliding hybrids for the same window starting time between the \texttt{SpinTaylorT1} and \texttt{SpinTaylorT4} \ac{PN} approximants.
The reference hybrids are used as a signal waveforms for \ac{PE} in the main study of the paper.
We compute the max-max overlaps as defined in App. B of~\cite{Babak:2016tgq}.
We show the resulting overlaps for SXS\_BBH\_0308 and SXS\_BBH\_0104 in Fig.~\ref{fig:hybridization_error}.
Both curves show that if one hybridizes early the mismatch is small and noisy. These mismatches are lower than the mismatches between different \ac{NR} resolutions quoted in Sec.~\ref{sub:numerical_relativity_waveforms}. Therefore hybridization errors are subdominant for these configurations.
For SXS\_BBH\_0308 the mismatch only rises beyond $10^{-4}$ for windows that start within $500 M$ of the merger, while for SXS\_BBH\_0104 the mismatches approach $10^{-3}$ already $1000 M$ before merger. 
In this regime \ac{PN} waveforms become inaccurate compared to \ac{NR} and differences between \ac{PN} approximants grow. 

We also want to briefly mention additional sources of errors.
Spin vectors are defined differently in \ac{PN} and \ac{NR}~\cite{Ashtekar:2004cn,Kyrian:2007zz,Santamaria:2010yb} 
and therefore, using the same spin values for both waveforms at the same time as we do in the hybrid construction will introduce an additional error that we do not quantify here.
The $m = 0$ ``memory'' modes may not be accurate without using \ac{CCE}~\cite{Taylor:2013zia}.
The waveforms used in this study, SXS\_BBH\_0308 and SXS\_BBH\_0104 do not use \ac{CCE}.

The configurations considered in this study are fairly easy to hybridize and one should not infer a general behavior of hybridization errors from them. 
For more challenging configurations (higher mass-ratios and spins) \ac{PN} and \ac{NR} are expected to show discrepancies further away from merger. 
How long \ac{NR} waveforms need to be so that hybridization errors are subdominant requires detailed study~\cite{Hannam:2010ky,MacDonald:2011ne,Boyle:2011dy,Ajith:2012az}.

\bibliography{paper}

\end{document}